\documentstyle[aps,preprint,tighten,epsf]{revtex}
\begin{document}
\title{Sea quark effects in $B$ Spectroscopy and Decay Constants}
\author{S.~Collins\thanks{On leave of absence from the University of Glasgow.}}
\address{Deutsches Elektron-Synchrotron DESY, D-15735 Zeuthen, Germany}

\author{C.~T.~H.~Davies\thanks{UKQCD Collaboration.}}
\address{University of Glasgow, Glasgow, Scotland G12 8QQ}

\author{U.~M.~Heller}
\address{SCRI, Florida State University, Tallahassee, Fl 32306-4052, USA}

\author{A.~Ali~Khan\thanks{Present address: Center for Computational Physics, University of Tsukuba, Ibaraki 305-8577, Japan.} and J.~Shigemitsu}
\address{The Ohio State University, Columbus, Ohio 43210, USA}

\author{ J.~Sloan}
\address{University of Kentucky, Lexington, KY 40506-0055, USA}

\author{ C.~Morningstar}
\address{University of California at San Diego, La Jolla, CA 92093, USA}
\maketitle

\narrowtext
\begin{abstract}
  
  We present comprehensive results for the spectrum and decay
  constants of hadrons containing a single $b$ quark. The heavy quark
  is simulated using an $O(1/M)$ NRQCD action and the light quark
  using the $O(a)$ tadpole-improved clover action on gauge
  configurations containing two degenerate flavours of sea quarks at
  $\beta^{n_f=2}=5.6$ provided by the HEMCGC collaboration. We present
  detailed results for the lower lying $S$ and $P$ wave $B$ meson
  states and the $\Lambda_b$ baryon.  We find broad agreement
  with experiment.
  In addition, we present results for the pseudoscalar and, for the
  first time, the vector decay constants fully consistent to
  $O(\alpha/M)$: $f_B =
  186(5)(stat)(19)(pert)(9)(disc)(13)(\mbox{{\small
  NRQCD}})(+50)(a^{-1})~\mbox{MeV}$, $f_B^* =
  181(6)$ $(stat)(18)(pert)(9)(disc)(13)(\mbox{{\small
  NRQCD}})(+55)(a^{-1})~\mbox{MeV}$ and $f_{B_s}/f_B =
  1.14(2)$ $(stat)(-2)(\kappa_s)$. 
   We present an investigation of sea quark effects in the $B$
  spectrum and decay constants. We compare our results with those from
  similar quenched simulations at $\beta^{n_f=0}=6.0$. For the
  spectrum, the quenched results reproduce the experimental spectrum
  well and there is no significant difference between the quenched and
  $n_f=2$ results.  For the decay constants, our results suggest that
  sea quark effects may be large. 
\end{abstract}
\section{Introduction}
\label{intro}
Hadrons containing a single $b$ quark hold the key to many
important questions facing particle physics. In particular, the weak
decays of these particles are being studied to look for
inconsistencies in the Standard Model and indications of new physics
beyond it.  Lattice calculations have a central and fundamental role
to play in this pursuit. Not only is lattice theory a first principles
approach but also offers the most reliable way to calculate the masses
of heavy-light hadrons and the low-energy QCD factors which are needed
to extract the electroweak physics from experiment.  An introduction
to $B$ physics and the theoretical advances in this field, for example Heavy
Quark Effective Theory~(HQET), as well as the contribution made from
lattice theory, can be found in reference~\cite{bphysics}.

The aim of a lattice calculation of $B$ meson phenomenology is to
produce reliable predictions where the systematic errors, such as
finite lattice spacing, finite volume and quenching, are understood
and under control. Our approach is to systematically improve the
actions used for the heavy and light quark and to use lattices with
physical volumes large enough to accommodate at least the lower lying
$B$ mesons. In particular, we simulate the $b$ quark on the lattice
using NRQCD: a cut-off of the order of the heavy mass is imposed and
we cannot extrapolate to the continuum. This requires the systematic
errors to be reduced to the order of the statistical errors at finite
$\beta$. In this paper, we argue that the uncertainty arising from
finite $a$, volume, and the truncation of the NRQCD series are under
control and it is now reasonable to investigate the effects of
quenching.  As a first step towards predicting $B$-meson properties in
full QCD, we perform a simulation of the $B$ spectrum and decay
constants including two degenerate flavours of dynamical quarks.  We
then make a comparison with our previous NRQCD calculations on
quenched configurations~\cite{arifaspec2,arifajunk} and study the
sensitivity of various quantities to the presence of sea quarks.

The paper is organised as follows: in the next section we describe the
details of the simulation. Each choice of quark action and simulation
parameter leads to an associated systematic error and the
corresponding effect on predictions for the $B$ spectrum and decay
constants are discussed in section~\ref{errors}. Results for the
spectrum are presented in section~\ref{spec} and compared with
experiment. We then study the effect of sea quarks on the $B$
spectrum and $\Lambda_b$ in section~\ref{nfspectrum}. 

The next section deals with the pseudoscalar and vector decay
constants. The lattice operators required to compute the decay
constants to $O(1/M)$ and the corresponding $1$-loop perturbative
matching factors are introduced and discussed. In addition, the
expectations from Heavy Quark Symmetry~(HQS) for the heavy quark mass
dependence of these quantities is outlined. Our results for $f_B$,
$f_{B_s}$, $f_{B^*}$ and $f_{B_s^*}$ are presented in
section~\ref{decayresults} and their dependence on the heavy quark
mass is investigated.  Finally, we present a detailed study of sea
quarks effects in the decay constants.

\section{Simulation Details}
\label{simdet}

The simulations were performed using 100 $16^3\times32$ gauge
configurations at $\beta=5.6$ with two flavours of staggered dynamical
sea quarks with a bare quark mass of $am_{sea}=0.01$, which roughly
corresponds to $m_\pi/m_\rho=0.525$.  These configurations were
generously made available by the \mbox{HEMCGC} collaboration; more
details can be found in~\cite{hemcgc}. We fixed the configurations to
the Coulomb gauge.

The light quark propagators were generated using the clover fermion
action at three values of the hopping parameter, $\kappa=0.1385$,
$0.1393$ and $0.1401$.  From an analysis of the light hadron
spectrum~\cite{sloanjap}, the second $\kappa$ value corresponds to a
quark mass close to strange, where $\kappa_s=0.1392(1)$ from $M_{K}$,
$\kappa_s=0.1394(1)$ from $M_{K^{*}}$ and $0.1392(1)$ from $M_\phi$;
$\kappa_c=0.1408$. The $O(a)$ improvement term in the clover action is
implemented with a tadpole-improved value for the clover coefficient,
$c_{sw}$. This amounts to dividing all the gauge links by $u_0$, where
we use $u_0=0.867$ measured from the plaquette, and setting the
coefficient to the tree-level value $c_{sw}=1$. In order to improve
matrix elements to the same order as the spectrum, we use the
prescription proposed by the Fermilab group~\cite{kronfeld} and
replace the quark field normalisation $\sqrt{2\kappa}$ with
$\sqrt{1 - 6\tilde{\kappa}}$, where $\tilde{\kappa}=u_0\kappa$.

In this simulation we truncate the NRQCD series at $O(1/M_0)$, where
$M_0$ is the bare heavy quark mass, and the
action takes the form:
\begin{equation}
S = \psi^{\dagger}(D_t + H_0 + \delta H) \psi
\end{equation}
where 
\begin{equation}
H_0 = -\frac{\Delta^{(2)}}{2M_0}\hspace{0.5cm} \mbox{and} 
\hspace{0.5cm} \delta H = -c_B\frac{\sigma\cdot B}{2M_0}.
\end{equation}
Tadpole improvement of the gauge links is used throughout and the hyperfine
coefficient is given the tree-level value $c_B=1$. We use the standard
clover-leaf operator for the $B$ field. The heavy quark propagators were
computed using the evolution equation~\cite{lepage}:
\begin{eqnarray}
G_{t+1} & = & \left(1-\frac{a\delta H}{2}\right)\left(1-\frac{aH_0}{2n}\right)^n 
U_4^{\dagger} \left(1-\frac{aH_0}{2n}\right)^n 
\left(1-\frac{a\delta H}{2}\right)G_t \hspace{0.5cm} 
\label{evol}
\end{eqnarray}
for all $t$, where $n$ is the stabilising parameter.

We generated heavy quark propagators at 10 values of ($aM_0$,n) corresponding
to (0.8,5), (1.0,4), (1.2,3), (1.7,2), (2.0,2), (3.0,2), (3.5,2), (4.0,2),
(7.0,1) and (10.0,1).  This roughly corresponds to a range of meson masses
from $M_B/3$ to $4M_B$ and is sufficient for a reasonable investigation of
heavy quark symmetry. Results in the static limit are not presented. We were
not able to improve on the quality of the signal compared to previous results
using Wilson light fermions~\cite{spect,decay}, and in this case the
extrapolation of the NRQCD results to the static limit was considered more
reliable than results at the static point itself. To improve statistics we
have also calculated the propagators on the time-reversed configurations.

Details of the fitting analysis and extraction of the spectral
quantities and decay constants from the heavy-light meson correlators
can be found in the appendix.

\section{Systematic errors}
\label{errors}
In this section we discuss the systematic errors associated with our
choice of quark actions and simulation parameters.  In previous work
we used Wilson light fermions which was estimated to introduce the
largest systematic error into the calculation of heavy-light
quantities. For tadpole-improved clover fermions the systematic errors
are now estimated to be a relative error of $O((a\Lambda_{QCD})^2)\sim
4\%$~(where we take $a\Lambda_{QCD}= a\Lambda_V = 0.185$ for these
configurations) in quantities of $O(\Lambda_{QCD})$.  However,
significant residual $O(\alpha a\Lambda_{QCD})$ errors may remain in
this simulation. An investigation of the scaling behaviour of quenched
light hadron masses using a nonperturbative determination of the
$O(a)$ improvement coefficient~\cite{luscher} in the clover action
concluded that these errors are negligible for
$\beta^{n_f=0}\raisebox{-.6ex}{$\stackrel{\textstyle{>}}{\sim}$}
6.0$~\cite{hartmut}. The remaining scaling violations at these
$\beta$'s are estimated to be a few percent~\cite{hartmut} in the
quenched results. In the present study tadpole-improved clover
fermions are used for the valence quarks and staggered fermions for
the dynamical quarks~(which introduce $O(\alpha a^2)$ errors).  Since
$\beta^{n_f=2}=5.6$ corresponds to roughly the same lattice spacing as
that at $\beta^{n_f=0}=6.0$~(see table~\ref{latt}) we do not expect
the associated scaling violations to be larger than in the quenched
case.

The truncation of the NRQCD series at $O(1/M)$ introduces an absolute error of
$O(\Lambda_{QCD}(\Lambda_{QCD}/M)^2)$.  This corresponds to approximately
$1\%$ errors in quantities of $O(\Lambda_{QCD})$ and $10\%$ errors in the
coefficients of the heavy quark expansion at $M_B$. In fact the significance
of the higher order terms omitted from the action depends on the quantity
considered. The additional terms to $O(1/M^2)$ and the leading $1/M^3$ terms
are:
\begin{eqnarray}
\delta H_{h.o.} & = & -c_3\frac{g}{8(M_0)^2}\sigma\cdot(\nabla\times E - E\times\nabla)
 +c_2\frac{ig}{8(M_0)^2}(\nabla\cdot E- E\cdot\nabla)
\nonumber\\
&  & -c_1\frac{(\Delta^{(2)})^2}{8(M_0)^3} + c_4\frac{a^2\Delta^{(4)}}{24M_0}
- c_5\frac{a(\Delta^{(2)})^2}{16n(M_0)^2},
\label{highterms}
\end{eqnarray}
where $\nabla$ is the symmetric gauge-covariant lattice derivative and
$\Delta^{(4)}$ is a discretised version of the continuum operator $\sum
D_i^4$.  The first two terms, of $O(1/M^2)$, represent the spin-orbit
interaction~(s.o.) and the Darwin term, respectively. The s.o. interaction
only contributes to states with non-zero angular momentum and is a
$1/M^2\sim1\%$ correction to the s.o. interaction arising from the light
quark, which dominates~(see the next section).  The Darwin term is a $\sim
1\%$ spin-independent shift which appears in both $S$ and $P$ states, but
which does not affect spin splittings, such as the hyperfine splitting.  The
next order terms at $O(1/M^3)\sim0.1\%$ are due to corrections to the kinetic
energy and discretisation corrections and should be even less significant.  Of
course, all the terms in $\delta H_{h.o.}$ will affect predictions through the
change in the meson mass. However, for example the $1/M^2$ terms represent a
$.1\%$ correction to the meson mass, and hence, the shift in the meson mass
has a less significant effect on predictions than those already mentioned.

Indeed, a comparison by Ishikawa et al~\cite{japanese} of the low
lying $S$-state heavy-light spectrum derived from an NRQCD action
correct to $O(1/M)$ and that including the additional terms in
equation~\ref{highterms} showed the higher order terms to be
insignificant. A more detailed analysis of the effect of higher order
terms on the $S$-state spectrum has been performed by Lewis and
Woloshyn~\cite{lewis}. These authors verified that the importance of
the terms in the NRQCD action follows the expectations of naive power
counting, i.e. $|O(1/M^3)|<|O(1/M^2)|<|O(1/M)|$, where all $O(1/M^3)$
terms were included. Terms of $O(1/M^2)$ and $O(1/M^3)$ were found to
have a negligible effect on the $S$-state $B$ meson mass splittings.

Ishikawa et al also compared the results for the pseudoscalar and
vector decay constant including tree-level terms to $O(1/M)$ and
$O(1/M^2)$~(for these quantities, corrections to the currents as well
as to the action must be considered). The decay constants differed
by approximately $3\%$ at the $B$ meson, and, $6\%$ is,
conservatively, estimated to be the corresponding error in $f_B$ if
$\delta H_{h.o.}$ and the $O(1/M^2)$ current corrections are omitted.

An additional uncertainty is introduced through the use of the
tree-level value~(with tadpole-improvement) for the hyperfine
coefficient, $c_B$. Tadpole improvement is expected to account for
most of the renormalisation of this factor. However, an error of
$O(\alpha\Lambda_{QCD}(\Lambda_{QCD}/M))$ will remain; for example
this corresponds to approximately $10{-}30\%$, for $\alpha\sim
0.1{-}0.3$ in the $B^*{-}B$ splitting, which is proportional to
$c_B$. The one-loop correction to $c_B$ has been
calculated~\cite{trottier} and for $M_0=1.9$, a bare quark mass close
to $M^b_0$ for this simulation, $c_B\sim 1.15{-}1.3$, depending on the
characteristic scale $q^*$ used.

For the determination of $f_B$ and $f_{B^*}$, the factors matching the
lattice currents to the full continuum QCD currents must be
calculated. Morningstar and Shigemitsu have computed these factors for
both the axial-vector~\cite{junkmorn1} and vector~\cite{junkmorn2}
currents.  The mixing at $O(\alpha)$ among heavy-light current
operators has been fully taken into account. In addition, an
$O(\alpha(a\Lambda_{QCD}))$ discretisation correction to the current
was found and corrected for in the case of clover light fermions.
This has been discussed in detail in~\cite{arifajunk,junkjap}, and we
present an outline in section~\ref{decaysection}. The largest
remaining perturbative error is $O(\alpha^2)\sim 1{-}10\%$.

Combining the tadpole-improved clover action for light quarks with the
$O(1/M^2)$ NRQCD action, Hein et al~\cite{joachim} found no
significant scaling violations in results for the quenched $B$
spectrum and decay constant between $\beta^{n_f=0}=6.0$ and $5.7$.
The results from a recent quenched calculation by Ishikawa et
al~\cite{japscaling} of $f_B$ at three $\beta$ values in the range
$5.7{-}6.1$ supports this finding. 

A further continuing uncertainty is the possibility of large finite
volume effects. A study by Duncan et al~\cite{eichten} in the static
limit in the quenched approximation found no significant finite volume
effects in the results for the lowest lying S-state $B$ mesons~(and in
particular for $f_B$) for a box size as small as $1.3$~fm. Finite
volume effects are likely to be larger when dynamical quarks are
included~\cite{finitesize}. However, since for our simulation,
$L=1.6$~fm, for a visible effect, finite volume problems would have to
be very different in the dynamical case, despite the fairly heavy sea
quark masses~($m_{sea}$ is around the strange quark mass) that we use.
A preliminary study by the SESAM collaboration~\cite{henning}, using
sea quark masses which correspond to $m_\pi/m_\rho=0.69$, found the
finite volume effects for the $\rho$ meson~(which is bigger physically
than the lowest lying $S$-state $B$ mesons) to be small, less than
$5\%$ for a lattice size similar to that used in this work.  However,
in the absence of a detailed investigation of finite volume effects
for $B$ mesons at $n_f=2$ this remains an unknown effect.

Table~\ref{syserrors} summarises the systematic errors expected to
dominate each mass splitting and combination of decay constants.  In
general, our statistical errors are of a comparable size. The only
significant exception is for the radial
and orbital excitations of the $B$ meson and the $\Lambda_b$
baryon. These particles are probably squeezed on a box of extent
$1.6fm$, although for $B(1P){-}B(1S)$ and $B(2S){-}B(1S)$ our
statistical errors are also large.

While within the $B$ system we are close to achieving our goal of
controlling the main systematic errors, a major uncertainty is
introduced when we convert from lattice numbers to physical
predictions. In particular, this uncertainty affects predictions
through the fixing of the quark masses~(discussed below) as well as
the final conversion from lattice units to MeV.  Ideally one would fix
the lattice spacing using a quantity within the $B$ system. However,
with the exception of the lowest lying $S$-states, the experimental
results for the $B$ spectrum are still preliminary and it is
unfeasible to fix $a$, for example, from the $P{-}S$ or $2S-1S$
splitting~(which have little dependence on $m_Q$). Thus, one must
consider determinations of $a$ from light and heavy spectroscopy;
table~\ref{latt} gives some examples for these configurations.

As in the discussion above for heavy-light hadrons the effect of
systematic errors on the value of the lattice spacing must be
considered. We computed $m_\rho$ using the tadpole-improved clover
action and results for this quantity from quenched studies are
consistent with minimal discretisation errors in this quantity for
$a^{-1}\sim 2$~GeV~\cite{hartmut}. Similarly, the $b\bar{b}$ spectrum
was computed with an $Mv^4$ NRQCD action, and in quenched studies the
spin-averaged splittings obtained using this action show no scaling
violations in the same range of lattice spacings~\cite{nrqcdscaling}.

Finite volume effects may affect $m_\rho$ for a box size of
$(1.6~\mbox{fm})^3$~(this is not expected to be an issue for heavy
quarkonia). These effects are not thought to be significant for a
calculation of $m_\rho$ with similar parameters in the quenched
approximation~(such as that detailed in
table~\ref{latt})~\cite{gottlieb}. However, finite volume problems are
expected to be larger when dynamical quarks are
introduced~\cite{finitesize}. Naively, one expects $m_\rho$ to be
overestimated, and hence $a^{-1}$ underestimated if the meson is
squeezed on a finite lattice. Although the preliminary SESAM study,
mentioned above, suggests this may not be significant, finite volume
dependence remains an uncertainty.

Table~\ref{latt} indicates there is a large discrepancy between
$a^{-1}$ determined from light and heavy spectroscopy for the HEMCGC
configurations. In the quenched approximation, different lattice
spacings are expected from quantities dominated by different physical
scales and one must choose the most relevant quantity with which to
fix $a$. When dynamical quarks are partially introduced some
convergence in the estimates of $a^{-1}$ is expected~(although with
$m_{sea}\sim m_{strange}$ significant residual quenching effects will
probably remain) and the choice of quantity to determine $a$ should be
less important.  However, the range in $a^{-1}$ is comparable to that
seen in the quenched approximation at $\beta^{n_f=0}=6.0$, also
detailed in the table~(the same light quark and NRQCD actions were
employed); $1P{-}1S/m_\rho=.43(1)$ and $.46(1)$ for $n_f=0$ and $2$
respectively, compared to $0.57$ from experiment.  One possibility is
that introducing two flavours of sea quarks with $m_{sea}\sim
m_{strange}$ produces little effect on $m_\rho/\Upsilon(1P{-}1S)$.
Alternatively, any effect of the sea quarks may be counteracted by
more significant finite volume problems~(or some other systematic
error dependent on the number of sea quarks) at $n_f=2$ for $m_\rho$
compared to $n_f=0$.

Until this issue has been clarified, we choose the lattice spacing from
$m_\rho$ to convert to physical units at $n_f=2$ and use
$a^{-1}=2.4$~GeV to give a~(conservative) indication of the
uncertainty in $a^{-1}$. This translates into an error in physical
predictions at least as large as those already mentioned. We comment
on how to compare our results for heavylight hadrons with those from
quenched simulations below.

To pin-point the $B$ and $B_s$ mesons from the range of meson masses
simulated in this study we must fix the bare quark masses
corresponding to the lightest quark~($u$, $d$), the strange quark and
the bottom quark. The bare mass of the lightest quark is fixed in the
standard way by extrapolating the pion mass to zero.
As indicated above, $m^0_s$ is derived from the
$\phi$, $K$ and $K^*$ mesons, and we use the difference in the results
as an estimate of the uncertainty in $\kappa_s$. The magnitude of this
error found in the $B_s{-}B$ splitting and the ratio $f_B/f_{B_s}$ is
given in table~\ref{syserrors}. In order to determine $M^0_b$ we must
first correct the meson simulation energies for the removal of the
mass term in the NRQCD action. As detailed in~\cite{spect} we use the
mass shift calculated from heavy quarkonia dispersion behaviour to
obtain the heavy-light meson masses from the simulation energies. Thus,
interpolating between our results for the pseudoscalar meson until we
obtain the $B$ meson mass, we find $aM_0^b=2.1{-}1.8$ for
$a^{-1}=2.0{-}2.4$~GeV.

In order to investigate sea quark effects we need to compare our
results with similar calculations in the quenched approximation. While
we believe the dominant systematic errors are under control, most of the
remaining residual uncertainties are minimised if we compare with a
simulation which uses a lattice with a similar lattice spacing and
physical volume. From table~\ref{latt} $\beta=6.0$ in the quenched
approximation roughly matches $\beta=5.6$ at $n_f=2$.
References~\cite{arifajunk} and~\cite{arifaspec2} report on NRQCD
results using tadpole-improved clover light quarks at this $\beta$, on
a lattice volume of $\sim (1.6fm)^3$. A higher-order NRQCD action is
employed for the heavy quarks~($H_{h.o.}$ is included), however,
considering the negligible effect these terms have at the $B$ meson
mass, these quenched results are suitable for comparison.

For the quenched simulation the lattice spacing derived from light
spectroscopy is considered most relevant for $B$ splittings and decay
constants which are expected to be dominated by the `brown muck'.
Thus, we initially compare the two simulations fixing the lattice
spacing from $m_\rho$ in both cases. The effect on the comparison of
the large uncertainty in $a^{-1}$ for $n_f=2$ is also considered.

Within our study it is premature to attempt an extrapolation of the
quenched and unquenched data to a physically relevant number of sea
quark flavours. We introduce uncertainties by implementing clover
light valence fermions while using staggered sea fermions. In
addition, we chirally extrapolate the valence light quarks to zero
light quark mass while keeping the sea quark mass fixed around the
strange quark mass. Thus, our simulation does not correspond to that
for a light and heavy quark bound state with a sea of two flavours of
physical quarks. However, inserting dynamical quarks allows the
sensitivity of various quantities to these effects to be
investigated. A more systematic approach to sea quark effects is
reserved for later work.

\section{Spectrum Results}
\label{spec}

\begin{figure}
\begin{center}
\setlength{\unitlength}{.025in}
\begin{picture}(130,100)(30,500)
\put(15,500){\line(0,1){100}}
\multiput(13,500)(0,50){3}{\line(1,0){4}}
\multiput(14,500)(0,10){11}{\line(1,0){2}}
\put(12,500){\makebox(0,0)[r]{{\large5.0}}}
\put(12,550){\makebox(0,0)[r]{{\large5.5}}}
\put(12,600){\makebox(0,0)[r]{{\large 6.0}}}
\put(12,570){\makebox(0,0)[r]{{\large GeV}}}
\put(15,500){\line(1,0){180}}

     \put(25,510){\makebox(0,0)[t]{{\large $B$}}}
     \put(26,528){\circle*{3}}
     \put(31,528){\circle{3}}
     \multiput(20,527.9)(3,0){4}{\line(1,0){2}}
     \put(26,588.3){\circle*{3}}
     \put(26,588.3){\line(0,1){8.3}}
     \put(26,588.3){\line(0,-1){8.3}}
     \put(37,595){\makebox(0,0)[t]{{\large $(2S)$}}}
     \multiput(20,586)(3,0){4}{\line(1,0){0.5}}
     \put(32,582){\circle{3}}
     \put(32,582){\line(0,1){12}}
     \put(32,582){\line(0,-1){12}}

     \put(55,510){\makebox(0,0)[t]{{\large $B^{*}$}}}
     \put(56,530.3){\circle*{3}}
     \multiput(50,532.6)(3,0){4}{\line(1,0){2}}
     \multiput(50,532.4)(3,0){4}{\line(1,0){2}}
       \put(63,530){\circle{3}}
       \put(63,530){\line(0,1){.6}}
       \put(63,530){\line(0,-1){.6}}
       \put(63,593){\circle{3}}
       \put(63,593){\line(0,1){12}}
       \put(63,593){\line(0,-1){12}}

     \put(80,510){\makebox(0,0)[t]{{\large $B_s$}}}
     \put(80,536.7){\circle*{3}}
     \multiput(75,537.0)(3,0){4}{\line(1,0){2}}
     \put(80,593){\circle*{3}}
     \put(80,593){\line(0,1){3.8}}
     \put(80,593){\line(0,-1){3.8}}
     \put(92,601){\makebox(0,0)[t]{{\large $(2S)$}}}
    \put(85,538){\circle{3}}
     \put(85,538){\line(0,1){1}}
     \put(85,538){\line(0,-1){1}}
    \put(85,588){\circle{3}}
     \put(85,588){\line(0,1){8}}
     \put(85,588){\line(0,-1){8}}

     \put(105,510){\makebox(0,0)[t]{{\large $B^{*}_s$}}}
     \put(106,539.){\circle*{3}}
     \multiput(100,541.4)(3,0){4}{\line(1,0){2}}
     \multiput(100,542.)(3,0){4}{\line(1,0){2}}
    \put(110,540.6){\circle{3}}
     \put(110,540.6){\line(0,1){1}}
     \put(110,540.6){\line(0,-1){1}}

     \put(142,510){\makebox(0,0)[t]{{\large $P-States$}}}
     \put(138,570.7){\circle*{3}}
     \put(138,572.2){\line(0,1){1.9}}
     \put(138,569.2){\line(0,-1){1.9}}
    \put(145,580){\circle{3}}
    \put(145,580){\line(0,1){4}}
    \put(145,580){\line(0,-1){4}}
     \put(130,576){\makebox(0,0){{\large $(\overline{B^*_1})$}}}
     \multiput(128,568.6)(3,0){8}{\line(1,0){2}}
     \multiput(128,571.0)(3,0){8}{\line(1,0){2}}
     \multiput(128,577.9)(3,0){8}{\line(1,0){0.5}}

     \put(182,510){\makebox(0,0)[t]{{\large $\Lambda_b$}}}
     \put(182,566.5){\circle*{3}}
     \put(182,565.0){\line(0,-1){7.1}}
     \put(182,568.0){\line(0,1){7.1}}
     \multiput(176,562.4)(3,0){4}{\line(1,0){2}}
     \put(189,583){\circle{3}}
     \put(189,583){\line(0,1){4}}
     \put(189,583){\line(0,-1){4}}

\end{picture}
\end{center}
\vspace{0.5cm}
\caption{The lower lying $B$ spectrum and $\Lambda_b$. The open
  circles denote our results.  The dashed lines denote the upper and
  lower bounds on the experimental results. The dotted lines indicate
  preliminary experimental signals.  For the $P$ states the dashed and
  dotted lines represent the $B^{(*)}\pi$ and the narrow $B\pi\pi$
  resonance respectively.  The errors shown are purely statistical.
  The filled circles represent preliminary results from a quenched
  simulation at $\beta^{n_f=0}=6.0$~\protect\cite{arifaspec2}. All
  results are converted to physical units using the scale from
  $m_\rho$, given in table~\protect\ref{latt}.}
\label{spectrum1}
\end{figure}
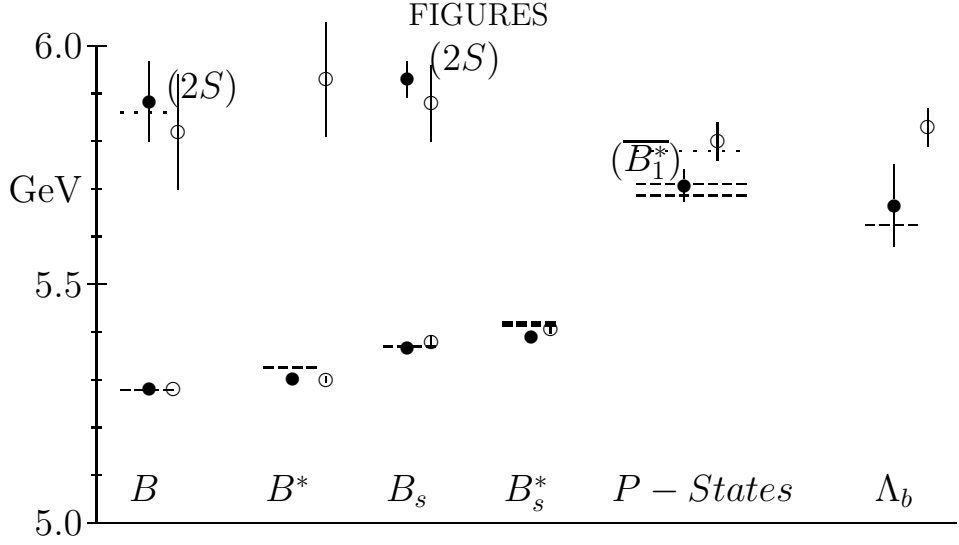

Figure~\ref{spectrum1} presents our results for the lower lying $B$
meson spectrum and the $\Lambda_b$ baryon compared to experiment.  At
this initial stage our results are in broad agreement with experiment.
The experimental results for all but the lowest states are still
uncertain.  In particular, there are only preliminary signals for the
$2S$ and $P$ states. For the latter, a signal has been found for a
$B\pi\pi$ resonance~\cite{data1}, which is probably has $j_l=3/2$, and
a $B^{(*)}\pi$ resonance~\cite{pdg}, which is likely to be a
superposition of various $P$ states. Our result for the $P$
state~($\bar{B}^{*}_1$) is obtained from an operator with quantum
numbers $^1P_1$ in the $^{2S+1}L_J$ nomenclature of quarkonia. Since
charge conjugation is not a symmetry in the heavy-light system the two
$l=1$, $J=1$ states mix and thus our operator has an overlap with both
states. We investigated this mixing by forming a matrix of correlators
with the $^3P_1$ operator. However, these operators were found to be
degenerate at this level of statistics.  In fact, we are unable to
resolve any splitting between the $P$ states in this study.

\begin{figure}
\setlength{\unitlength}{.03in}
\begin{picture}(350,110)(0,930)
\put(8,940){\line(0,1){115}}
\put(8,940){\line(1,0){160}}

\put(15,963){\makebox(0,0)[t]{$1S$}}
\put(29,935){\makebox(0,0)[t]{$B$}}
\multiput(23,955)(3,0){5}{\line(1,0){3}}

\put(13,1034){\makebox(0,0)[t]{$2S$}}
\multiput(23,1026)(3,0){5}{\line(1,0){3}}

\put(63,935){\makebox(0,0)[t]{$B^*$}}
\multiput(53,963)(3,0){5}{\line(1,0){3}}

\multiput(53,1034)(3,0){5}{\line(1,0){3}}

\put(47,950){\makebox(0,0)[t]{$j_l=\frac{1}{2}$}}

\put(82,963){\makebox(0,0)[t]{\}$1/m$}}
\put(82,1034){\makebox(0,0)[t]{\}$1/m$}}

\put(102,1005){\makebox(0,0)[t]{$1P$}}
\put(135,935){\makebox(0,0)[t]{$B^{**}$}}

\put(133,1029){\makebox(0,0)[t]{$j_l=\frac{3}{2}$}}
\put(120,1008){\makebox(0,0)[t]{\small{J=1}}}
\put(120,1016){\makebox(0,0)[t]{\small{J=2}}}
\multiput(130,1007)(3,0){5}{\line(1,0){3}}
\multiput(130,1012)(3,0){5}{\line(1,0){3}}
\put(160,1013){\makebox(0,0)[t]{$\}1/m$}}

\put(133,984){\makebox(0,0)[t]{$j_l=\frac{1}{2}$}}
\put(120,993){\makebox(0,0)[t]{\small{J=0}}}
\put(120,999){\makebox(0,0)[t]{\small{J=1}}}
\multiput(130,995)(3,0){5}{\line(1,0){3}}
\multiput(130,991)(3,0){5}{\line(1,0){3}}
\put(160,996){\makebox(0,0)[t]{$\}1/m$}}

\end{picture}
\vspace{0.5cm}
\caption{The theoretical prediction for the structure of the $B$
spectrum~\protect\cite{isgurwise}. $j_l$ and $J$ denote the
total spin of the light quark and the heavy-light meson respectively.}
\label{spectrumt}

\end{figure}
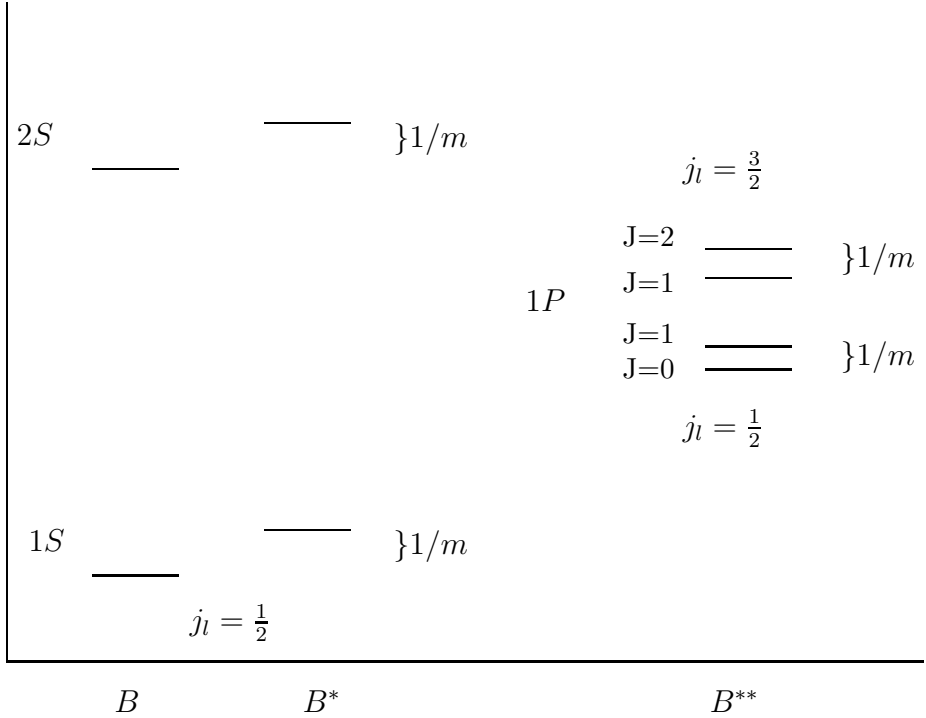

In addition, we can compare with the theoretical expectations from
Heavy Quark Symmetry~(HQS), shown in
figure~\ref{spectrumt}~\cite{isgurwise}.  The picture of a heavy quark
surrounded by a light quark cloud, where the heavy quark acts merely
as a colour source in the heavy quark limit, predicts a gross spectrum
determined by the light quark degrees of freedom; this gives rise to
meson mass splittings of  $O(\Lambda_{QCD})$ which are independent of
the heavy quark at lowest order.  In particular, the large scale
features of the spectrum are due to the radial~($2S$) and
orbital~($1P$) excitation of the light quark. At the next order are
the $j_l=3/2$ and $1/2$ doublets of the $P$ states, which are split
due to the spin-orbit interaction.  The $2S{-}1S$ and $P{-}S$
splittings in figure~\ref{spectrum1} are $300{-}600$~MeV or
$O(\Lambda_{QCD})$, in agreement with this naive picture.  Similarly,
heavy-light hadrons which only differ from the $B$ meson in the light
quark flavour, for example $B_s$, or the number of
light quarks, for example $\Lambda_b$, will also give rise to
splittings independent of $m_Q$ at lowest order.

At finite $m_Q$, flavour and spin symmetry are broken by the kinetic
energy of the heavy quark and the hyperfine interaction. The latter
removes the degeneracy in the heavy quark spin, for example in the
$j_l=1/2$ S-states and the $j_l=3/2$ and $j_l=1/2$ P-states. These
hyperfine doublets are expected to be split by
$O(\Lambda_{QCD}^2/M)\sim 50$ MeV; this compares well with the
experimental hyperfine splitting, $B^*{-}B=46$~MeV. These splittings
vanish as $1/M$ in the static limit.

Our results are compared with experiment in more detail, taking into
account the systematic errors, in table~\ref{prediction}.  If we
assume, initially, $a^{-1}=2.0$~GeV is a reasonable estimate of the
lattice spacing for these configurations the only significant
disagreement with experiment is found in the $\Lambda_b{-}B$
splitting. Finite volume effects, which are not included in the
estimates of the systematic errors, may well account for the rather
high value for the $\Lambda_b$ mass. 
For all splittings, the error from setting
$a^{-1}$ is at least as large as the statistical or other systematic errors.
However, since this uncertainty leads to a positive shift in the
splittings it does not lessen the disagreement with experiment found
for the $\Lambda_b{-}B$ splitting. There is also less agreement
for the $B_s{-}B$ and $B^{**}{-}B$ splittings using $a^{-1}=2.4$~GeV.

\begin{figure}
\centerline{
\setlength{\epsfxsize}{80mm}\epsfbox[10 60 640 570]{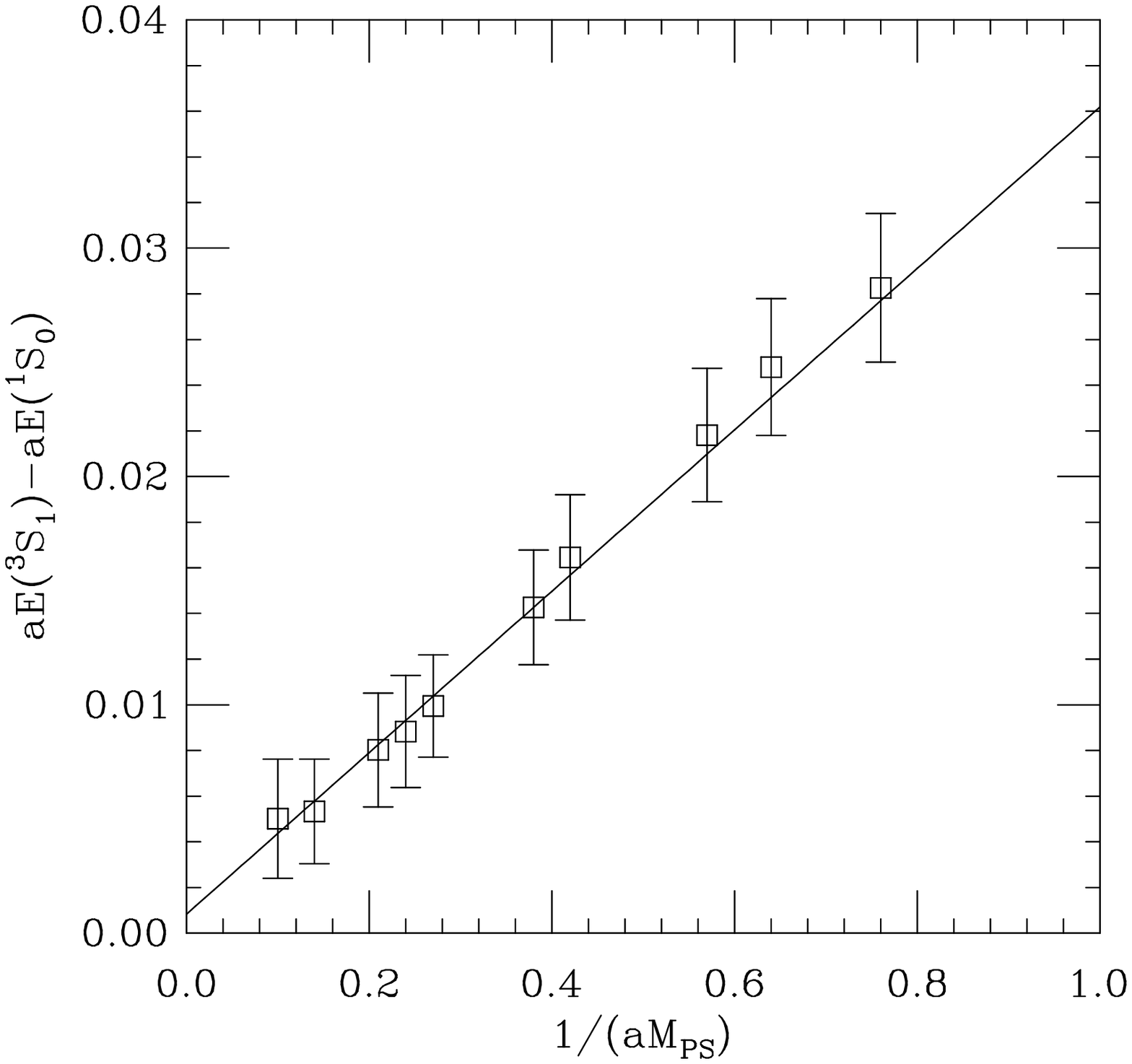}
\setlength{\epsfxsize}{80mm}
\epsfbox[10 60 640 570]{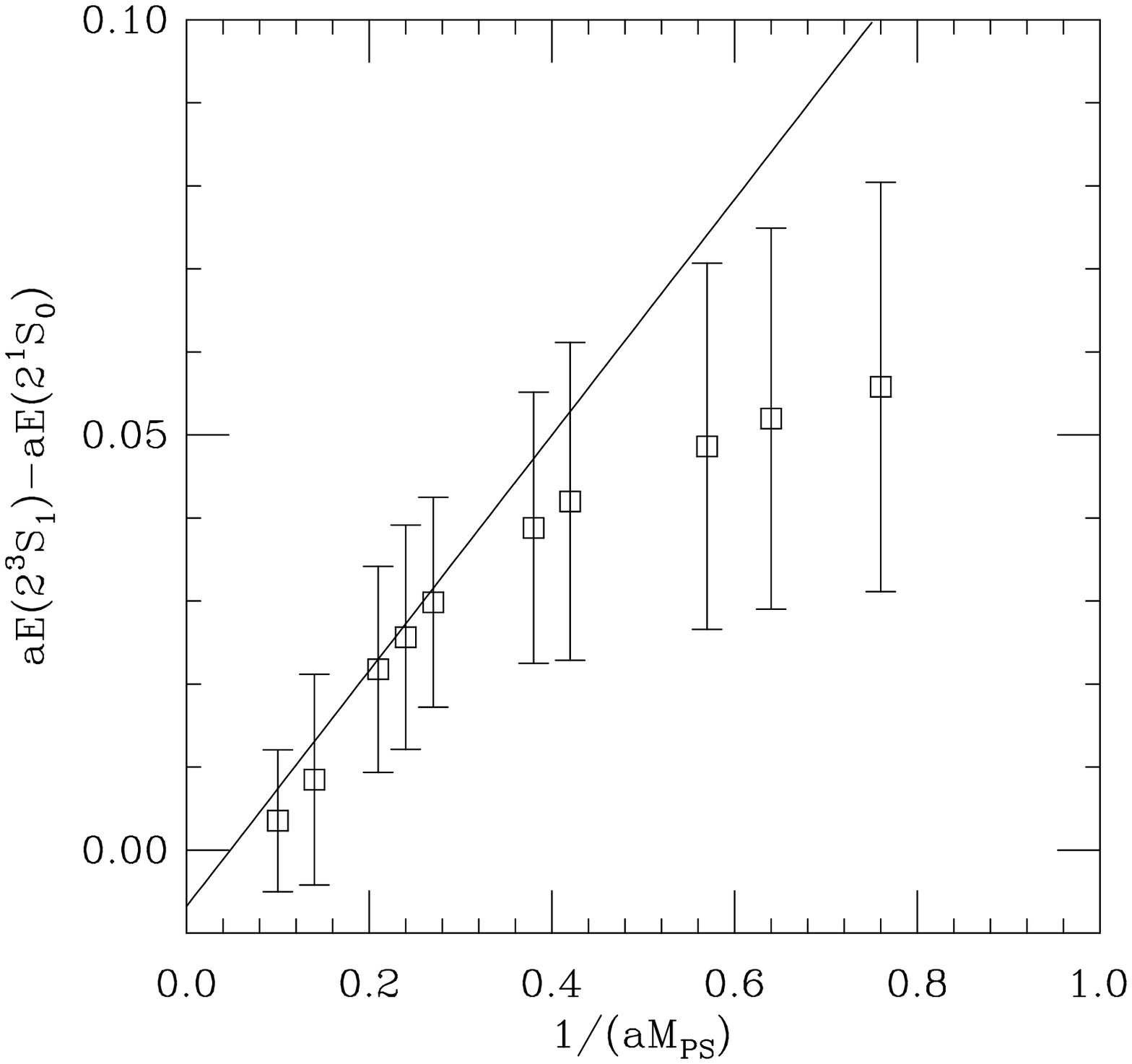} }
\vspace{0.5cm}
\centerline{
\setlength{\epsfxsize}{80mm}\epsfbox[10 60 640 570]{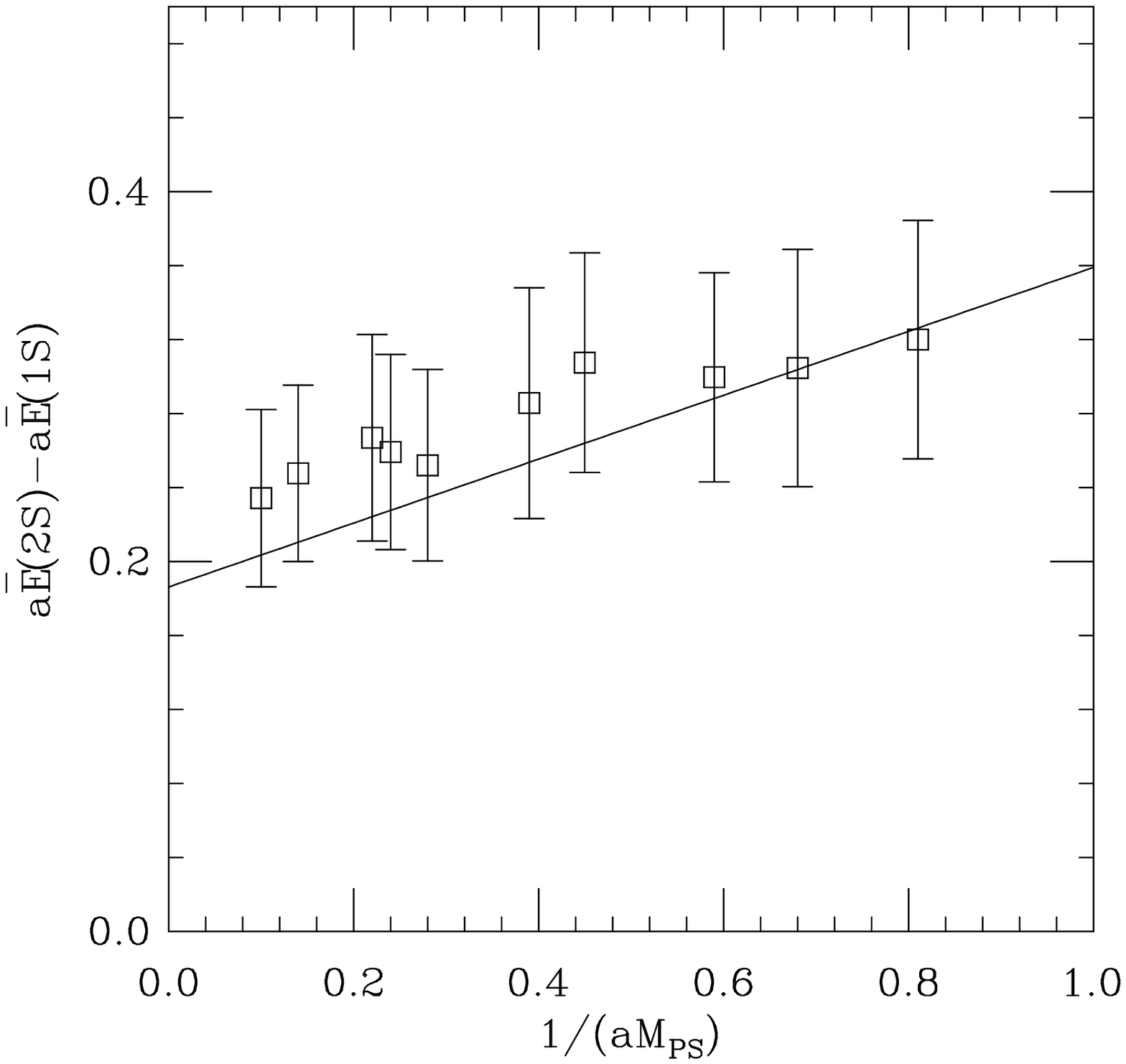}}
\vspace{0.5cm}
\caption{The dependence of the mass splittings of the ground and first
excited state $PS$ and $V$ mesons with the inverse pseudoscalar meson
mass in lattice units. $\kappa_l=\kappa_c$ for the
$\bar{E}(2S){-}\bar{E}(1S)$ splitting, while $\kappa_l=0.1385$ for the
$1S$ and $2S$ hyperfine splittings.\label{hqdep}}
\end{figure}

\begin{figure}
\centerline{\setlength{\epsfxsize}{80mm}\epsfbox[10 60 640 570]{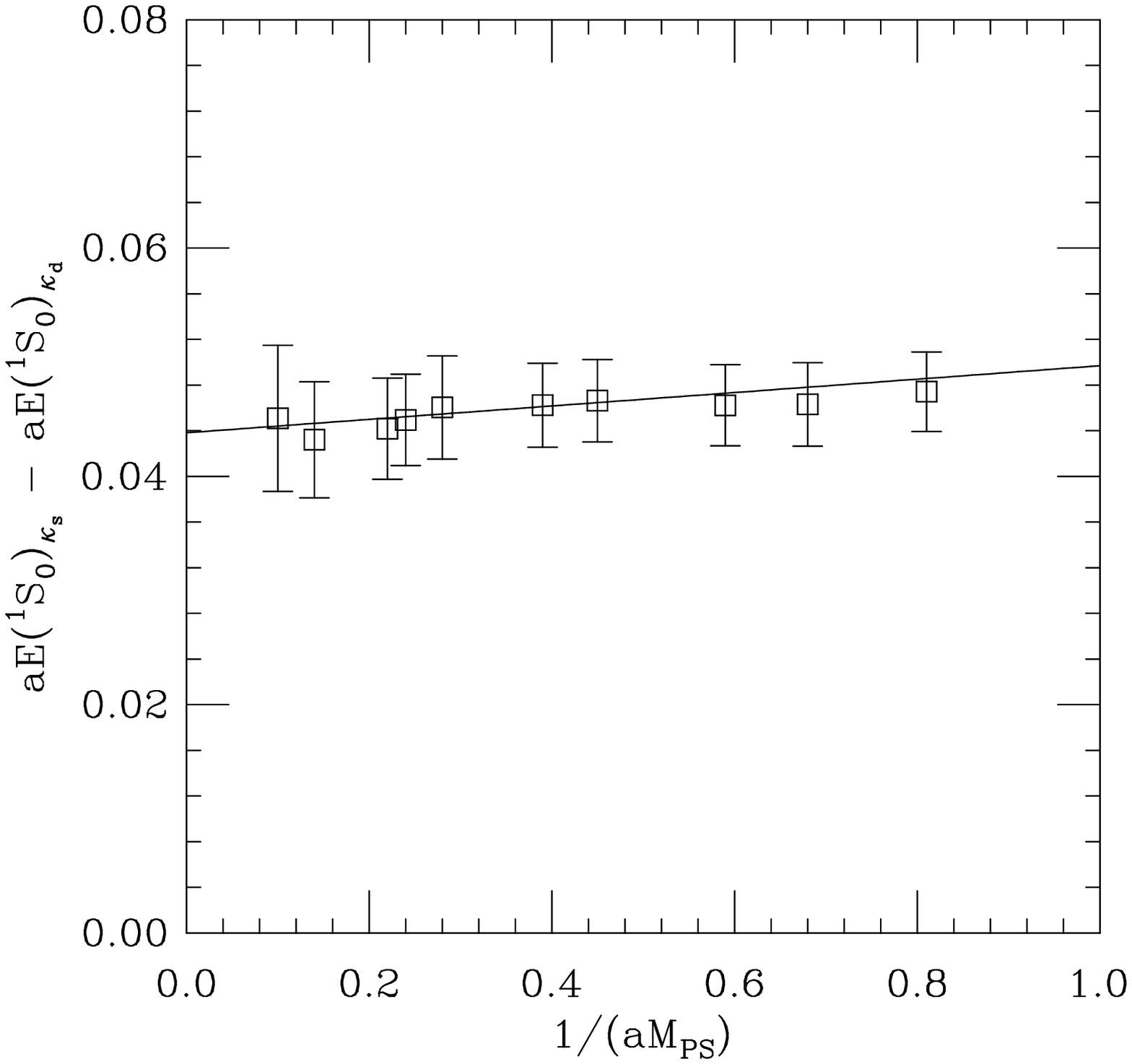}
\setlength{\epsfxsize}{80mm}\epsfbox[10 60 640 570]{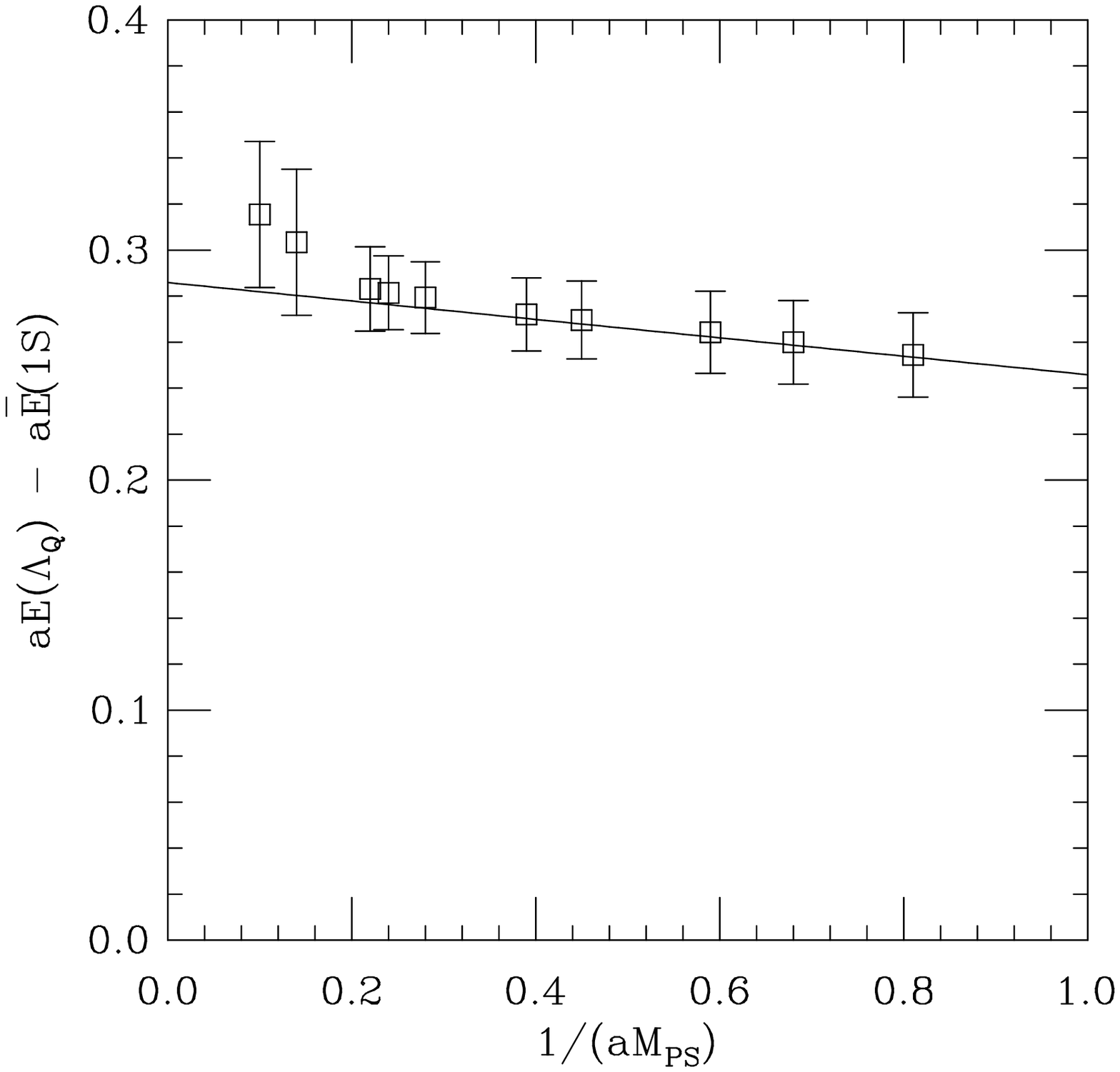}
}
\centerline{\setlength{\epsfxsize}{80mm}\epsfbox[10 60 640 570]{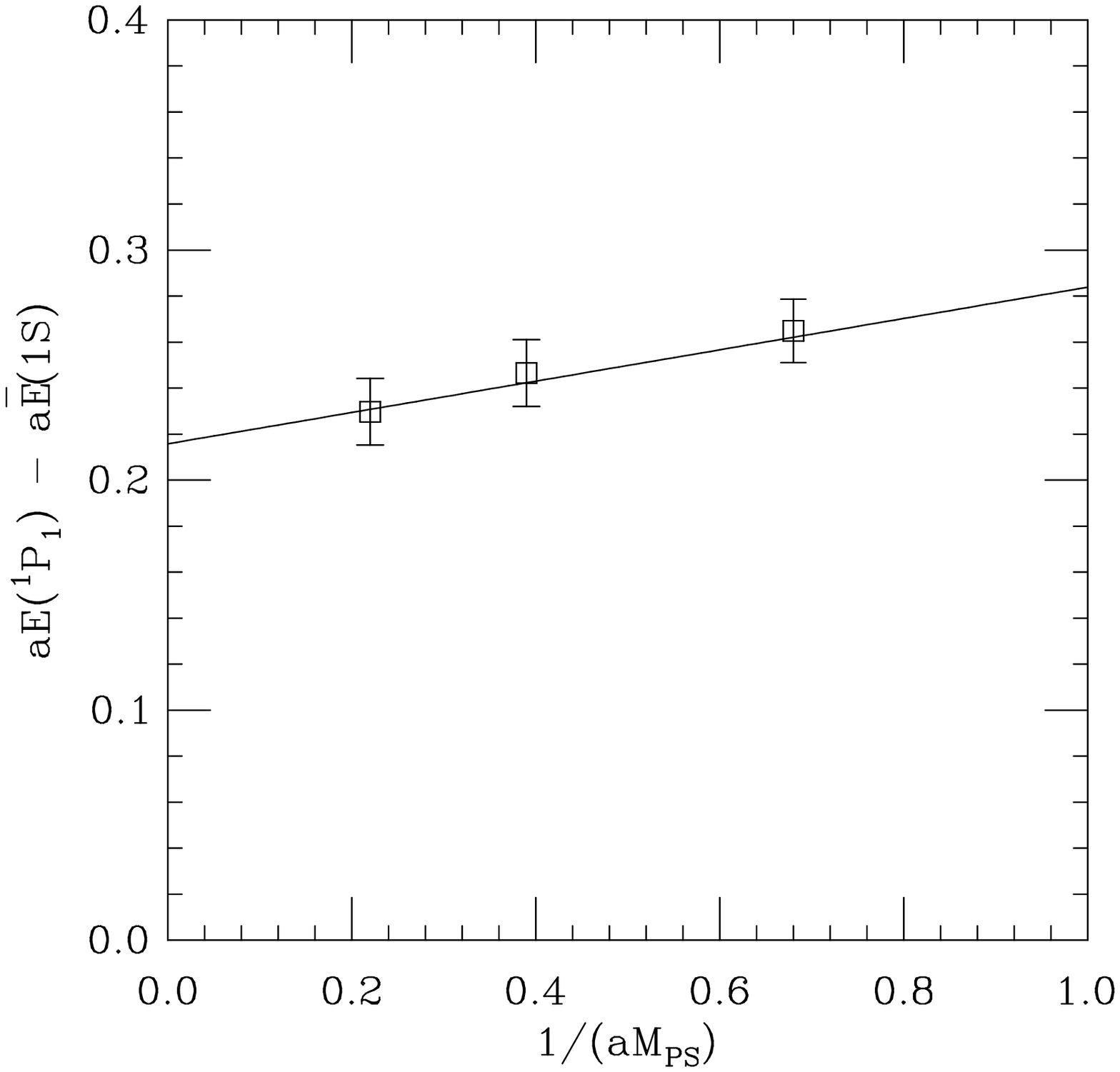}
}
\vspace{0.5cm}
\caption{The dependence of the
$E(^1S_0)_{\kappa_s}{-}E(^1S_0)_{\kappa_d}$,
$E(\Lambda_Q){-}\bar{E}(1S)$ and $E(1P)-E(1S)$ splittings on
$1/(aM_{PS})$ in lattice units, where
$\kappa_l=\kappa_c$.
\label{hqdep_cont}}
\end{figure}

With the mass splittings calculated for a wide range of heavy quark
masses, not just in the region of the $B$ meson, we are able to
investigate violations of HQS at finite $m_Q$. The behaviour of these
splittings as a function of $1/M_{PS}$ is shown in figures~\ref{hqdep}
and~\ref{hqdep_cont}, in lattice units. The
$\bar{E}(2S){-}\bar{E}(1S)$,
$E(^1S_0)_{\kappa_s}{-}E(^1S_0)_{\kappa_d}$,
$E(\Lambda_Q){-}\bar{E}(1S)$ and $E(^1P_1){-}\bar{E}(1S)$ splittings
are very weakly~(and linearly) dependent on $1/m_Q$, consistent with
behaviour dominated by the light quark, with small corrections due to
the kinetic energy of the heavy quark and the hyperfine interaction;
$E(^1S_0)$ denotes the simulation energy of a pseudoscalar meson,
while $\bar{E}(1S)$ denotes the spin-average of the simulation energy
with the corresponding vector meson. $E(^1S_0)_{\kappa_s}$ indicates
that the light quark mass has been interpolated to that of the strange
quark mass. Note that by taking the spin-average the dependence of
that mass on the hyperfine term is removed. Similarly, the hyperfine
splitting in the $1S$ and $2S$ states depends linearly on $1/m_Q$ over
the range of masses studied, vanishing in the static limit, as
expected theoretically.

To quantify the heavy quark mass dependence we fit the mass splittings
to the form:
\begin{equation}
\Delta M = C_0 + C_1/M + \ldots
\end{equation}
where $M$ is the pseudoscalar meson mass, and extract the coefficients
$C_0$ and $C_1$, which are expected to be of $O(\Lambda_{QCD})$ and
$O(\Lambda_{QCD}^2)$ respectively~(with the exception of the hyperfine
splittings where $C_0=0$).  In fact, the coefficients can be related
to the binding energy and expectation values of the kinetic energy of
the heavy quark and the hyperfine interaction in the static limit for
each meson or baryon.  These quantities are needed in analytical
approaches such as HQET.

If we define the heavy-light hadron binding energy,
$\bar{\Lambda}_H^{m_Q}$, using:
\begin{equation}
M_H = m_Q + \bar{\Lambda}_H^{m_Q}
\label{bind}
\end{equation}
where $M_H$ is the mass of the hadron, and expand using first order
perturbation theory,
\begin{equation}
\bar{\Lambda}_H^{m_Q} = \bar{\Lambda}_H + \frac{1}{M}\left< H \right|{\cal O}_{kin}\left| H \right> + \frac{1}{M}\left< H \right|{\cal O}_{hyp} \left| H \right>\label{expmass}
\end{equation}
where $\left< H \right|$ and $\bar{\Lambda}_H$ represent the hadron state and
the binding energy, respectively, in the infinite mass limit; ${\cal
O}_{kin} = -\psi^\dagger (D^2/2) \psi$ and ${\cal O}_{hyp} = -\psi^\dagger(
\sigma\cdot B/2)\psi$. In principle, the separation of the meson mass
into the heavy quark mass and binding energy in equation~\ref{bind}
must be clearly defined.  However, we only consider physical mass
splittings, for which the differences in the definitions of
$\bar{\Lambda}_H^{m_Q}$ cancel. This has the additional benefit that
the $O(1/M)$ coefficients of the mass splittings correspond to the
physical difference in expectation values of the kinetic
energy~(and/or hyperfine interaction) of the particles in the
splitting.  Thus, we consider, for example, the $2S{-}1S$ mass
splitting of the spin-average S-states, which can be expressed as
\begin{equation}
\bar{E}(2S){-}\bar{E}(1S) = \bar{\Lambda}_{2S} - \bar{\Lambda}_{1S} +
\frac{1}{M}\left<2S\right|{\cal O}_{kin} \left|2S\right>_{phys} - \frac{1}{M}\left< 1S\right|{\cal O}_{kin} \left| 1S\right>_{phys}
\end{equation}
where $\left< 1S\right|{\cal O}_{kin} \left| 1S\right>_{phys}$ denotes the
physical expectation value of the kinetic energy in the $B$ meson.
The results are given in table~\ref{Etab} in physical units and
compared with the theoretical expectation.

Given the weak dependence of most splittings on the heavy quark mass,
an estimate of the intercept and slope for some splittings can be made
from the experimental results for the $B$ and $D$ spectrum and $\Lambda_b$
and $\Lambda_c$ baryons. The hyperfine splittings are strongly
dependent on $m_Q$. However, the intercept is zero, and this can be
used along with the splitting for the $B$ meson to estimate the slope.
The corresponding estimates of the intercept and slope of various
meson and baryon mass splittings are also shown in the table.

Table~\ref{Etab} shows that for quantities which are weakly dependent
on $1/m_Q$, we can reliably extract $C_0$. However, it is difficult to
reliably extract the slope and an increase in statistics is required
in order to provide quantitative predictions.  Qualitatively, the
results are consistent with the theoretical predictions of only small
violations of HQS, $O(\Lambda_{QCD}/M)\sim 10\%$, around the $B$
meson. Comparing with the estimates derived from experiment we find
our value for the slope of the hyperfine splitting is low, reflecting
the low value we obtain for the $B^*{-}B$ splitting.  The absolute
statistical error in the splitting is more or less constant with
$m_Q$, while the splitting rises as the mass decreases. Thus, a low
value of the slope is a clearer indication than the $B$ hyperfine
splitting itself, that our results are inconsistent with experiment.
The hyperfine coefficient, $c_B$, depends on the heavy quark mass and 
a determination of this coefficient for the range of masses studied
here is needed to clarify the source of any remaining discrepancy with
experiment, for example residual quenching effects. Reasonable
agreement is found between the lattice results for the
$E(^1S_0)_{\kappa_s}{-}E(^1S_0)_{\kappa_d}$ splitting and the estimates from
experiment. This is also the case for the slope of the
$E(\Lambda_b){-}\bar{E}(1S)$ splitting, while the intercept is too high~(consistent with finite volume problems mentioned previously). 
Our results for the intercepts and slopes of spectral quantities are
consistent with our previous estimates using Wilson light quarks
detailed in reference~\cite{spect}.

\subsection{$n_f$ dependence}
\label{nfspectrum}

An indication of sea quark effects can be found by comparing the
$n_f=2$ results with our previous spectrum calculation in the quenched
approximation at $\beta=6.0$, detailed in reference~\cite{arifaspec2}.
The results for the splittings which have been calculated in both
cases are detailed in figure~\ref{spectrum1} and table~\ref{pred_pev},
where $a^{-1}$ is fixed from $m_\rho$ in both cases.  With the
exception of the hyperfine splitting, the quenched simulation
reproduces the experimental spectrum well to this level of statistical
and systematic uncertainty. Furthermore, comparing with the $n_f=2$
results, we see consistency to within $1{-}2\sigma$. However,
switching to $a^{-1}=2.4$~GeV for the dynamical configurations,
differences of $\sim3\sigma$ appear for the $B_s{-}B$, $B^{**}{-}B$
and $\Lambda_b{-}B$ splittings. The uncertainty in the lattice spacing
must be reduced before any clear indications of sea quark effects can
be found.

The hyperfine splitting is a quantity where sea quark effects are
expected to be seen. However, in practice this is not easy to
investigate due to the difficulty in measuring such a small
quantity. The quenched results for $B^*{-}B$ and $B^*_s{-}B_s$, with
lower statistical errors than at $n_f=2$, are approximately half the
experimental values. The splittings may be boosted by as much as
$30\%$ to $\sim 31$ and $\sim 35$~MeV respectively when the 1-loop
corrections to $c_B$ are included.
The statistical errors for $B_s^*{-}B_s$ are small
enough that the splitting is still significantly below experiment,
suggesting quenching effects.  The corresponding $n_f=2$ result 
is also increased to $\sim 34$~MeV~(for $a^{-1}=2.0$~GeV), when $c_B$
is corrected.  However, the statistical error is large and must be
reduced significantly before we can see if partially including
dynamical quarks increases the hyperfine splitting. Using
$a^{-1}=2.4$~GeV for the $n_f=2$ results only leads to a $1\sigma$
difference between quenched and $n_f=2$ and thus does not
significantly change the comparison.

\section{Decay Constants}
\label{decaysection}
In the continuum, the pseudoscalar~(PS) and vector~(V) decay constants
are defined by
\begin{eqnarray}
\left<0\right| A_0 \left|PS\right> & = & \left<0\right| \bar{q}\gamma_5\gamma_0 h \left|PS\right> =
f_{PS}M_{PS}, \label{pseqn}\\
\left<0\right| V_k \left|V_k\right> & = & \left<0\right| \bar{q}\gamma_k h \left|V_k\right> = \epsilon_{k} f_V M_V,\label{veqn}
\end{eqnarray}
where $q$ and $h$ represent 4-component light and heavy quark fields
respectively. In lattice NRQCD, $A_0$ and $V_i$ are given by power
series of operators in $1/M$.
To a given order, all operators with the
appropriate quantum numbers and power of $1/M$ appear. This includes
operators representing discretisation corrections, which vanish in the
continuum. The corresponding matrix elements for these operators are
combined with the renormalisation factors matching the lattice matrix
elements to $\left<0\right|A_0\left|PS\right>$ and $\left<0\right|V_k\left|V\right>$ in full QCD,
where mixing between the operators under renormalisation must be taken
into account.

In this study we truncate the NRQCD action at $O(1/M)$, and match the
decay constants to full QCD through $O(\alpha/M)$.  The axial-vector
current is then~\footnote{In the limit of zero light quark mass.}:
\begin{equation}
\left< A_0 \right> = \sum_{j=0}^2 C_j(\alpha, aM) \left<J_L^{(j)}\right> + O(1/M^2,\alpha^2, a^2, \alpha a/M),\label{opsps}
\end{equation}
where
\begin{eqnarray}
\left<J_L^{(0)}\right> & = & \left<0\right|\bar{q}\gamma_5\gamma_0 Q\left|PS\right>, \\
\left< J_L^{(1)}\right> & = & \left<0\right|-\bar{q} \label{psone}
\gamma_5\gamma_0 \frac{\vec{\gamma}\cdot \vec{D}}{2M_0}Q \left|PS\right>, \\
\left<J_L^{(2)}\right> & = & \left<0\right| \bar{q}\frac{ \stackrel{\leftarrow}{D} \cdot \stackrel{\leftarrow}{\gamma}}{2M_0}\gamma_5\gamma_0 Q \left|PS\right>. \label{pstwo}
\end{eqnarray}
$q$ is now the light quark field in the lattice theory, and $Q$ is
related to the 2-component heavy quark field $\psi$ in lattice NRQCD by
\begin{equation}
Q = \left(\psi\atop 0\right).
\end{equation}
$J_L^{(0)}$ and $J_L^{(1)}$
are the tree-level operators which are obtained through the $O(1/M)$
inverse Foldy-Wouthuysen transformation connecting $h$ to
$Q$. $J_L^{(2)}$ appears at $1$-loop in perturbation theory. Thus,
$C_0$ and $C_1$ are $O(1)$,
\begin{eqnarray}
C_0 = 1 + \alpha\rho_0, \hspace{1cm} C_1 = 1 + \alpha\rho_1,
\end{eqnarray}
while $C_2$ is $O(\alpha)$,
\begin{equation}
C_2 = \alpha(\rho_2 - \zeta_A),
\end{equation}
where we use the nomenclature of reference~\cite{junkmorn1}. At this
order one must also consider an $O(\alpha a)$ lattice artifact, which
appears through the mixing between $J_A^{(0)}$ and $J_A^{(2)}$ and
which does not vanish in the infinite heavy quark mass
limit~\cite{junkmorn1}:
\begin{eqnarray}
\left< J_L^{disc}\right> & = & \left<0\right|a\bar{q} \stackrel{\leftarrow}{D} \cdot \stackrel{\leftarrow}{\gamma}\gamma_5\gamma_0 Q \left|PS\right>.
\label{psthree}
\end{eqnarray}
This factor can be viewed as a discretisation correction to
$J_L^{(0)}$, and hence, we define an improved
operator~\cite{junkmorn1}:
\begin{equation}
J_L^{(0)imp} = J_L^{(0)} + C_A J_L^{disc}.
\end{equation}
where $C_A=\alpha(1+\zeta_A/(2aM_0))$. $\zeta_A$ reflects the freedom
in the definition of the improved operator and is cancelled by the
term $-\alpha\zeta_A$ appearing in $C_2$. In practice we set it to
zero, which one is free to do at $1$-loop accuracy.

In the lattice simulation, the number of matrix elements that must be
calculated can be reduced by noting that
\begin{equation}
\frac{1}{2M_0} J_L^{(disc)} = J_L^{(2)},
\label{equality1}
\end{equation}
and, also, that at zero momentum on the lattice
\begin{equation}
\left<J_L^{(1)}\right> = \left<J_L^{(2)}\right>.
\label{equality2}
\end{equation}
Thus, it is sufficient to compute the matrix elements corresponding to
the tree-level operators.  Since these matrix elements are generated
separately, the contribution to the decay constant from each matrix
element in equation~\ref{opsps} can be analysed. For
this purpose, we define,
\begin{eqnarray}
f_{PS}^{(j)}\sqrt{M_{PS}} & = & \frac{1}{\sqrt{M_{PS}}} \left<0\right| J_L^{(j)} \left|PS\right>.
\end{eqnarray}
We use $f_{PS}\sqrt{M_{PS}}$ to denote the total decay constant at both
tree-level and $1$-loop.

For the vector current:
\begin{equation}
\left<V_k\right> = \sum_{j=0}^{4} C_{Vj}(\alpha,aM) \left<J_k^{(j)}\right> + O(1/M^2,\alpha^2,a^2,\alpha a/M)\label{opsv}
\end{equation}
where,
\begin{eqnarray}
\left<J_k^{(0)}\right> & = & \left<0\right|\bar{q} \gamma_k Q \left|V_k\right>\label{first},\\ 
\left<J_k^{(1)}\right> & = & \left<0\right| -\bar{q} \gamma_k \frac{\vec{\gamma} \cdot \vec{D}}{2M_0}Q \left|V_k\right>, \label{vone}\\
\left<J_k^{(2)}\right> & = & \left<0\right| -\bar{q}\frac{\stackrel{\leftarrow}{D} \cdot \stackrel{\leftarrow}{\gamma}}{2M_0} \gamma_0\gamma_k Q \left|V_k\right>, \\
\left<J_k^{(3)}\right> & = & \left<0\right| -\bar{q}\frac{\vec{D}_k}{2M_0} Q \left|V_k\right>, \\
\left<J_k^{(4)}\right> & = & \left<0\right| \bar{q} \frac{\stackrel{\leftarrow}{D}_k}{2M_0} Q \left|V_k\right>.
\label{last}
\end{eqnarray}
$J_k^{(0)}$ and $J_k^{(1)}$ correspond to tree-level operators, while
the rest appear at one-loop in perturbation theory; $C_{V0}$ and
$C_{V1}$ are $O(1)$,
\begin{equation}
C_{V0} = 1 + \alpha\rho_0^V, \hspace{1cm} C_{V1} = 1 + \alpha\rho_1^V,
\end{equation}
while $C_{V2}$, $C_{V3}$ and $C_{V4}$ are
$O(\alpha)$,
\begin{eqnarray}
C_{V2} = \alpha (\rho_2^V - \zeta_V), \hspace{1cm} C_{V3} = \alpha\rho_3^V, \hspace{1cm} C_{V4} = \alpha\rho_4^V.
\end{eqnarray}
Analogous to the axial-vector case, there is a discretisation correction
which can be absorbed into a redefinition of the zeroth order
operator:
\begin{eqnarray}
\left< J_k^{disc}\right> & = & \left<0\right|-a\bar{q}\stackrel{\leftarrow}{D} \cdot \stackrel{\leftarrow}{\gamma} \gamma_0\gamma_k Q \left|V_k\right>.\label{laster}\\
J_k^{(0)imp} & = & J_k^{(0)} + C_V J_k^{disc},
\end{eqnarray}
where $C_V=\alpha(1 +\zeta_V/(2aM_0))$; we set $\zeta_V=0$ when
computing $C_V$ and $C_{V2}$.  Note that,
\begin{eqnarray}
J_k^{(2)} & = &  - J_k^{(1)} + 2J_k^{(3)}, \label{equvec1}\\
\frac{1}{2M_0} J_k^{(disc)}& = & J_k^{(2)},\label{equvec2}
\end{eqnarray}
and at zero momentum on the lattice,
\begin{eqnarray}
\left<J_k^{(3)}\right> & = & \left<J_k^{(4)}\right>.
\label{equvec3}
\end{eqnarray}
Thus, the 6 vector matrix elements can be reconstructed from the
subset $j=0,1,3$. For the purpose of analysing the individual
contributions to the vector decay constant of the matrix
elements~\ref{first}{-}~\ref{last} and~\ref{laster}, we define,
\begin{eqnarray}
f_{V}^{(j)}\sqrt{M_{V}} & = & \frac{1}{\sqrt{M_V}} \left<0\right| J_k^{(j)} \left|V_k\right>.
\end{eqnarray}
We use $f_V\sqrt{M_V}$ to denote the total decay constant at both
tree-level and $1$-loop.

The matching coefficients $C_j$ and $C_A$, and $C_{Vj}$ and $C_V$,
have been evaluated to $1$-loop in perturbation theory by Morningstar
and Shigemitsu~\cite{junkmorn1,junkmorn2}. These coefficients depend
on $M$ and $\alpha$, where we take $\alpha_V(q^*)$~\cite{macklep} for
the strong coupling. The scale $q^*$ also depends on the heavy
quark mass. This scale is unknown at present, although in the static
limit Hernandez and Hill found that, $q^*=2.18/a$~\cite{hill}. At this
stage, we assume that $q^*$ for each mass lies somewhere between $1/a$ and
$\pi/a$, and compute $f_{PS}$ and $f_V$ for these two limits.

The coefficients $C_0$ and $C_1$ in the pseudoscalar case, and
$C_{V0}$, $C_{V1}$, $C_{V2}$ and $C_{V4}$ in the vector case, include a term
$\alpha ln(aM)/\pi$. When computing $f_B$ and $f_{B^*}$ we insert
$aM_0$ for the heavy quark mass. There are no problems with large
logarithms for the quark masses used in this study, however, for
larger $M$, resumming the logarithms would be more appropriate.  A
different approach is needed when extracting the slope of the decay
constants and comparing with the static limit. In this case, it is
assumed that the meson simulated at each $M_0$ is the $B$ meson~(to a
better or worse approximation), and thus, $aM_0^b$ must be inserted in
the logarithms for all heavy quark masses.

A possible concern is that the $\alpha/(aM)^n$ terms in the coefficients
combine with nonperturbative lattice errors, $\propto
(a\Lambda_{QCD})^n$ in the matrix elements $\left<J_L^{(0)}\right>$ or
$\left<J_k^{(0)}\right>$. This, then, leads to contributions that look like
physical $(\Lambda_{QCD}/M)^n$ corrections to $f_B$ or
$f_{B^*}$. However, since $\left<J_L^{(0)}\right>$ and $\left<J_k^{(0)}\right>$
are designed to match full QCD through $O(\alpha a)$, these errors
appear at $O(\alpha^2(a\Lambda_{QCD}))$ or
$O((a\Lambda_{QCD})^2)$. The leading contribution to $f_B$ or
$f_{B^*}$ from these errors is beyond the order in the NRQCD series
considered here. These issues have been discussed previously, in more
detail, in references~\cite{arifajunk} and~\cite{junkmorn1}.

Theoretically, within HQET, the decay constants in the combinations,
$f_{PS}\sqrt{M_{PS}}$ and $f_V\sqrt{M_V}$, are non-zero and equal in
the static limit, reflecting spin and flavour symmetry. Away from this
limit, $O(1/M)$ corrections take the form:
\begin{equation}
f\sqrt{M} = (\alpha(M))^{-1/(2\beta_0)}(f\sqrt{M})^\infty \left( 1 + C/M + \ldots\right)
\label{hqsym}
\end{equation}
where the decay constant in the static limit, $(f\sqrt{M})^\infty$,
and the coefficient of the linear slope, $C$, are expected to be of
$O(\Lambda_{QCD})$. Thus, the ratio of the vector to pseudoscalar decay
constant can be expressed as:
\begin{equation}
f_V\sqrt{M_V}/f_{PS}\sqrt{M_{PS}} = 1 + (C_V-C_{PS})/M + \ldots.
\end{equation}
Another quantity to consider is the ratio of the pseudoscalar decay
constants for the heavy-light mesons containing a $u$ light quark and
that containing a $s$ quark,
\begin{equation}
\frac{f_{PS}\sqrt{M_{PS}}|_{\kappa_s}}{f_{PS}\sqrt{M_{PS}}|_{\kappa_c}}
= \left(\frac{ f_{PS}\sqrt{M_{PS}}|_{\kappa_s}}{f_{PS}\sqrt{M_{PS}}|_{\kappa_c}}\right)^\infty( 1 + (C_{PS}^{\kappa_s}-C_{PS}^{\kappa_c})/M + \ldots).
\end{equation}
Since the heavy quark is almost a spectator in the heavy-light meson,
the shift in the decay constant when the light quark mass is changed
is expected to be roughly independent of $M$. Thus, $C_{PS}|_{\kappa_s}$ is approximately equal to $C_{PS}|_{\kappa_c}$, and 
the ratio should have little dependence on the heavy quark mass.

The contributions to the slope of the decay constants from the
$O(1/M)$ terms in the NRQCD action and corrections to the current can
be identified. From first order perturbation theory in $1/M$ about the
static limit:
\begin{eqnarray}
C & = & G_{kin} + 2d_MG_{hyp} + d_MG_{corr}/6,
\label{slope}
\end{eqnarray}
where $d_M=3$ and $-1$ for pseudoscalar and vector mesons
respectively. Thus,
\begin{eqnarray}
C_{PS}-C_{V} & = &  8G_{hyp} + 2G_{corr}/3.
\label{slopehyp}
\end{eqnarray}
The kinetic and hyperfine terms
contribute to the decay constant through the correction to the meson
wavefunction.
\begin{eqnarray}
\lefteqn{\left<0\right| q^\dagger \Gamma^{(2)} \psi \left|P\right>_\infty G_{kin}  = }\hspace{2.4cm} \nonumber\\
& & \left<0\right| \int dy T\{q^\dagger \Gamma^{(2)}\psi(0),{\cal O}_{kin}(y)\} \left|P\right>_{\infty},\label{gkin}
\end{eqnarray}
\vspace{-0.8cm}
\begin{eqnarray}
\lefteqn{\left<0\right| q^\dagger \Gamma^{(2)} \psi \left|P\right>_\infty 2d_MG_{hyp}  = } \hspace{3.2cm}\nonumber\\
& & \left<0\right| \int dy T\{q^\dagger \Gamma^{(2)}\psi(0),{\cal O}_{hyp}(y)\}\left|P\right>_{\infty},\label{ghyp}
\end{eqnarray}
while $G_{corr}$ is directly related to the tree-level current correction:
\begin{eqnarray}
\left<0\right| q^\dagger \Gamma^{(2)} \psi \left|P\right>_\infty d_MG_{corr}/6 & = & \left<0\right| q^\dagger\Gamma^{(2)} \frac{\sigma\cdot \vec{D}}{2} \psi \left|P\right>_\infty.\hspace{3.8cm}\label{cooreqn}
\end{eqnarray}
$\left|P\right>_\infty$ represents the meson in the limit of infinite
heavy quark mass and $\Gamma^{(2)}=1$ and $\sigma$ for the
pseudoscalar and the vector, respectively. Note that
equations~\ref{gkin}-\ref{cooreqn} are tree level expressions.  This
analysis is discussed in more detail in reference~\cite{decay}.

\subsection{Results}
\label{decayresults}
The bare lattice matrix elements computed in the lattice simulation,
\mbox{$f^{(0,1)}_{PS}\sqrt{M_{PS}}$} and \mbox{$f^{(0,1,3)}_{V}\sqrt{M_{V}}$},
are given in the appendix in tables~\ref{decaytps} and~\ref{deltapstree} for
the pseudoscalar, and tables~\ref{decaytv},~\ref{deltavcor}
and~\ref{deltavloop} for the vector, for the range of masses studied.
Combined with the appropriate matching factors, we obtain the tree-level and
renormalised decay constants shown in tables~\ref{deltaps} and~\ref{deltav},
for the pseudoscalar and vector respectively.

Considering, initially, $aM_0=2.0$, a quark mass in the region of the
$b$ quark mass~($aM_0^b=2.1$ for $a^{-1}=2.0$~GeV),
table~\ref{deltaps} shows that $f_{PS}\sqrt{M_{PS}}$ is reduced from
the tree-level value by $10\%$ for $aq^*=\pi$ and $16\%$ for
$aq^*=1.0$. Thus, the combined 1-loop corrections to the decay
constant, $O(\alpha)$, $O(\alpha a)$ and $O(\alpha/M)$, are
significant but not larger than expected.  In addition, the
uncertainty in $f_B$ due to the unknown $q^*$ is not unduly large,
$\sim 6\%$, although it is bigger than the statistical errors of $\sim
3\%$.

We study the tree-level and renormalised currents further by analysing
the individual contributions to the $PS$ decay amplitude from the various
operators and matching factors.  Table~\ref{contribs}, details the
percentage correction of these contributions to the zeroth order
tree-level matrix element, $\left<J_L^{(0)}\right>$; $aM_0=2.0$ and $aq^*=1.0$
is used for the 1-loop terms. Note that the contributions from
$J^{(2)}_L$ and $J^{disc}_L$ are reconstructed from $\left<J_L^{(1)}\right>$.

From the table we see that the $O(1/M)$ tree-level correction,
$\left<J^{(1)}\right>$, reduces the decay constant by $13\%$. The
1-loop correction for this matrix element is small, $\sim 1\%$,
however, the other 1-loop terms, the $O(\alpha)$ correction to
$\left<J^{(0)}\right>$, $\alpha\rho_0$, and the $O(\alpha/M)$ matrix
element and the $O(a\alpha)$ discretisation correction,
$\alpha\rho_2\left<J^{(2)}\right>+\alpha\left<J^{disc}\right>$, are
all of a similar magnitude to the tree-level correction, around
$10\%$.  However, none of the terms are unduly large and there is no
indication that NRQCD is breaking down.  Since the terms contribute
with different signs there is some cancellation between them and the
combined 1-loop correction is $-14\%$ for $aq^*=1.0$, the same size as
the $O(1/M)$ tree-level correction. Thus, the total correction to
$f^{(0)}_{PS}\sqrt{M_{PS}}$ is $-27\%$; this falls to $-21\%$ for
$aq^*=\pi$, of which $-8\%$ is due to the 1-loop terms.

The picture is fairly similar for the vector decay
constant. Table~\ref{deltav} shows $f_V\sqrt{M_V}$ at $aM_0=2.0$ is
reduced from the tree-level value by $14\%$ and $22\%$ when the
matching factors are included with $aq^*=\pi$ and $1.0$, respectively;
the dependence on $q^*$ is moderate. In the tree-level case,
$f_{PS}\sqrt{M_{PS}}$ is slightly below $f_V\sqrt{M_V}$, while for the
same $q^*$ they are roughly equal. However, there is no reason for the
characteristic scale for the vector to be the same as that for the
pseudoscalar, and thus, it is difficult to compare the two at this
stage.

The contributions to the vector decay constant from the individual
matrix elements and matching coefficients are also detailed in
table~\ref{contribs}.  We see that the 1-loop terms, $\alpha\rho_0$ and
$\alpha\rho_2\left<J^{(2)}\right>+\alpha\left<J^{disc}\right>$
are all the same order of magnitude as in the $PS$ case, and combine
to give a $-20\%$ correction to the zeroth order vector current. The
remaining 1-loop terms, all $O(\alpha/M)$, are very small, of order
$1\%$.  However, since the $O(1/M)$ tree-level term is smaller and
with opposite sign, compared to the $PS$ case, the overall correction
to $\left<J^{(0)}\right>$ is smaller: $-17\%$~($-10\%$) for
$aq^*=1.0$~($\pi$).  The relative size of the individual corrections
compared to those for the $PS$ is discussed later.

Interpolating the results for $f_{PS}\sqrt{M_{PS}}$ and
$f_{V}\sqrt{M_{V}}$ to  $M_0=M_0^b$, and converting into physical
units we obtain the values for $f_B$ and $f_{B^*}$ shown in
table~\ref{work}; the results for $f_{B_s}$ and $f_{B_s}/f_B$ are also
given. Table~\ref{work} presents the predictions for the decay
constants derived using both $a^{-1}=2.0$ and $2.4$~GeV. We take the
difference in the decay constants for the different choices of lattice
spacing as an estimate of the error in $f_B$ due to uncertainty in the
scale.

Combined with estimates of the other dominant systematic errors,
detailed in table~\ref{syserrors}, our predictions for the
pseudoscalar and vector decay constants are:
\begin{eqnarray}
f_B &=&  186(5)(19)(9)(13)(+50)~\mbox{MeV}  \\
f_{B_s} &=&  215(3)(22)(9)(15)(+49)(-5)~\mbox{MeV}  \\
 f_{B^*} & =&  181(6)(18)(9)(13)(+55)~\mbox{MeV} \\
 f_{B^*_s} & =& 213(4)(21)(9)(15)(+60)(-4)~\mbox{MeV} 
\end{eqnarray}
The first error is statistical. The second is the estimate of the
$O(\alpha^2)$ perturbative error $\sim 10\%$. The third
and fifth errors are the estimates of uncertainty arising from the
light quark discretisation errors and the uncertainty in
$a^{-1}$. Note that the latter includes the corresponding change in
$aM_0^b$.  For $f_{B_s}$ there is an additional error due to the
uncertainty in $\kappa_s$.

The fourth error is due to the truncation of the NRQCD series.  To
estimate this, one must consider the next order corrections,
$O(1/M^2)$, to the current and to the NRQCD action~(given in
equation~\ref{highterms}). Naively, one expects these terms to be of
order $\sim 1\%$.  As mentioned in section~\ref{errors} Ishikawa et
al~\cite{japanese} performed a tree-level analysis of the $O(1/M^2)$
terms for $f_{PS}$ and $f_V$.  These authors found that in the region
of the $B$ meson $\delta H_{h.o.}$ had no effect on the $PS$ or $V$
matrix elements. For lighter values of $aM_0$ the higher order terms
in the action tended to increase the matrix elements.

The individual tree-level current corrections are,
\begin{eqnarray}
J^{(a)} & = & \frac{1}{8M_0^2} \bar{q} \Gamma D^2 Q,\\
J^{(b)} & = & \frac{1}{8M_0^2} \bar{q} \Gamma g\vec{\Sigma}\cdot \vec{B} Q,\\
J^{(c)} & = & -\frac{1}{8M_0^2} \bar{q} \Gamma 2ig\vec{\alpha}\cdot \vec{E} Q,
\end{eqnarray}
where $\vec{\alpha}=\gamma_0\vec{\gamma}$ and
$\vec{\Sigma}=diag(\vec{\sigma},\vec{\sigma})$.
$\Gamma=\gamma_5\gamma_0$ and $\gamma_k$ for the pseudoscalar and
vector respectively. These terms were each found to be less than $2\%$
in magnitude. The corrections contribute with differing signs and
there is a cancellation between them which leads to a small overall
decrease in $f_B\sqrt{M_B}$ of $\sim 3\%$.  Since at $O(1/M)$ the
1-loop terms are as large as those at tree-level, as discussed
previously, we estimate the overall uncertainty in the decay constants
due to the omitted $O(1/M^2)$ and $O(\alpha/M^2)$ terms as $6\%$.

In total, adding the statistical and systematic errors in quadrature,
there is a $14\%$ uncertainty in the decay constants.  This is much
smaller than the $\sim 30\%$ error due to the uncertainty in $a^{-1}$.
Thus, a determination of $f_B$ of around $\sim 10\%$ is possible, but
only if the systematic error in $a^{-1}$ is reduced significantly.

In contrast, $f_{B_s}/f_B$ and $f_{B_s^*}/f_{B^*}$ are quantities for
which most systematic errors cancel out. These ratios are independent
of the heavy quark mass~(see below), while table~\ref{work} indicates
that it is insensitive to the $1$-loop corrections. Furthermore,
$f_{B_s}/f_B$ and $f_{B_s^*}/f_{B^*}$ are not dependent on the lattice
spacing. The only significant error is that due to the uncertainty in
$\kappa_s$. We find,
\begin{eqnarray}
f_{B_s}/f_{B} & = &1.14(2)(-2), \\
f_{B_s^*}/f_{B^*} & = & 1.14(2)(-2)
\end{eqnarray}
where the first error is statistical and the second is due to
$\kappa_s$.

\begin{figure}
\centerline{
(a)\setlength{\epsfxsize}{80mm}\epsfbox[10 60 640 570]{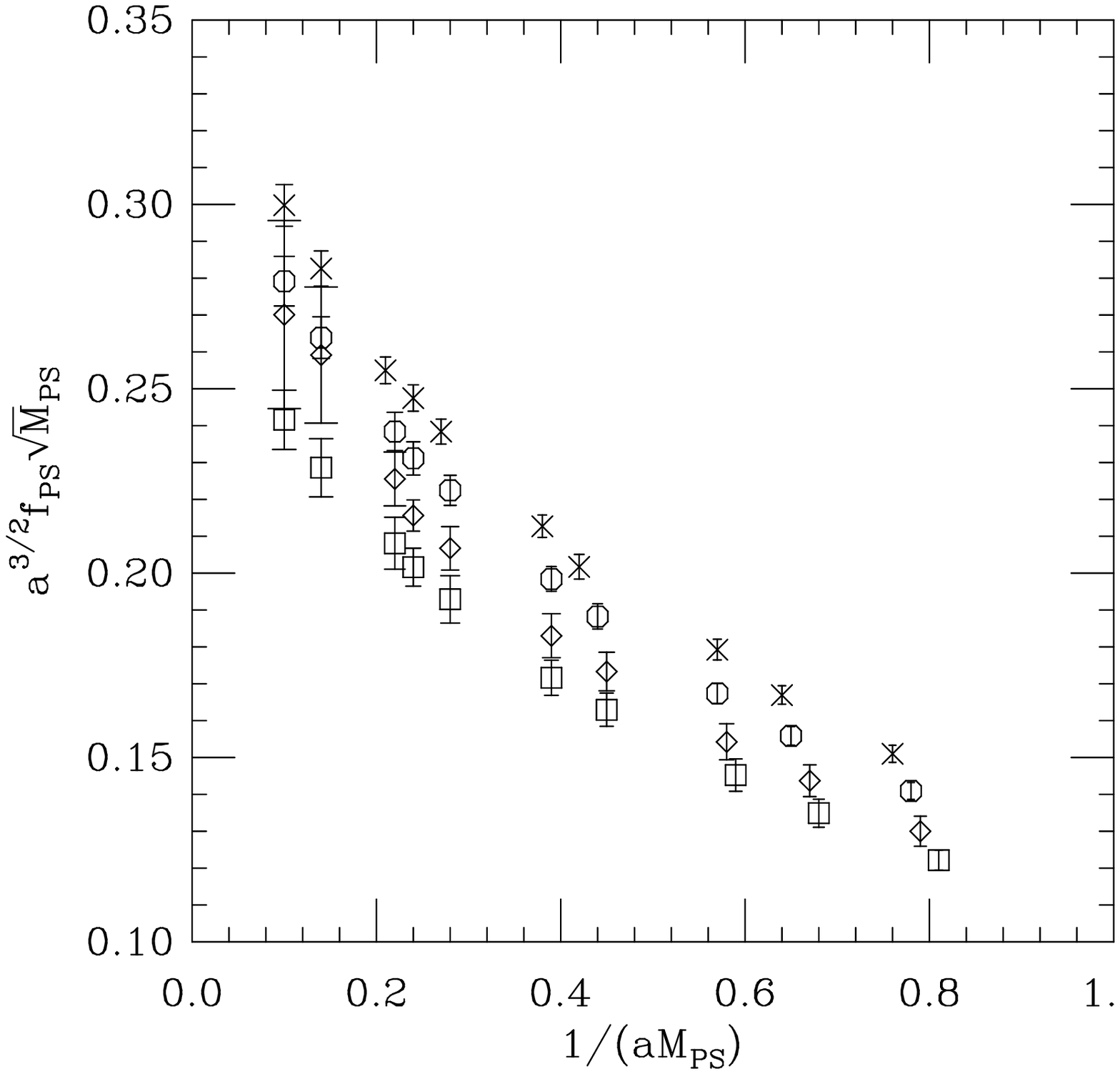}
(b)\setlength{\epsfxsize}{80mm}\epsfbox[10 60 640 570]{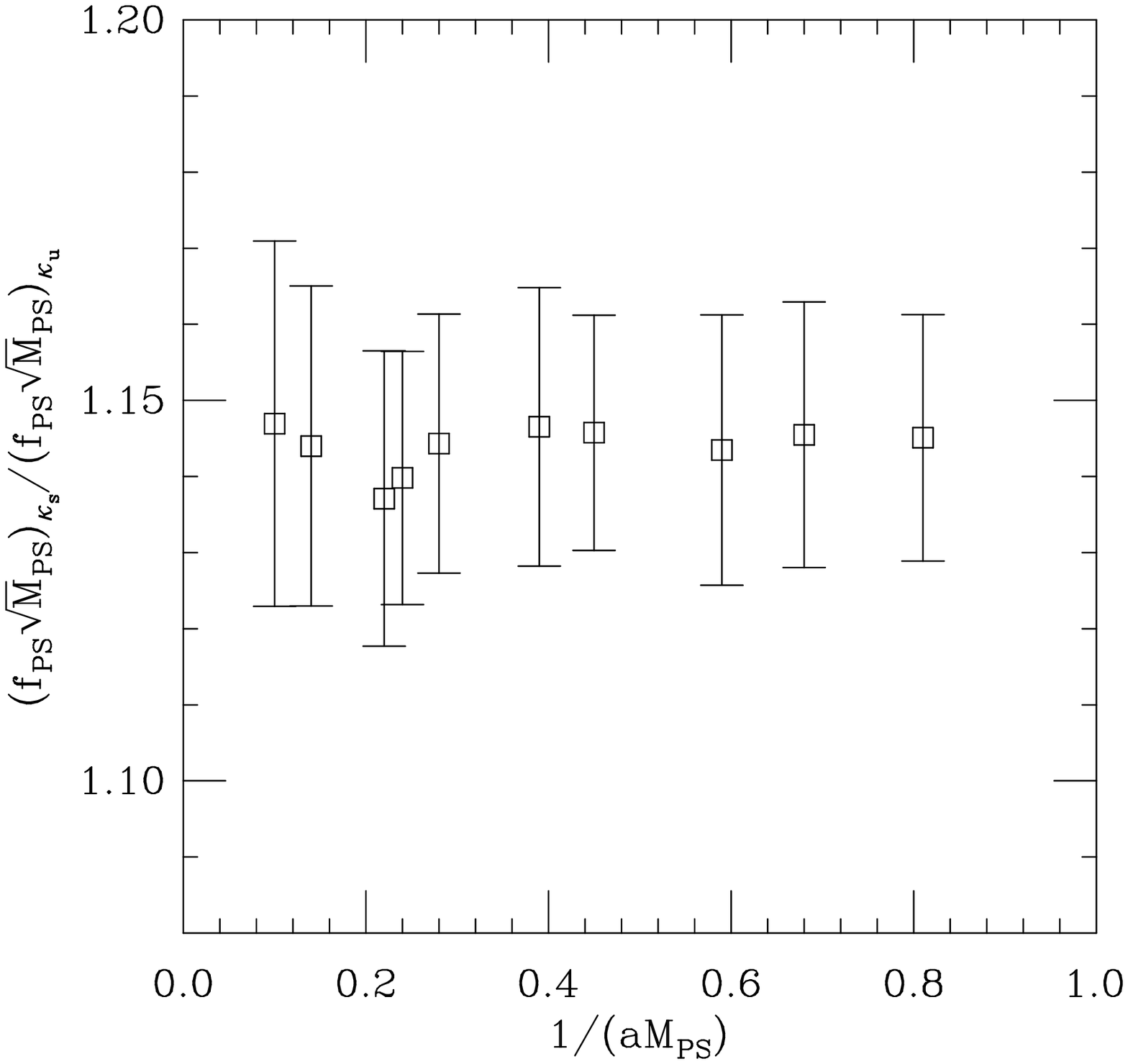}}
\vspace{0.5cm}
\caption{ (a) shows the tree-level $a^{3/2}f\protect\sqrt{M_{PS}}$ for the
  three values of $\kappa_l=0.1385$~(crosses), $0.1393$~(octagons)
  and~$0.1401$~(diamonds), and extrapolated to the chiral limit~(squares) as a
  function of $1/(aM_{PS})$.  (b) shows the heavy quark mass dependence of the
  ratio of $(f\protect\sqrt{M_{PS}})_{\kappa_s}$ to
  $(f\protect\sqrt{M_{PS}})_{\kappa_c}$, where $\kappa_s$ is fixed using the
  $K$ meson.  
  \label{decaylight}}
\end{figure}

We now study the dependence of $f_{PS}\sqrt{M_{PS}}$ on the light and heavy
quark mass. Considering the light quark mass dependence initially, we find the
decay constant is linearly dependent on $m_q$ as shown in an example of the
chiral extrapolation of this quantity in the appendix for $aM_0=1.0$. As
discussed in section~\ref{decaysection} the dependence on $am_q$ is expected
to be roughly the same for all $aM_0$, i.e. $C_{PS}$ is insensitive to the
light quark mass. Figure~\ref{decaylight} gives the tree-level values for
$f_{PS}\sqrt{M_{PS}}$ for all heavy quark masses and $\kappa_l$.

We perform a fit to the decay constant for fixed $\kappa_l$ using a
functional form motivated by equation~\ref{hqsym},
\begin{equation}
f_{PS}\sqrt{M_{PS}}= C_0 + C_1/M + \ldots
\label{decayfit}
\end{equation}
where $C_0=(f\sqrt{M})^\infty$ and $C_1/C_0=C_{PS}|_{\kappa_l}$.  The
slopes for each $\kappa_l$ and the chiral limit are given in
table~\ref{fcoef}. We find the slope for the lighter two
$\kappa$'s~(corresponding to $m_q\sim m_s$ and smaller) are consistent
with the chiral limit, $C_{PS}\sim 1$ in lattice units.  However, from
the results for $\kappa_l=0.1385$, we see there is a much stronger
dependence of the decay constant on $1/(aM_{PS})$ for $m_q$ heavier
than $m_s$, although with such large errors it is only a $2\sigma$
effect. This behaviour may be due to discretisation errors in
$C_{PS}$~(which is a light quark quantity) arising
from the light quark action, which increase with the mass of the light
quark: $O(\alpha a m_q)$ and $O((am_q)^2)$ for the clover
action. Since $C_{PS}|_{\kappa_s}\approx C_{PS}|_{\kappa_c}$, the
ratio
$(f_{PS}\sqrt{M_{PS}})_{\kappa_s}/(f_{PS}\sqrt{M_{PS}})_{\kappa_c}$,
has almost no dependence on the heavy quark mass, as seen in
figure~\ref{decaylight}~(b).

\begin{figure}
\centerline{\setlength{\epsfxsize}{100mm}\epsfbox[10 60 640 570]{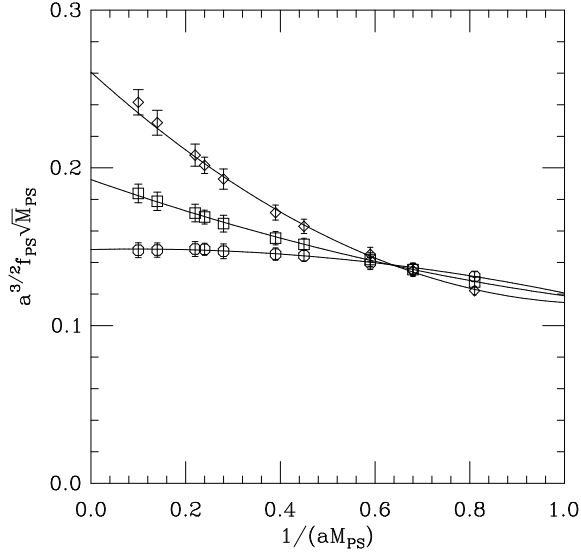}}
\vspace{0.5cm}
\caption{The pseudoscalar decay constant in lattice units fully
  consistent to $O(\alpha/M)$ as a function of $1/(aM_{PS})$. The results
  shown as circles are obtained using $aq^*=1.0$ for the renormalisation
  factors, while the squares use $aq^*=\pi$; the tree-level results are also
  shown as diamonds. $aM_0^b=2.1$, for $a^{-1}=2.0$~GeV, has been used for the
  argument of the logarithms appearing in the matching coefficients, for all
  $aM_0$. The results are for $\kappa_l=\kappa_c$.
\label{pertcomp}}
\end{figure}

The heavy quark mass dependence of $f_{PS}\sqrt{M_{PS}}$ at tree-level and
1-loop is compared in figure~\ref{pertcomp} for $\kappa_l=\kappa_c$. In the
tree-level case there is a steep dependence on the heavy quark mass, and this,
initially suggested higher orders in the heavy quark expansion would be needed
at the $B$ meson~\cite{decay}; $f_{PS}\sqrt{M_{PS}}$ is reduced from the
static limit by $\sim 35\%$ at the $B$ meson.  However, figure~\ref{pertcomp}
shows the slope is dramatically reduced when the 1-loop corrections are
included; $f_{PS}\sqrt{M_{PS}}$ at the $B$ meson is reduced by $\sim 17\%$
compared to the static limit for $aq^*=\pi$, while there is almost no
dependence on $M$ for $aq^*=1.0$.

We perform a fit to the data sets using equation~\ref{decayfit}, the
results are detailed in table~\ref{sloperenorm}. The relative slope at
tree-level is $\sim -1$ in lattice units or $\sim -2$~GeV, using
$a^{-1}=2.0$~GeV. This is well above the naive expectation of $-300$
to $-500$~MeV~($O(\Lambda_{QCD})$). However, at $1$-loop this is
reduced to anywhere between $-1$~GeV and $0$, depending on $q^*$.
Note that $q^*$ depends on the mass, and this dependence must be
determined before the slope can be extracted correctly. However, the
results with fixed $q^*$ provide a rough bound on the size of the
slope. Previous calculations in the quenched approximation using the
clover action and the Fermilab approach~\cite{kronfeld} find the slope
to be close to the upper limit around
$-1$~GeV~\cite{ukqcd,bernard,jim,hash}. Note that the values of $q^*$
are also not known for this method.

There are clear advantages to simulating at finite heavy quark mass
with an inverse lattice spacing around $2.0$~GeV. The combined $1$-loop
corrections grow as $M$ tends towards the static limit~(with the
$\alpha \ln(aM)/\pi$ terms in the matching coefficients fixed to
$\alpha \ln(aM_0^b)$), as does the accompanying uncertainty due to the
omission of higher orders in the matching coefficients. Thus, for a
coarser lattice spacing, i.e.  with a larger value for $aM_0^b$, (and
indeed for the static limit) a nonperturbative determination of the
matching factors becomes essential. Certainly this is required for a
determination of the slope of the decay constant. For a finer lattice,
$aM_0^b$ becomes smaller, and at some point NRQCD breaks down: the
bare lattice matrix elements grow larger, i.e. increasingly higher
orders in the NRQCD expansion become important, while the
$\alpha/(aM)$ terms in the matching coefficients also become large and
perturbation theory is not well controlled.

\begin{figure}
\centerline{\setlength{\epsfxsize}{100mm}\epsfbox[10 60 640 570]{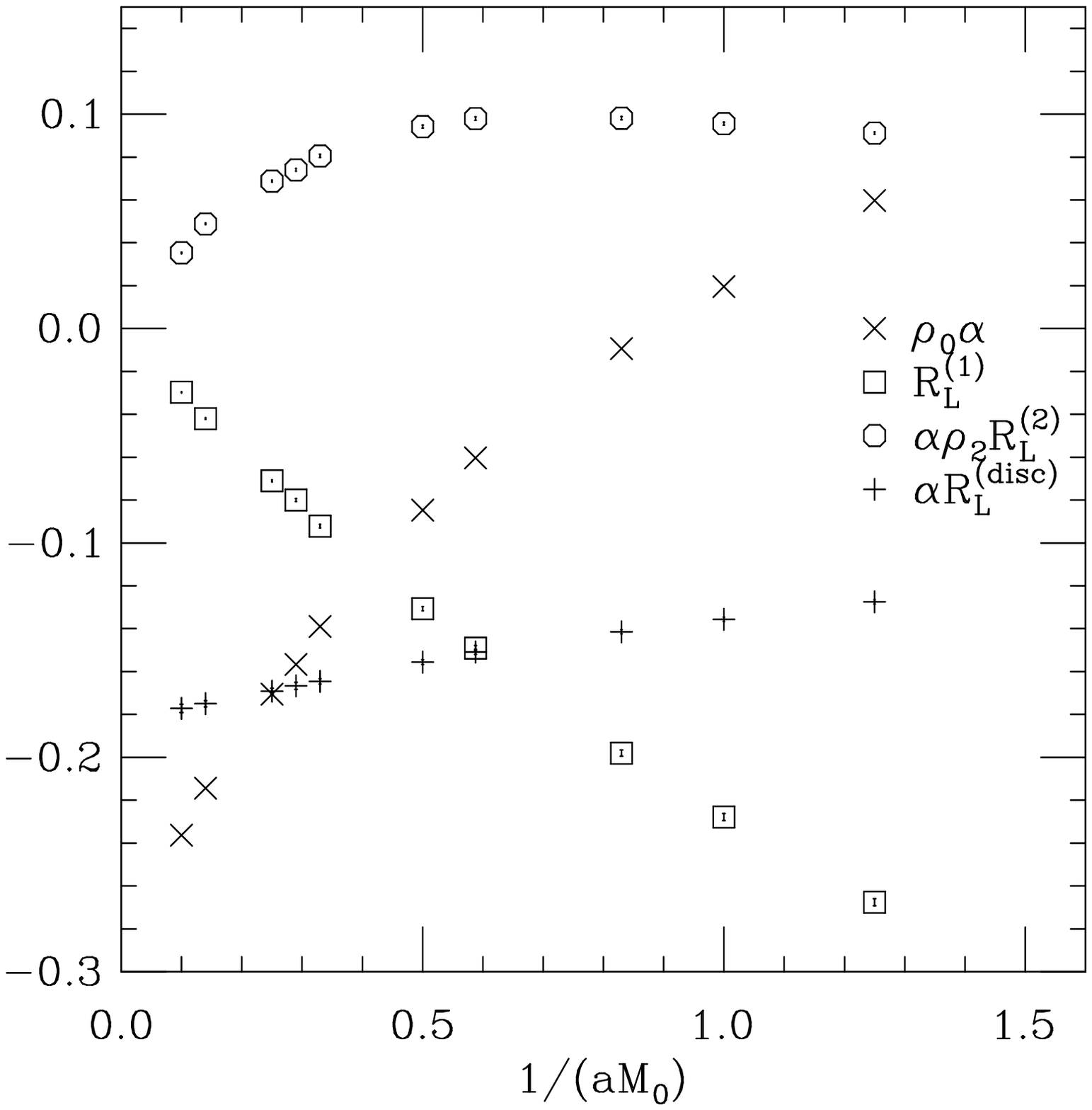}
\setlength{\epsfxsize}{100mm}\epsfbox[10 60 640 570]{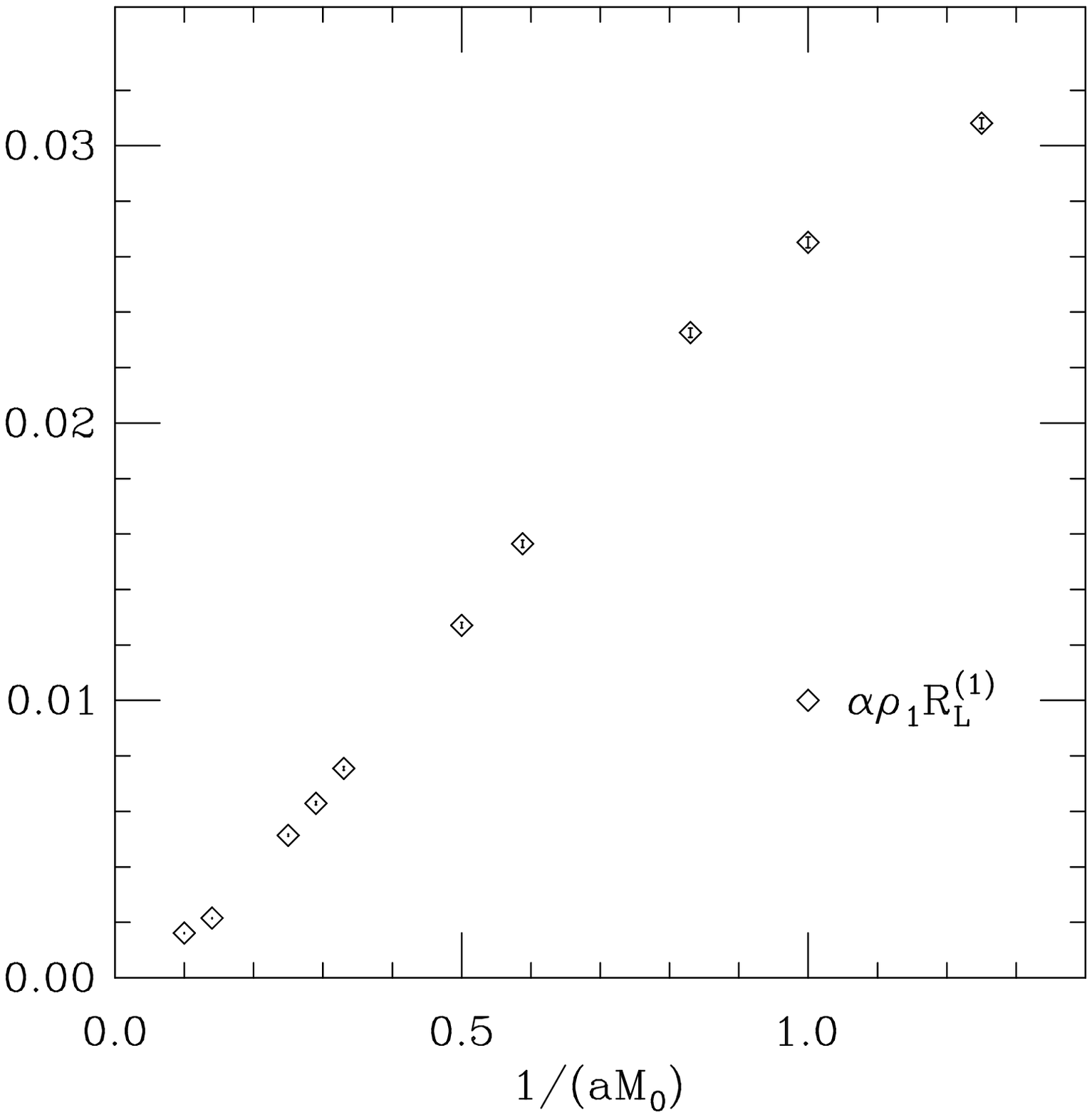}}
\vspace{0.5cm}
\caption{The ratio of individual tree-level and $1$-loop  current corrections
  to the zeroth order $PS$ decay constant, $f^{(0)}\protect\sqrt{M_{PS}}$, 
  as a function of $1/(aM_0)$.
 $R_L^{(i)}=\left<J_L^{(i)}\right>/\left<J_L^{(0)}\right>$. $\kappa_l=\kappa_c$
  and $aq^*=1.0$.
$aM_0=aM_0^b$ has been used for the argument of the
logarithms appearing in the matching coefficients.
\label{pertcontrib1}}
\end{figure}

The interplay between the individual $PS$ matrix elements and their
effect on the slope of the decay constant can be seen in
figure~\ref{pertcontrib1}. The tree-level and 1-loop corrections to
the zeroth order matrix element are shown~(as percentages) as a
function of $1/(aM_0)$ for $aq^*=1.0$.  The $O(1/M)$ and $O(\alpha/M)$
terms, $\left<J^{(1)}\right>$ and $\alpha\rho_2\left<J^{(2)}\right>$
and $\alpha\rho_1\left<J^{(1)}\right>$, respectively, vanish in the
static limit and grow as $aM_0$ becomes small. The discretisation
correction has very little dependence on the heavy quark mass, while
the $O(\alpha)$ correction to $\left<J^{(0)}\right>$ decreases
dramatically from the static limit, and goes through zero as $aM_0$
becomes lighter.

The corrections contribute with differing signs leading to some
cancellation between them. Towards the heavy mass limit,
$1/(aM_0)<0.5$, the $O(1/M)$ and $O(\alpha/M)$ terms approximately
cancel each other. This leaves the $O(\alpha)$ and $O(a\alpha)$ terms
which are both large in this region, $\sim-20\%$ each, leading to a
very large 1-loop correction to $f_{PS}^{(0)}\sqrt{M_{PS}}$. In
contrast, for the lighter values of $aM_0<1.2$, $\alpha\rho_0$ is much
smaller, while the $O(\alpha/M)$ terms are now large enough to
approximately cancel the discretisation correction. Thus, surprisingly,
at $1/aM_{PS}\sim 0.6$, the decay constant is roughly equal to the
tree-level value. Note, however, that in this region the 1-loop terms
$\alpha\rho_2\left<J^{(2)}\right>$ and $\left<J^{disc}\right>$ are not
small at around $10\%$.

Thus, the almost complete removal of any heavy quark mass dependence when the
$1$-loop terms are included using $aq^*=1.0$, is due to the large $1$-loop
correction in the large mass limit~(where $O(1/M)$ terms are small), $\sim
-40\%$ in the static limit, and a large tree-level $O(1/M)$ correction for the
lighter meson masses~(where the $1$-loop terms cancel), $\sim -30\%$ for
$aM_0=0.8$. Since the tree-level correction is so large for the lighter meson
masses, higher orders in the NRQCD expansion should be considered in this
region.

\begin{figure}
\centerline{\setlength{\epsfxsize}{100mm}\epsfbox[10 60 640 570]{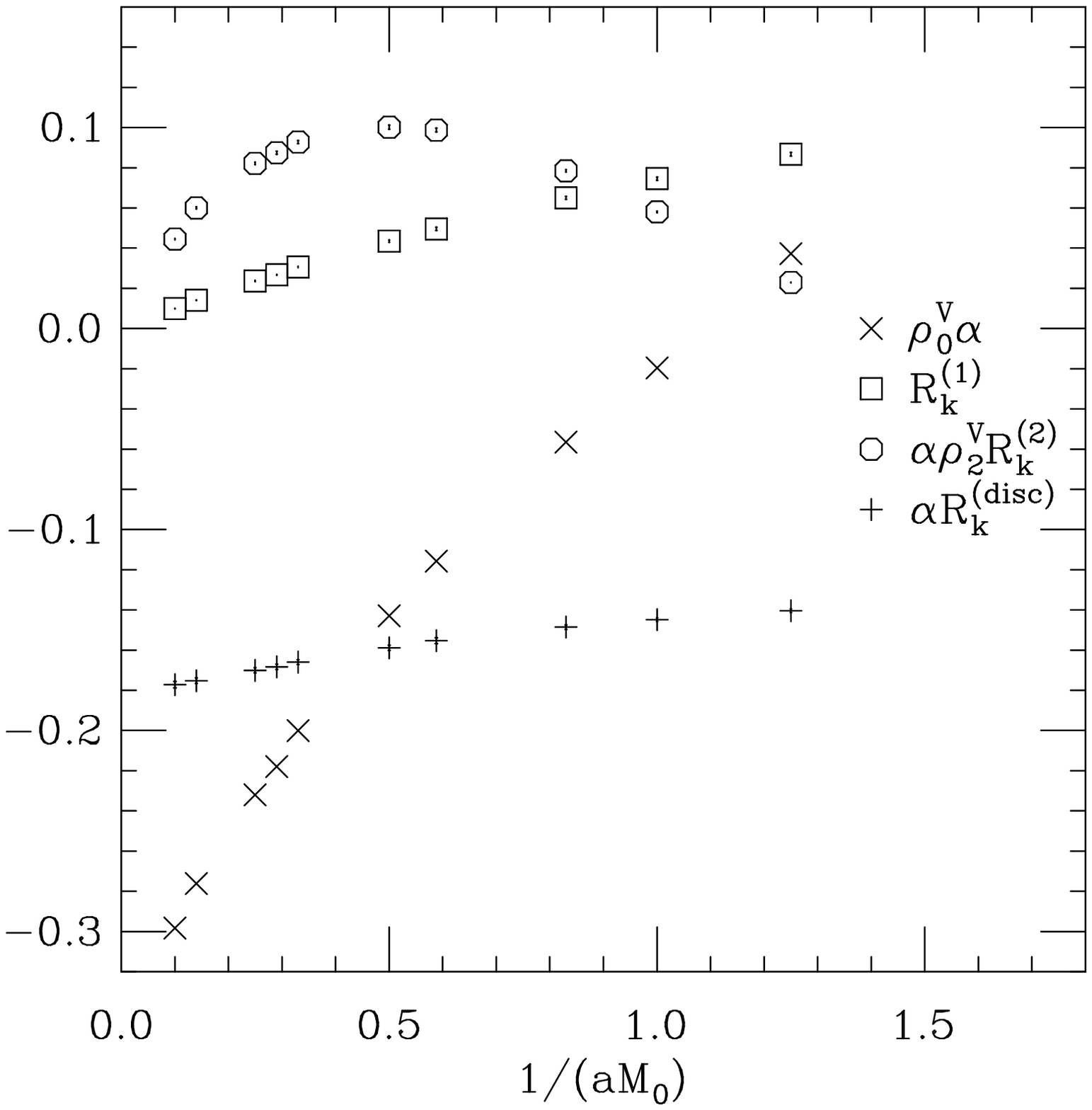}
\setlength{\epsfxsize}{100mm}\epsfbox[10 60 640 570]{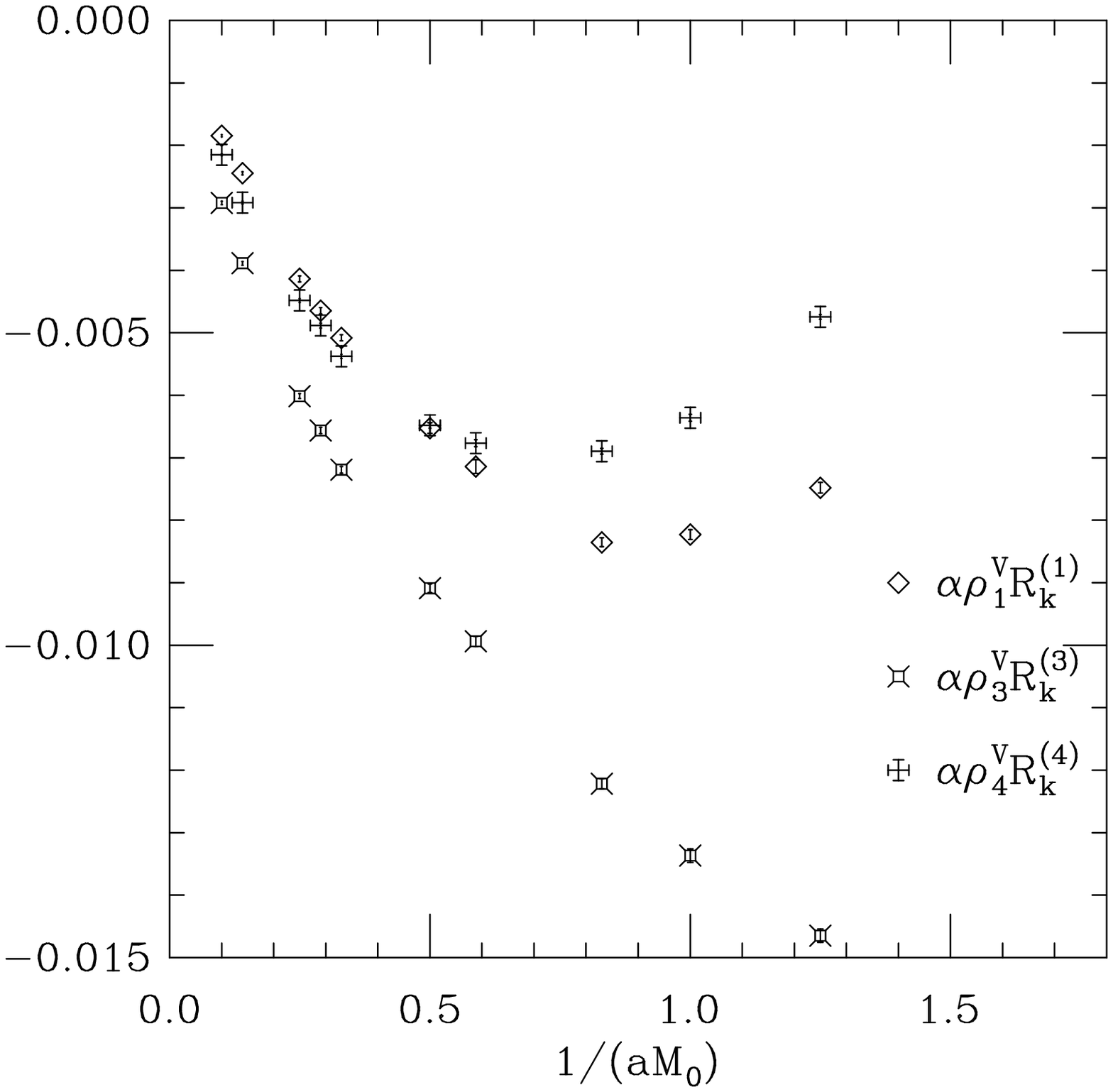}}
\vspace{0.5cm}
\caption{The individual tree-level and $1$-loop current corrections to the
  zeroth order $V$ decay constant, $f^{(0)}_V\protect\sqrt{M_{PS}}$ as a
  function of $1/(aM_0)$.
 $R_k^{(i)}=\left<J_k^{(i)}\right>/\left<J_k^{(0)}\right>$. $\kappa_l=\kappa_c$ and
  $aq^*=1.0$. $aM_0=aM_0^b$ has been used for the argument of the logarithms
  appearing in the matching coefficients.
\label{pertcontrib2}}
\end{figure}

The vector decay constant behaves in a similar way as a function of
the heavy quark mass to the pseudoscalar. Figure~\ref{pertcontrib2}
details the individual percentage corrections to $f^{(0)}_V\sqrt{M_V}$
versus $1/(aM_0)$. The bare matrix elements, $\left<J^{(i)}_k\right>, i=1,4$
are related to each other as given in
equations~\ref{equvec1}{-}~\ref{equvec3}. In the heavy quark limit,
these expressions simplify and furthermore one finds additional
relations between the vector and pseudoscalar current corrections.
If we re-write $J^{(1)}_k$ and $J^{(3)}_k$ in the form,
\begin{eqnarray}
J^{(1)}_k& =& -\frac{1}{2M_0} \bar{q} \left(- \gamma_{\perp}\cdot D_{\perp} + \gamma_k D_k\right) \gamma_k Q\\
J^{(3)}_k& =& -\frac{1}{2M_0} \bar{q}  \gamma_k D_k \gamma_k Q
\end{eqnarray}
and take $\left|V_k\right>=[Q^{\dagger} \gamma_k q] \left|0\right>$, then by assuming
that the contraction of $QQ^{\dagger}$ is spin diagonal~(in the large M
limit) we can deduce,
\begin{eqnarray}
\left<0\right| \bar{q} \left(\gamma_i D_i\right) \gamma_k Q \left|V_k\right>\hspace{0.3cm} (i \ne k) &  = & \left<0\right| \bar{q} \left(\gamma_k D_k\right) \gamma_k Q\left|V_k\right>,\label{vecrel}
\end{eqnarray}
i.e. $\left<J^{(1)}_k\right>=-\left<J^{(3)}_k\right>$. Tables~\ref{deltavcor}
and~\ref{deltavloop} show that our results agree with this
relation for the heavier values of $M\raisebox{-.6ex}{$\stackrel{\textstyle{>}}{\sim}$} 3.0$.  Thus,
equation~\ref{equvec1} reduces to $\left<J^{(2)}_k\right> =
3\left<J^{(3)}_k\right>=-3\left<J^{(1)}_k\right>$.

Following the same argument, and taking $\left|PS\right> = [Q^{\dagger}
\gamma_5 q] \left|0\right>$, one can show that $\left<J^{(2)}_L\right>$ for the
PS is equal to $\left<J^{(2)}_k\right>$ for the V,
\begin{eqnarray}
\left<0\right| \bar{q} (\gamma_j \stackrel{\leftarrow}{D}_j) \gamma_0\gamma_k Q \left|V_k\right>  
& = &  \left<0\right| \bar{q} (\gamma_j \stackrel{\leftarrow}{D}_j) \gamma_5\gamma_0 Q \left|PS\right>\label{psrel}
\end{eqnarray}
for any $j$. Thus the corresponding discretisation corrections are
also equal, $\left<J^{(disc)}_k\right>=\left<J^{(disc)}_L\right>$, and
$\left<J^{(1)}_L\right>=\left<J^{(2)}_L\right>=-3\left<J^{(1)}_k\right>$; the latter
clearly comes from spin-symmetry. The results in
figures~\ref{pertcontrib1} and~\ref{pertcontrib2} in the large mass
limit agree with these expressions.  In general the terms which spoil
equations~\ref{vecrel} and~\ref{psrel} at finite $M$ are due to the
$O(1/M)$ terms in the action modifying the meson wavefunction from the
static limit~(as in equation~\ref{ghyp} and~\ref{gkin} for
the PS decay constant), and are small since they appear as an $O(1/M)$
correction to the matrix element.

\subsection{Making contact with HQET}

As discussed in section~\ref{decaysection}, one can attempt to analyse the 1/M
corrections to the static limit within the framework of HQET.
Equations~\ref{slope} and~\ref{slopehyp} suggest that by taking appropriate
combinations of $f_{PS}\sqrt{M_{PS}}$ and $f_V\sqrt{M_V}$, $G_{kin}$ and
$G_{hyp}$ can be separated; since $\left<J^{(1)}\right>$ is calculated separately
$G_{corr}$ can easily be obtained.  In particular, the spin average of the
$PS$ and $V$ decay constants~(without $\left<J^{(1)}\right>$),
$\overline{f\sqrt{M}}$, cancels the hyperfine contribution, and the slope of
this quantity is purely determined by $G_{kin}$. Conversely, the ratio of
these quantities cancels $G_{kin}$ and a determination of the slope gives
$G_{hyp}$.

Note that the renormalisation factors calculated by Morningstar and
Shigemitsu are not those required for a detailed investigation of spin
and flavour symmetry within HQET.  In references~\cite{junkmorn1}
and~\cite{junkmorn2} the lattice NRQCD matrix elements are matched
directly to full QCD while, the factors matching lattice NRQCD to
continuum HQET are the ones that are needed.  In particular, only the
tree-level decay constants will obey the spin symmetry relation that
the ratio of $V$ to $PS$ is $1$ in the static limit.  Thus, we
restrict ourselves to a tree-level analysis of spin symmetry and the
origin of the slope $C$.  One-loop matching to full QCD would
introduce a short distance correction factor to the ratio
$f_V\sqrt{M_V}/f_{PS}\sqrt{M_{PS}}$ of $\eta_{static} = [ 1 - \alpha
(8/(3\pi)) ]$.

\begin{figure}
\centerline{(a)\setlength{\epsfxsize}{80mm}\epsfbox[10 60 640 570]{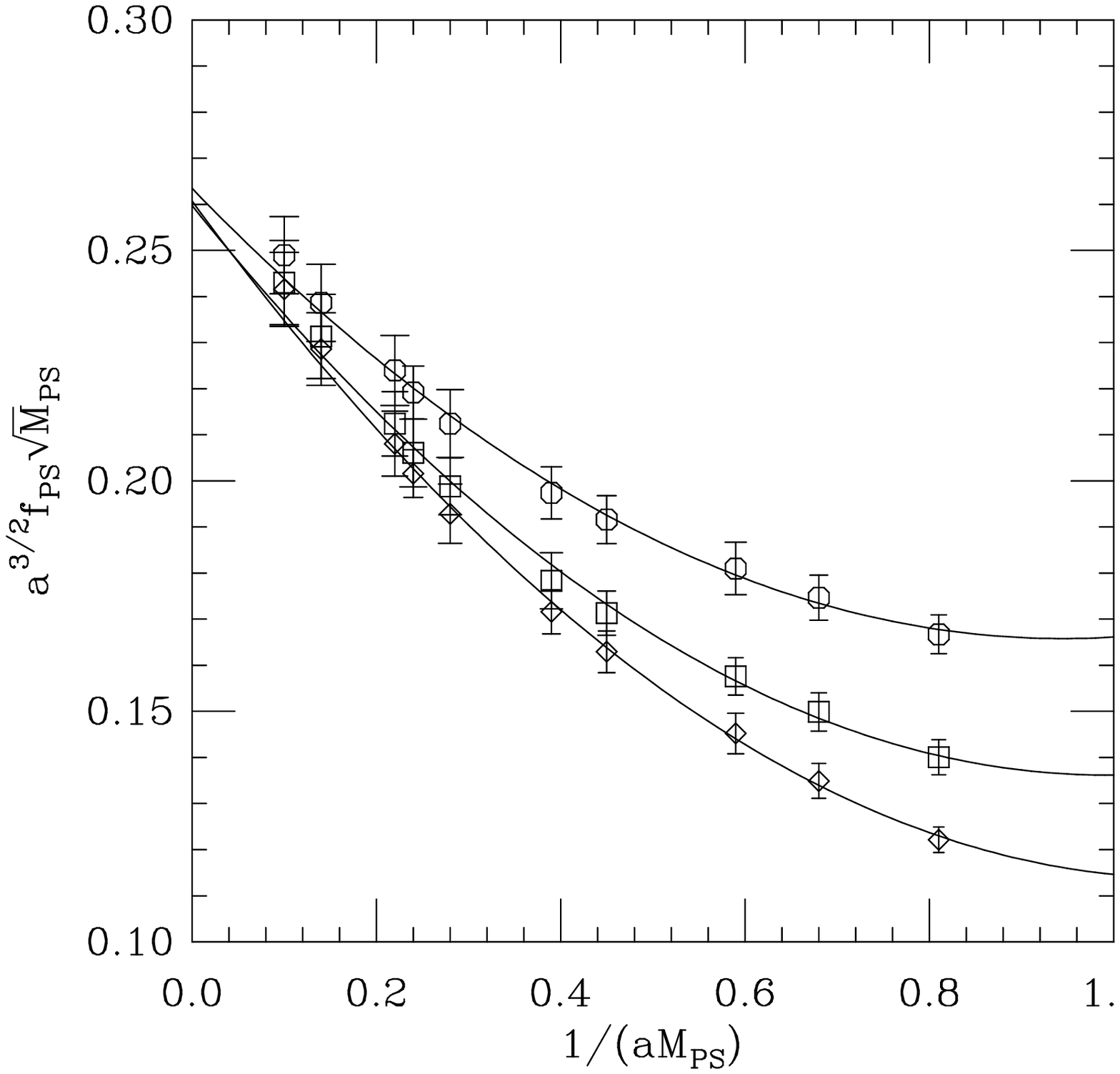}
(b)\setlength{\epsfxsize}{80mm}\epsfbox[10 60 640 570]{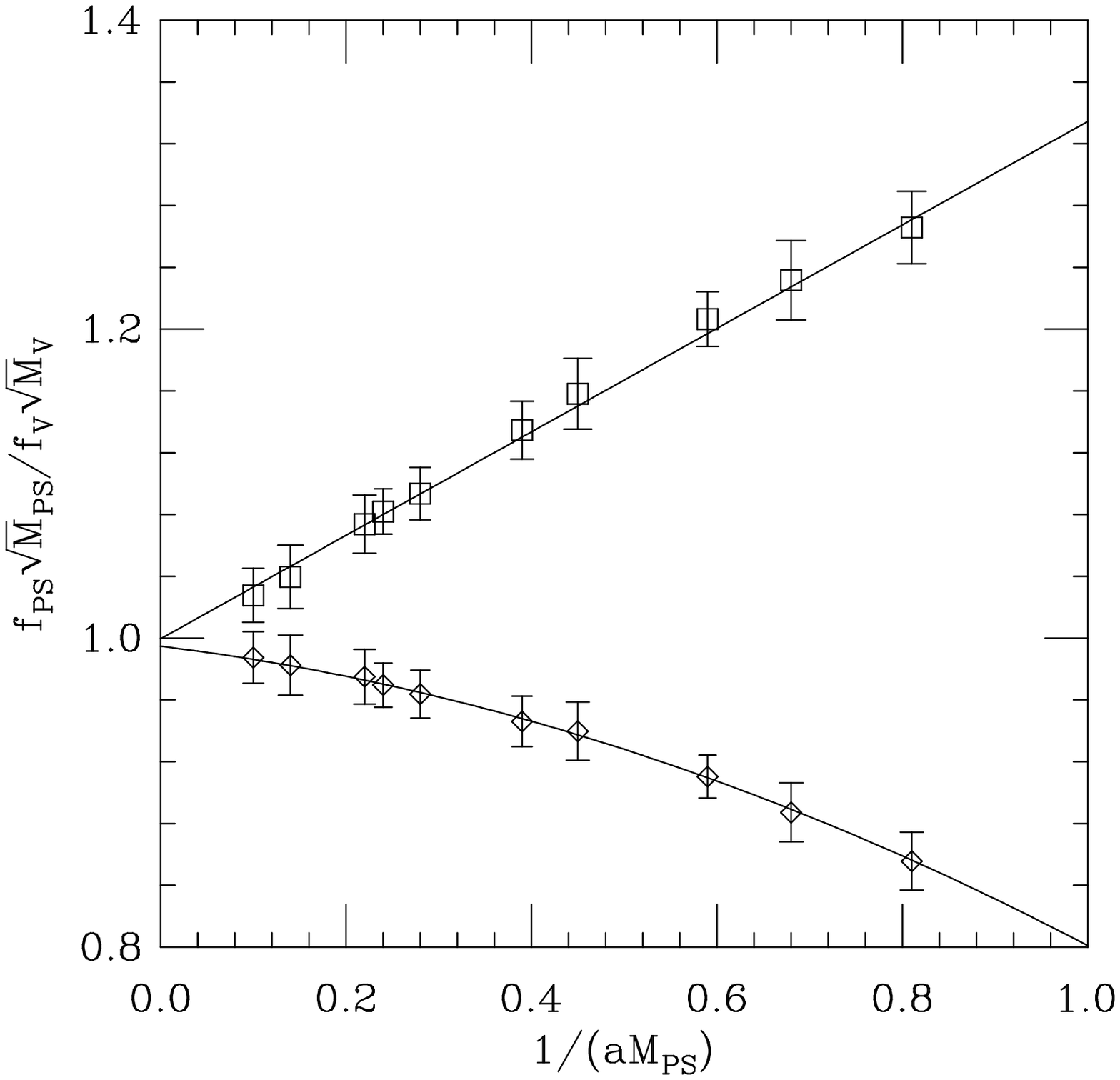}}
\vspace{0.25cm}
\centerline{(c)\setlength{\epsfxsize}{80mm}\epsfbox[10 60 640 570]{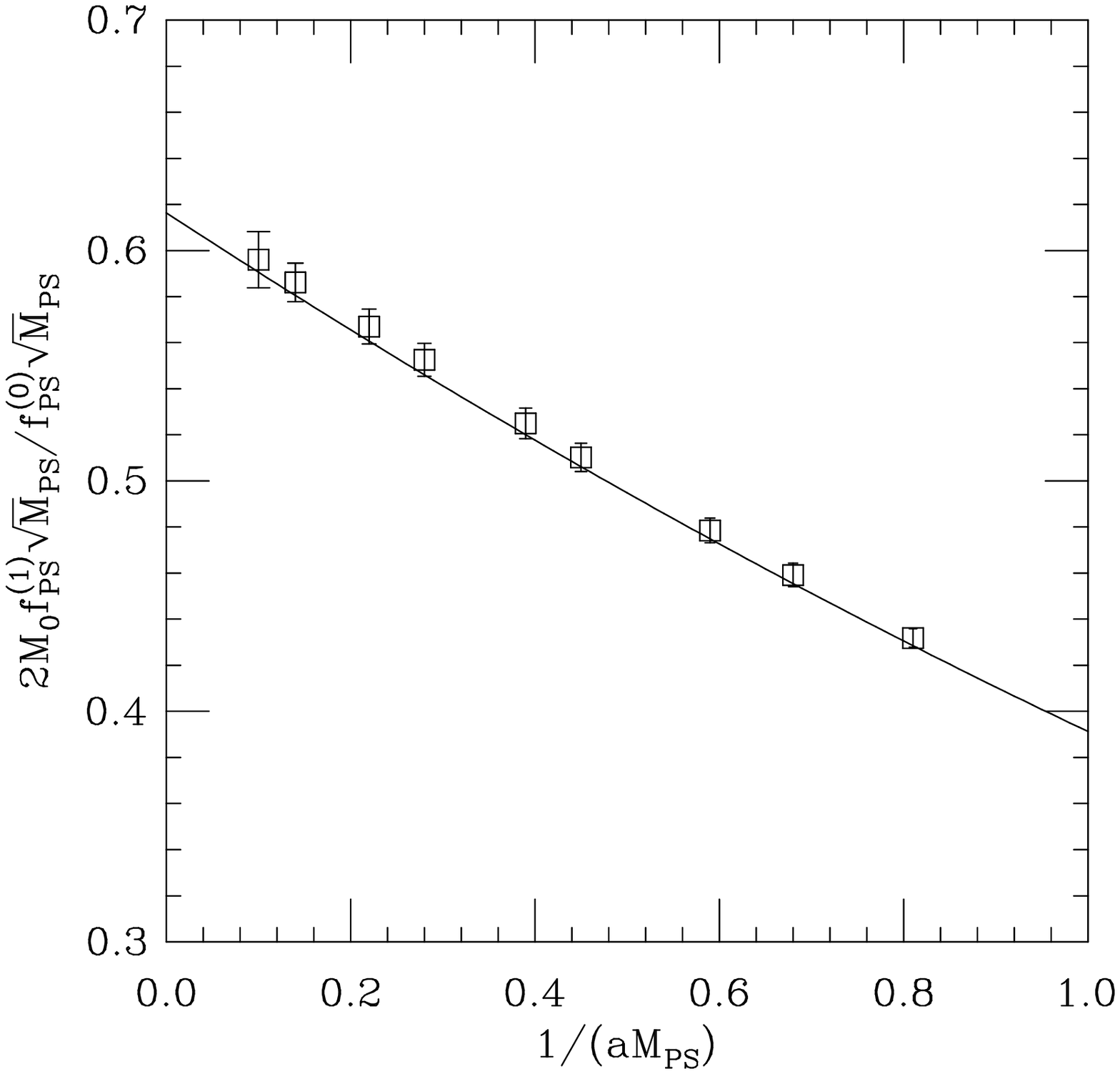}
}
\vspace{0.5cm}
\caption{Heavy quark dependence of various combinations of the
  tree-level $PS$ and $V$ decay constants in lattice units for
  $\kappa_l=\kappa_c$.  (a) presents $f\protect\sqrt{M_{PS}}$~(diamonds),
  $f^{(0)}\protect\sqrt{M_{PS}}$~(octagons) and the spin-average of the $PS$
  and $V$ decay constants~(squares). (b) shows the
  ratio of the $PS$ to the $V$ tree-level decay constant with~(diamonds) and
  without~(squares) the $O(1/M)$ current corrections.
  (c) gives the contribution to the slope of the decay constants from the
  $O(1/M)$ current correction.
\label{hqdep_fp}
}
\end{figure}

Figure~\ref{hqdep_fp} presents the results for $\overline{f\sqrt{M}}$,
compared to the $PS$ decay constant, with and without the tree-level $O(1/M)$
current correction. In the heavy mass limit, all three quantities converge
indicating the dominant contribution to the tree-level $C_{PS}$ is from
$G_{kin}$. The figure also shows $f_{PS}\sqrt{M_{PS}}/f_{V}\sqrt{M_{V}}$.
Without $\left<J^{(1)}\right>$ this ratio is positive~(indicating $G_{hyp}$ is
positive) and tends to $1$ in the static limit, consistent with spin-symmetry;
when the current correction is included $f_V\sqrt{M_V}<f_{PS}\sqrt{M_{PS}}$.
$G_{corr}$ can be very accurately determined by extrapolating
$2M_0\left<J^{(1)}\right>$ to the static limit, displayed in
figure~\ref{hqdep_fp}~(c).

We perform fits to these combinations of $PS$ and $V$ decay constants and
extract $G_{kin}$, $G_{hyp}$ and $G_{corr}$. The results are given in
table~\ref{ftab}. Note that only the slopes of physical combinations of the
decay constants, i.e.  $f_{PS}\sqrt{M_{PS}}$, $\overline{f\sqrt{M}}$ and
$f_{PS}\sqrt{M_{PS}}/f_{V}\sqrt{M_{V}}$, are expected to be
$O(a\Lambda_{QCD})$. The ratio of decay constants is roughly in agreement with
this picture; the slope of the ratio is small, while $G_{corr}$ and $G_{hyp}$
are individually much larger.  However, the physical decay constants
themselves, dominated by $G_{kin}$, have a much larger slope.  Considering the
strong dependence on $M$ of the $O(\alpha)$ correction to
$f^{(0)}_{PS}\sqrt{M_{PS}}$ seen in the results of Morningstar and Shigemitsu,
$G_{kin}$ is likely to be significantly affected by renormalisation.
Similarly, the $1$-loop corrections to $G_{corr}$ are likely to be large.

The results in table~\ref{ftab} through the use of the clover action for the
light quarks and extrapolating to the chiral limit, are an improvement on our
previous results with Wilson light fermions published in
reference~\cite{decay}. Numerically, the difference between the results is
mainly due to the extrapolation to $\kappa_c$ in the clover case~\footnote{The
  Wilson results in reference~\cite{decay} correspond to a light quark mass
  greater than $m_s$ and thus the slope of the decay constant is much larger
  than in the chiral limit, as discussed previously.}, since at tree-level the
decay constant does not change significantly between the two light quark
actions.

\subsection{$n_f$ dependence}
\label{nfdecaycomp}
We compare our results with those calculated as part of the $O(1/M^2)$
quenched simulation at $\beta^{n_f=0}=6.0$ already mentioned in
section~\ref{nfspectrum}. In reference~\cite{arifajunk} we quote,
\begin{eqnarray}
f_B &= & 147(11)(^{+8}_{-12})(9)(6)~\mbox{MeV}\label{respap}\\
f_{B_s}& = & 175(08)(^{+7}_{-10})(11)(7)(^{+7}_{-0})~\mbox{MeV}\label{respap1}\\
f_{B_s}/f_B & = & 1.20(4)(^{+4}_{-0})
\end{eqnarray}
for the quenched result. The errors are calculated in a similar way to
those in this paper and, with the exception of the uncertainty due to
fixing $a^{-1}$, are of a similar size. The first error corresponds to
statistical errors and those errors due to extrapolation/interpolation
in $\kappa$ and $aM_0$. The second error indicates the uncertainty in
$a^{-1}$ calculated using $a^{-1}=2.0$~GeV and $1.8$~GeV. This range
represents the spread in $a^{-1}$ derived from various light
spectroscopic quantities. Higher order perturbative and relativistic
uncertainties are given by the third error and the fourth error is due to
discretisation corrections. For $f_{B_s}$ the error in fixing
$\kappa_s$ is found by fixing the strange quark mass from the $K$
meson~(for the central value) and the $\phi$ meson; the $K^*$ meson
gave the same value of $\kappa_s$ as the $\phi$ meson.

The quenched calculation includes $\delta H_{h.o.}$ in the action and
the tree-level $O(1/M^2)$ current corrections. Since the $O(1/M^2)$
current corrections were calculated separately we can easily omit them
in order to make a better comparison with our present $O(1/M)$ $n_f=2$
results. Apart from this the quenched simulation and analysis remains
the same as detailed in reference~\cite{arifajunk}. The modified
quenched results are presented in table~\ref{comp_pev}. As mentioned
in section~\ref{decayresults}, the contributions from $\delta
H_{h.o.}$ have a minimal effect on the pseudoscalar matrix elements in
the region of the $B$ meson. Thus, there should only be a very small
uncertainty in the comparison between quenched and $n_f=2$ results due
to the differing actions. In fact, since $\delta H_{h.o.}$ tends to
increase $f_B$, removing it would increase rather than decrease the
sea quark effect that we see.

The comparison is made at tree-level, and at $1$-loop for $aq^*=\pi$
and $aq^*=1.0$ in table~\ref{comp_pev}. In perturbation theory the
effects of sea quarks appear beyond $1$-loop.  Thus, $aq^*$ is likely
to be very similar for the same $Ma$ values for quenched and $n_f =
2$. Since the values of $a$ and hence $aM_0^b$ are closely matched for
the two simulations, it is reasonable to compare at fixed $aq^*$.  

Considering, in addition, that the quenched and unquenched simulations
are very similar in method and analysis we expect the systematic
errors to be similar~(and correlated) in both cases. Thus, initially,
we make a comparison considering only the statistical errors.
Table~\ref{comp_pev} shows that $f_B$ is a quantity which is sensitive
to the presence of sea quarks and that the quenched value for $f_B$
may be substantially lower than the value in full QCD.  We find a
$23{-}25\%$ or $>3\sigma$~\footnote{The significance of the shift in
  the decay constant is found by comparing the difference between
  $f_B$ at $n_f=2$ and $0$ with the statistical errors added in
  quadrature.} increase in $f_B$, when two flavours of sea quarks are
included.  The increase is roughly the same at tree-level and at
$1$-loop for $aq^*=\pi$ and $aq^*=1.0$.  If the actual value of $aq^*$
for $n_f=0$ turns out to be higher than that for $n_f=2$ then the sea
quark effect becomes smaller.  In the reverse case the effect would be
larger.

Alternatively, we can ignore the fact that the systematic errors
between the quenched and unquenched results are probably highly
correlated. Combining the statistical and systematic errors in
quadrature, we obtain $f_B=149(19)$~MeV at $n_f=0$, compared to
$186(25)$~MeV at $n_f=2$; the lattice spacing is fixed using $m_\rho$
in both cases, i.e. the uncertainty arising from fixing $a$ is ignored, and
the central value is obtained from the average of the results for
$aq^*=1.0$ and $\pi$. In this case, the $27$~MeV shift in $f_B$ is a
$1\sigma$ effect.  Note that, considering the uncertainty in $a^{-1}$
for $n_f=2$, using $a^{-1}=2.4$~GeV will lead to a larger sea quark
effect. This increase includes any effect due to finite volume
problems for the determination of $a^{-1}$ from $m_\rho$ at $n_f=2$.
As mentioned in section~\ref{errors} $a^{-1}$ is expected to be
underestimated if the $\rho$ meson is squeezed.

Comparing the central values for $f_{B_s}$ the increase with $n_f$ is
slightly smaller, $\sim 20\%$, but statistically more significant at
between $5$ and $6.5\sigma$.  The error in $f_{B_s}$ due to the
uncertainty in $\kappa_s$ clouds the comparison for this quantity
slightly. For the $n_f=2$ configurations, $m_{\phi}$ and $m_{K}$ give
consistent values for $\kappa_s$, while $m_{K^*}$ gives a slightly
higher value~(see section~\ref{simdet}). In contrast, for the quenched
results the error in $\kappa_s$ is larger, and as mentioned
previously, is computed using from using $m_K$ and $m_\phi$; $m_{K^*}$
and $m_\phi$ give consistent results.  Thus, if $m_\phi$ is used the
quenched decay constant is increased by $7$~MeV, while the $n_f=2$
value is unchanged. The difference between the two values for
$f_{B_s}$ is reduced to $14{-}16\%$ or a $4{-}5.5\sigma$ effect~(a
similar decrease is found using $m_{K^*}$). A detailed study of how to
fix $\kappa_s$ is needed to improve the comparison of quenched and
unquenched results for $f_{B_s}$. If the systematic errors are taken
in account, $f_{B_s}=180(20)$~MeV at $n_f=0$ and $215(29)$~MeV for
$n_f=2$; the central value is found using $\kappa_s$ from $m_K$, but
the associated uncertainty in $\kappa_s$ is included in the error.
Assuming the errors are independent, the $35$~MeV shift corresponds to
$1\sigma$.

The effect of sea quarks in the decay constants have also been studied
by the MILC collaboration~\cite{milc}. $f_B$ was found to increase by
$\sim 13\%$ from the quenched to the $n_f=2$ result. The Wilson action
was employed for the heavy and light quark at several $\beta$ values
at both $n_f=0$ and $2$. Extrapolating the results to the continuum
limit these authors find $f_B = 159
(11)(stat)(+22-9)(sys)(+21-0)(quench)$~MeV.  Clearly systematic
errors~(of which discretisation errors are the dominant source) are as
large an effect as quenching. In this situation the size of the sea
quark effects is very sensitive to the interpolation of the $n_f=0$
results to a lattice spacing which coincides with that found at
$n_f=2$ to enable a comparison. However, until the systematic errors
are explored thoroughly at $n_f=2$, the difference between our results
and those from the MILC collaboration are not significant.

\begin{figure}
\centerline{\setlength{\epsfxsize}{80mm}\epsfbox[10 60 640 570]{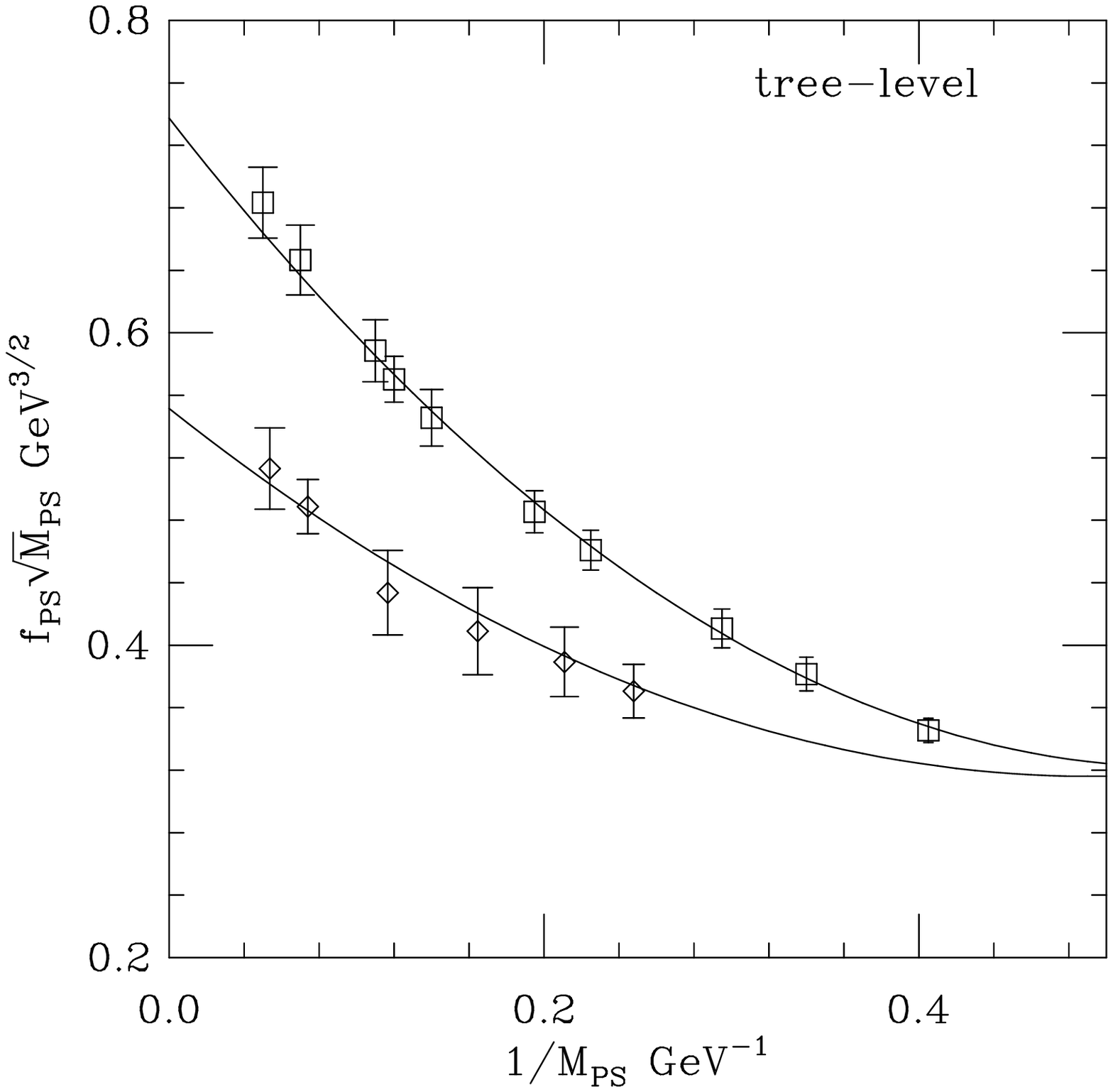}
\setlength{\epsfxsize}{80mm}\epsfbox[10 60 640 570]{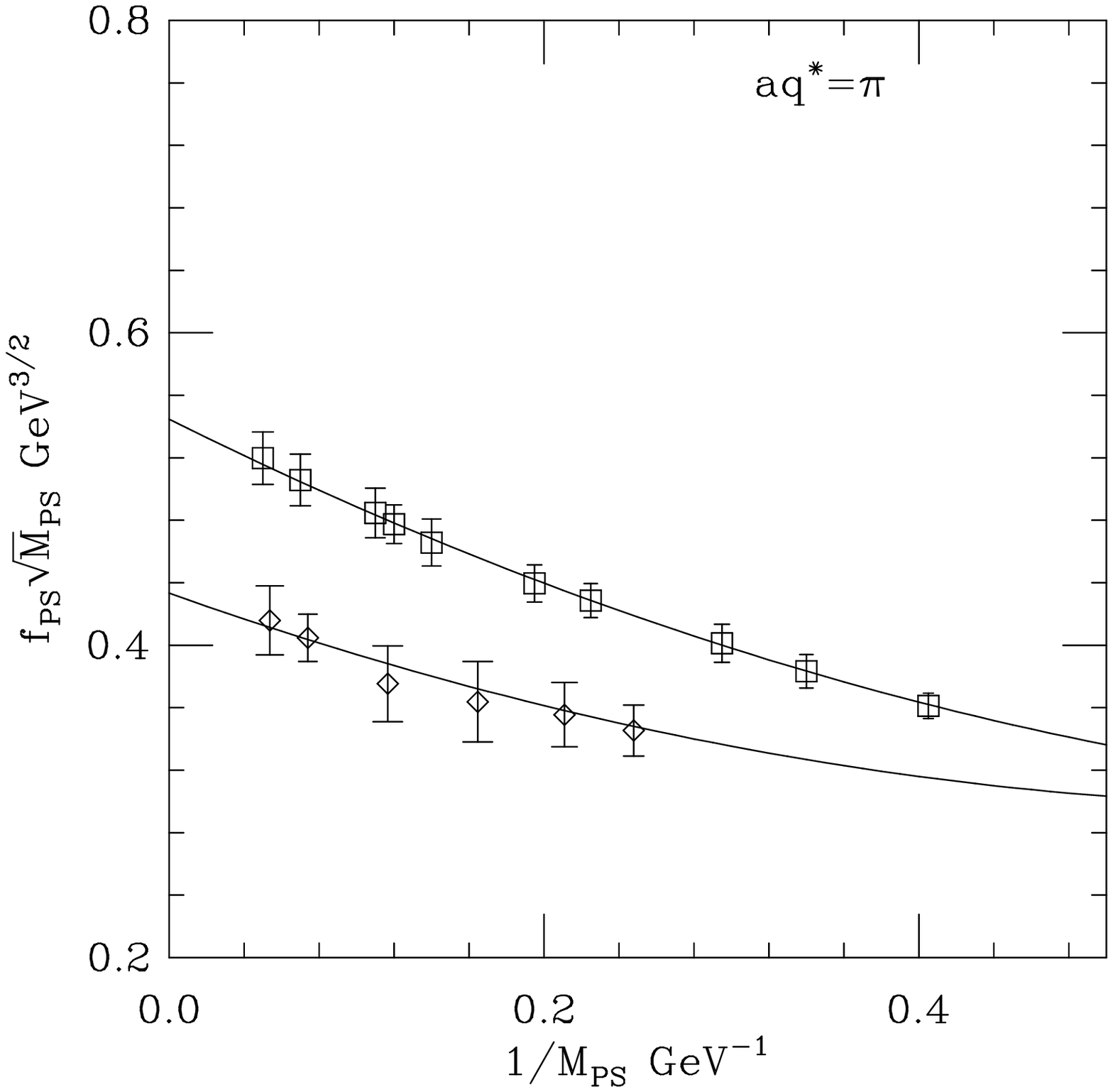}}
\centerline{\setlength{\epsfxsize}{80mm}\epsfbox[10 60 640 570]{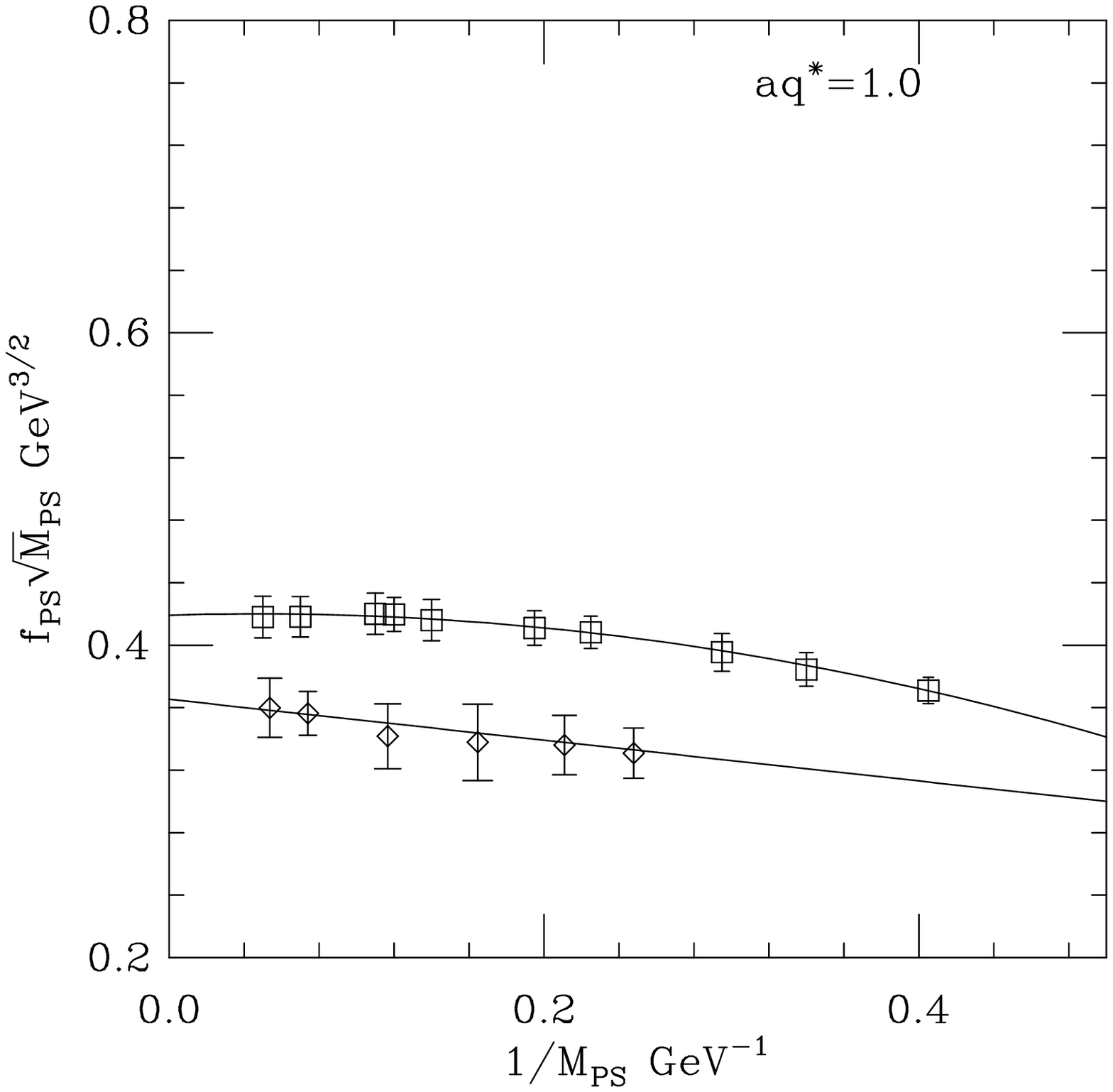}}
\vspace{0.5cm}
\caption{The renormalised pseudoscalar decay constant as a function of
  $1/M_{PS}$ in physical units for $\beta^{n_f=0}=6.0$~(diamonds
  $1/M^2$) and $\beta^{n_f=2}=5.6$~(squares) for tree-level and
  $1$-loop for $q^*=\pi$ and $q^*=1.0$. $aM_0^b$ has been used for
  the argument of the logarithms appearing in the matching
  coefficients.
\label{pertnf}}
\end{figure}

Since the matching factors have been calculated over a wide range of
$M$, we can also investigate the mass dependence of the sea quark
effects. Figure~\ref{pertnf} presents the results for tree-level and
$1$-loop for two values of $aq^*$. The roughly constant shift between
$n_f=0$ and $n_f=2$ for the whole range of meson masses from $\sim
M_B/3$ to $4M_B$ indicates the sea quark effects are associated with the
light quark rather than dependent on the heavy quark.  We also compare
the results for the ratio $f_{B_s}/f_B$.  The large sea quark effects
we see in $f_{B_s}$ and $f_B$ individually appear to cancel in the
ratio. There is only a $\sim5\%$ or $1.3\sigma$ decrease in
$f_{B_s}/f_B$ when two flavours of sea quarks are introduced. If
$\kappa_s$ from $m_\phi$ is used the results are consistent to within
less than $0.5\sigma$.

\section{Conclusions}
\label{conc}

We presented comprehensive results for the spectrum and decay
constants of the $B$ meson including the effects of two flavours of
dynamical quarks.  In addition, an investigation of the effects of sea
quarks through a comparison with quenched simulations was performed.
At present, we do not attempt to extrapolate in $n_f$ but look for
initial indications of quenching effects.

Results, summarised in table~\ref{prediction}, were presented for the lower
lying $S$ and $P$ meson states and the $\Lambda_b$ baryon. With the exception
of the $\Lambda_{b}{-}B$ splitting, the estimates of the systematic errors for
all mass splittings were comparable to or less than the statistical errors.
The $\Lambda_b{-}B$ splitting represented the only significant disagreement
with experiment. This discrepancy is probably due to finite volume effects.

A detailed analysis of the heavy quark mass dependence of physical
spectral mass splittings was performed.  We found the behaviour with
$M$ to fulfill the naive expectation: $O(\Lambda_{QCD})$ quantities
are weakly dependent on $M$, while spin-dependent splittings vanish as
$1/M$ in the large mass limit.  Estimates of the slope and intercept
of these quantities were extracted and agreement was found with the
theoretical expectation from HQS.  A comparison was also made with
experiment. This showed that the slope for the hyperfine splitting was
too low in our simulation. Residual quenching effects are a possible
explanation for this discrepancy. However, we have underestimated the
splitting through the use of the tree-level value of the coefficient
of the hyperfine term in our simulation~(this underestimate is not
significant within statistics at the $B$ meson).  $c_B$ depends on $M$
and improved estimates of this coefficient must be included for all
$aM_0$ before the origin of the discrepancy is clear.

We presented a comparison of our results with a similar quenched
calculation at $\beta^{n_f=0}=6.0$ using a $O(1/M^2)$ action detailed
in~\cite{arifaspec2}.  A comparison with a higher order NRQCD action
is possible as the additional terms in the action do not significantly
affect the mass splittings considered. The quenched results reproduce
the experimental spectrum well to the present level of accuracy, and
there is no significant difference between the quenched and $n_f=2$
results.

In addition, results were presented for the $f_B$, $f_{B_s}$,
$f_{B^*}$ and $f_{B^*_s}$ decay constants at tree-level and fully
consistent to 1-loop in perturbation theory. We found the combined
$1$-loop corrections are moderate and reduce the tree-level value of
$f_B$ by $10{-}16\%$ depending on $q^*$.  Our predictions for the $PS$
and $V$ decay constants are
\begin{eqnarray}
f_B &=&  186(5)(stat)(19)(pert)(9)(disc)(13)(\mbox{{\small NRQCD}})(+50)(a^{-1})~\mbox{MeV},  \nonumber\\
f_{B_s} &=&  215(3)(stat)(22)(pert)(9)(disc)(15)(\mbox{{\small NRQCD}})(+49)(a^{-1})(-5)(\kappa_s)~\mbox{MeV}, \nonumber \\
 f_B^* & =&  181(6)(stat)(18)(pert)(9)(disc)(13)(\mbox{{\small NRQCD}})(+55)(a^{-1})~\mbox{MeV},\nonumber \\
f_{B_s^*} & = &  213(4)(stat)(21)(pert)(9)(disc)(15)(\mbox{{\small NRQCD}})(+60)(a^{-1})(-4)(\kappa_s)\mbox{MeV}.\nonumber
\end{eqnarray}
Adding the statistical and systematic errors in quadrature~(errors
$1{-}4$), there is a combined uncertainty of $14\%$ in the decay
constants. This indicates a determination of $f_B$ to around $10\%$ is
possible using an $O(1/M)$ NRQCD action with clover light fermions.
However, first of all the error due to the uncertainty in the lattice
spacing~($\sim 30\%$) must be significantly reduced.  Conversely, for
$f_{B_s}/f_B$ and $f_{B_s^*}/f_{B^*}$ most systematic errors cancel,
and these quantities can be more easily determined accurately. We find
\begin{eqnarray}
f_{B_s}/f_B & = & 1.14(2)(stat)(-2)(\kappa_s)\nonumber,\\
f_{B_s^*}/f_{B^*} & = & 1.14(2)(stat)(-2)(\kappa_s)  \nonumber .
\end{eqnarray}

We presented a detailed study of the heavy quark mass dependence of the decay
constants. We find the steep dependence of the tree-level $PS$ decay constant,
for which $C_{PS}\sim 2$~GeV, is dramatically reduced when the $1$-loop terms
are included; $C_{PS}=1{-}0$~GeV depending on $q^*$. We show how this occurs
through the cancellation between the various $1$-loop and tree-level current
corrections as the individual terms vary with the heavy quark mass.

We have shown that simulating directly at the $B$ with $a^{-1}$ such
that $aM_0^b$ is in the region of $2$ is advantageous. In particular,
this value of $aM_0$ sits in the window between large perturbative
corrections for larger $aM_0$ and significant contributions from
higher orders in NRQCD for smaller $aM_0$.  Nevertheless, the $1$-loop
matching factors should be checked by a nonperturbative
determination; this is essential for a determination of the slope.
Similarly, for the smaller meson masses studied here, higher orders in
the NRQCD expansion should be considered.  However, there are no
indications that NRQCD is breaking down for the range of masses
studied.

In addition, we present a tree-level analysis of the origin of the slope
$C_{PS}$, and separate the contributions from the hyperfine and kinetic energy
terms in the NRQCD action and the tree-level current correction. We find
$G_{kin}\sim 1.8$~GeV dominates $C_{PS}$, while the slope of the ratio
$f_{PS}\sqrt{M_{PS}}/f_{V}\sqrt{M_{V}}$, related to $G_{hyp}$ and $G_{corr}$
is $O(\Lambda_{QCD})$, consistent with naive expectation from HQS. However,
considering the results for the $1$-loop corrections discussed previously
(which are not suitable for a study of HQS), $G_{kin}$ is likely to be reduced
significantly by renormalisation and $1$-loop corrections to $G_{corr}$ are
also likely to be large.

With this extensive analysis of the decay constants we are able to
perform a detailed comparison of our results with those from the
$O(1/M^2)$ quenched simulation at $\beta=6.0$.  We find that $f_B$
appears to be considerably larger at $n_f=2$ compared to that at
$n_f=0$. Clearly, $f_B$ is a quantity which is sensitive to internal
quark loops, and further work is needed to investigate the dependence
of the decay constant on the mass and the number of flavours of sea
quarks.  We find the sensitivity to $n_f$ cancels in the ratio
$f_{B_s}/f_B$.

\section{Acknowledgements}
The computations were performed on the CM-2 at SCRI.  We thank the
HEMCGC collaboration for use of their configurations, and J.~Hein and
G.~P.~Lepage for useful discussions.  This work was supported by PPARC
under grant GR/L56343 and the U.S.~DOE under grants DE-FG02-91ER40690,
DE-FG05-85ER250000 and DE-FG05-96ER40979. We acknowledge support by
the NATO under grant CRG~941259 and the EU under contract
CHRX-CT92-0051. SC has been funded as a research fellow by the Royal
Society of Edinburgh and the Alexander von Humboldt Foundation.

\section{Appendix}
\label{appendix}
In this section we illustrate the fitting analysis and extraction of
the masses and amplitudes of the heavy-light meson correlators. The
fitting method is described in detail in reference~\cite{spect}, and
since the data is of similar quality in terms of efficacy of the
smearing functions etc. we only present additional fits not shown
in~\cite{spect} and~\cite{decay}. 

\subsection{Spectrum}

Considering the spectrum initially, the simplest quantities to extract
are the $^1S_0$ and $^3S_1$  ground state energies, $E(^1S_0)$ and
$E(^3S_1)$ respectively. A multiple exponential `vector' fit was
performed to the $C(l,1)$ and $C(l,2)$ correlators, where
$C(l,1)$~($C(l,2)$) denotes a correlator with the heavy quark smeared
at the source with a hydrogen-like ground~(first excited) state
wavefunction and local at the sink. The operators inserted at the
source and sink constructed from heavy and light quarks are given in
table~\ref{ops}.

\begin{figure}
\centerline{\setlength{\epsfxsize}{80mm}\epsfbox[10 60 640 570]{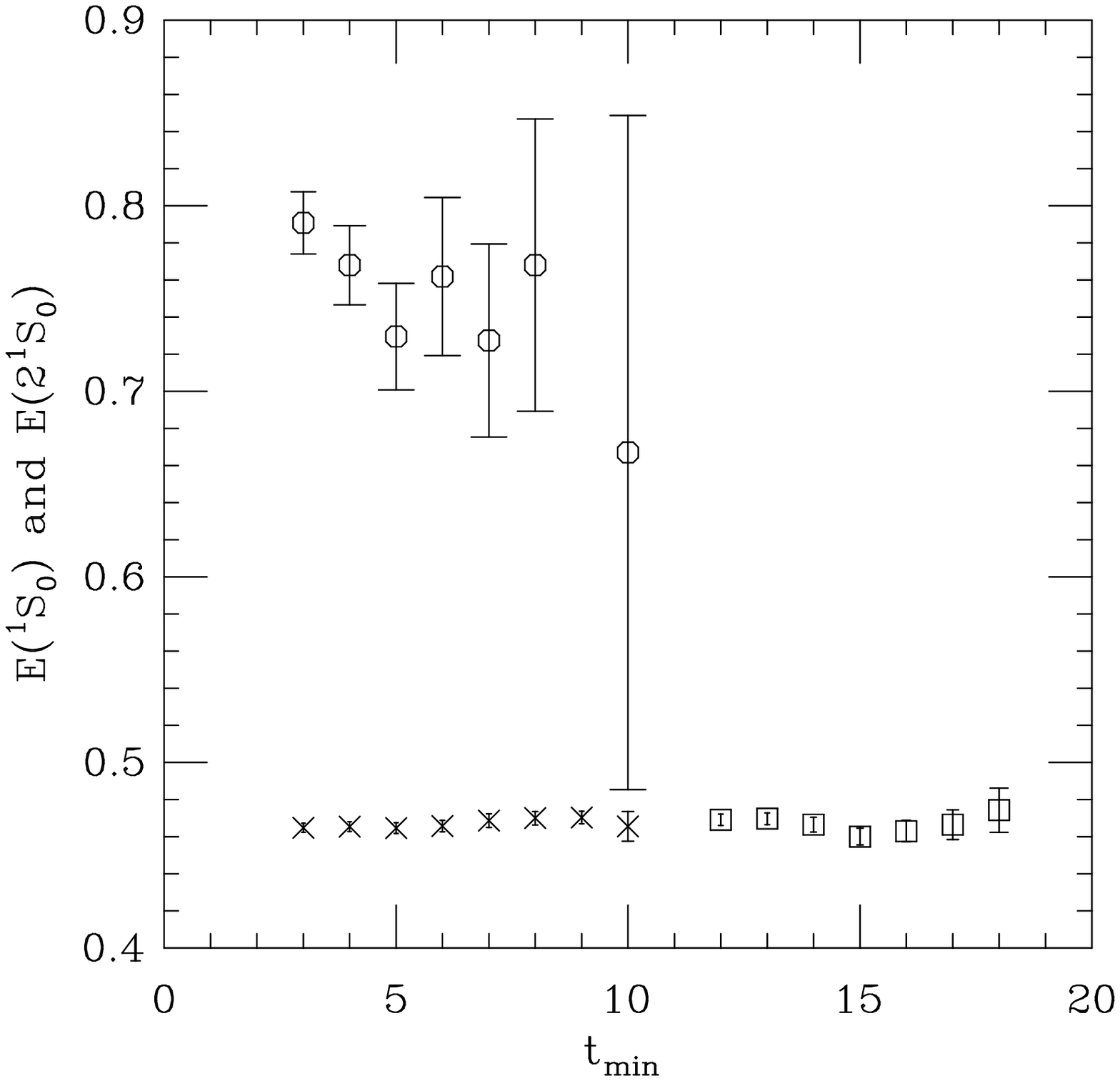}
\setlength{\epsfxsize}{80mm}\epsfbox[10 60 640 570]{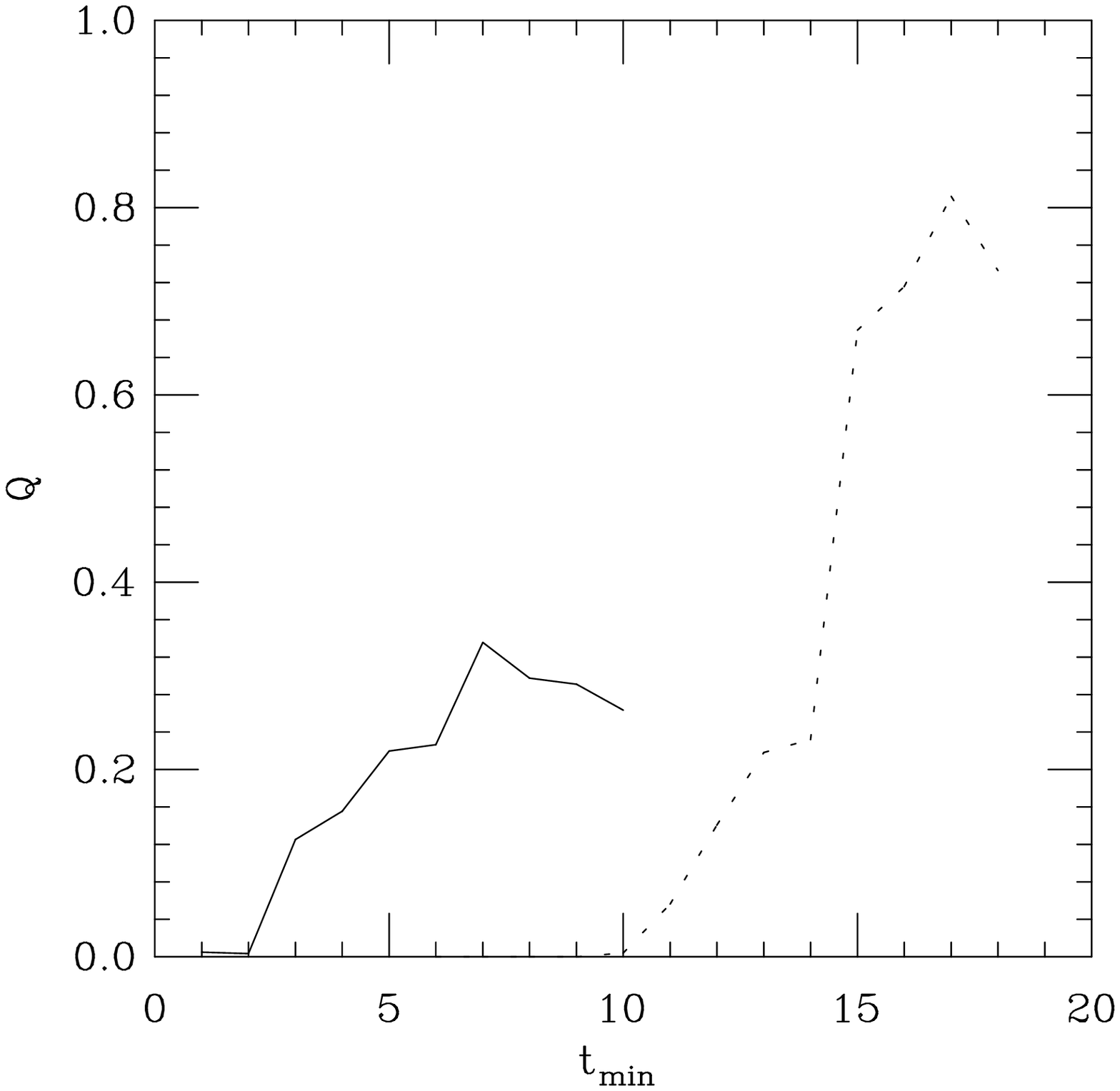}
}
\vspace{0.5cm}
\caption{The ground and first excited
  state energies extracted as a function of $t_{min}$ from a vector
  fit to the $C(l,1)$ and $C(l,2)$ pseudoscalar correlators for
  $aM_0=1.0$ and $\kappa_l=0.1385$; $t_{max}$ is fixed to 20.
  $E(^1S_0)$ is shown as squares~(crosses) for a one~(two) exponential
  fit and $E(2^1S_0)$ as circles for the two exponential fit. The
  quality of fit parameter, $Q$, is also shown as a dotted and solid
  line for a one and two exponential fit respectively, where $Q>0.1$
  defines a 'good' fit.
\label{fitexp}
}
\end{figure}

An example of the quality and stability of the fits is shown
in figure~\ref{fitexp} for $aM_0=1.0$ and $\kappa_l=0.1385$. The
ground state energy extracted from a single exponential
fit~($n_{exp}=1$) in the range $t_{min}>12$ is in agreement with
$E(^1S_0)$ from a $n_{exp}=2$ fit for $3<t_{min}<10$, and we feel
confident that we have minimal excited state contamination. The
excited state energy is stable with $t_{min}$, however, a $n_{exp}=3$
fit is needed to confirm this state. Similar results are found for all
$M_0$ and $\kappa_l$ and we chose a fitting range of $5-20$ with
$n_{exp}=2$ for both $^1S_0$ and $^3S_1$ mesons.

Tables~\ref{energies} and~\ref{energiesV} detail the energies
extracted. The results are chirally extrapolated to zero light quark
mass and we find only linear dependence in $1/\kappa_l$;
figure~\ref{fitchiral} shows the chiral extrapolation of
$E_{sim}^{PS}$ for $aM_0=1.0$.

The hyperfine splittings, $E(^3S_1){-}E(^1S_0)$ and
$E(2^3S_1){-}E(2^1S_0)$, are calculated from the differences of 100
bootstrap samples of the $n_{exp}=2$ fits. In order to study the dependence of
these and other splittings on the heavy quark mass, the meson mass~($M_{PS}$)
is calculated using the mass shifts, $\Delta=M_{PS}-E(^1S_0)$, obtained
from heavy quarkonia dispersion behaviour, given in reference~\cite{spect}.
The mass shifts are given in table~\ref{energies}.

\begin{figure}
\centerline{\setlength{\epsfxsize}{80mm}\epsfbox[10 60 640 570]{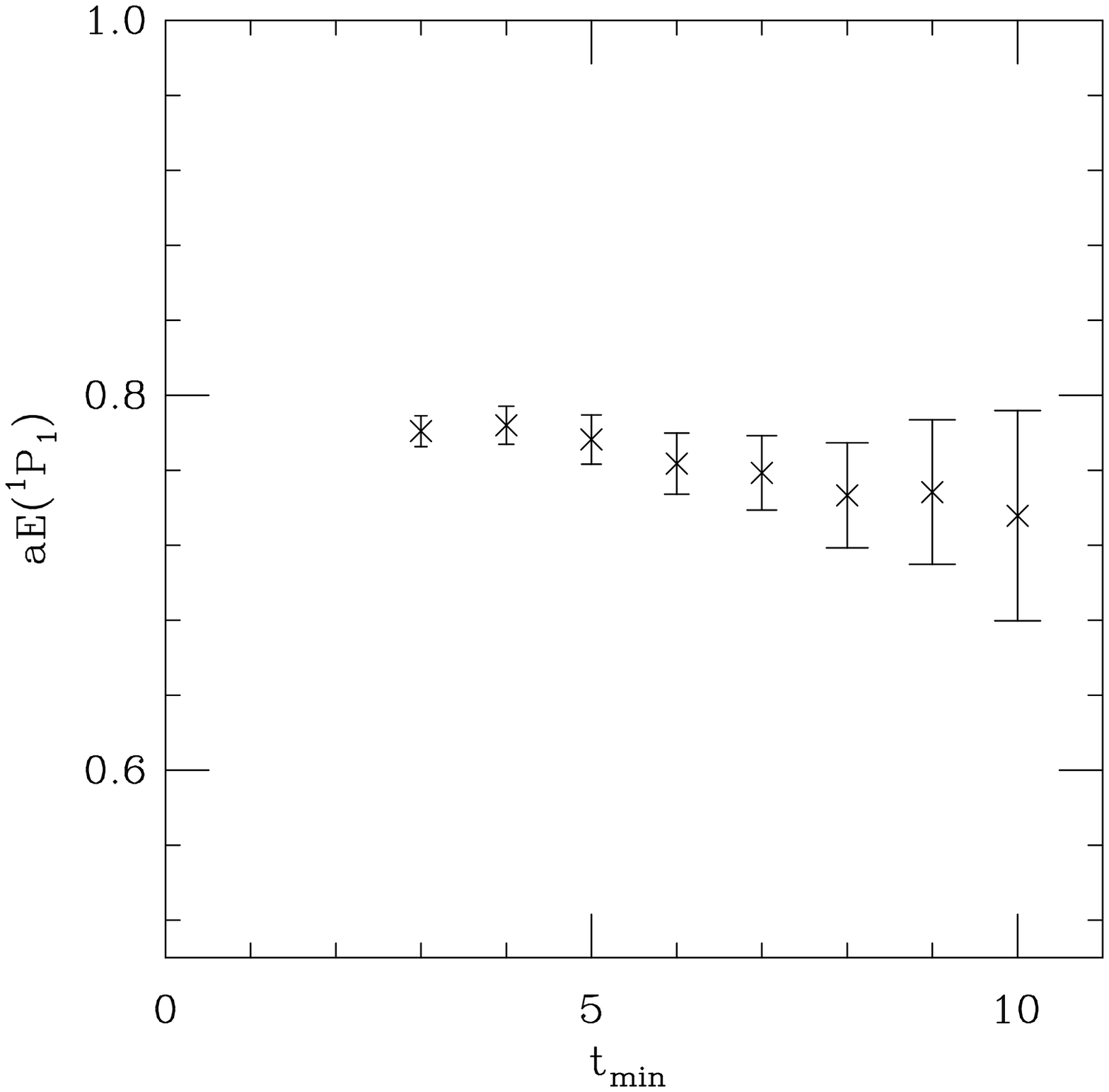}
\setlength{\epsfxsize}{80mm}\epsfbox[10 60 640 570]{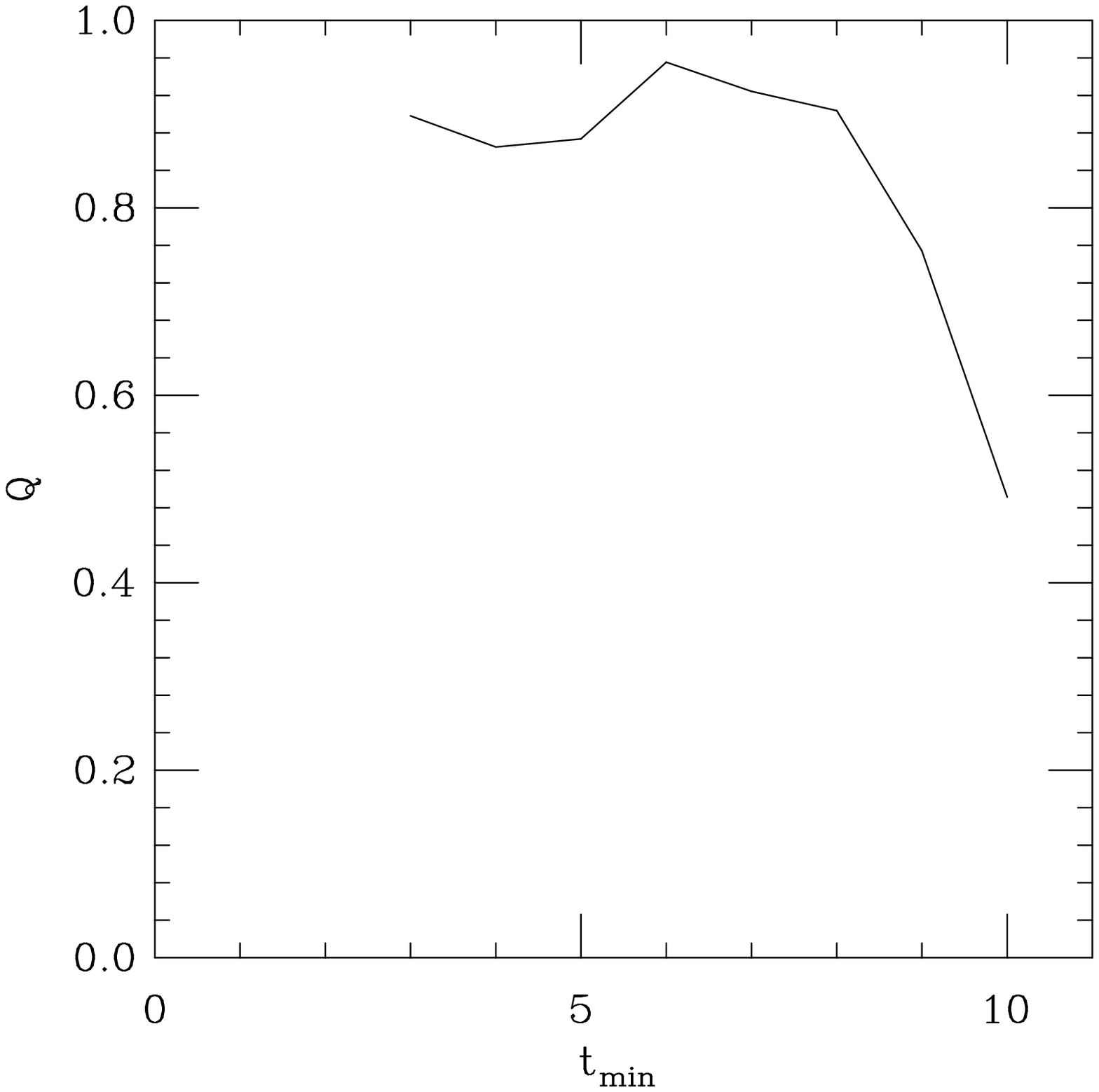}
}
\vspace{0.5cm}
\caption{The ground state energy of the $^1P_1$ state extracted from a
$n_{exp}=1$ fit to $C(l,2)$ as a function of $t_{min}$ for $aM_0=1.0$
and $\kappa_l=0.1385$; $t_{max}=10$. The corresponding values of $Q$
are also shown.
\label{fit1p1}
}
\end{figure}

The singlet and triplet $l=1$ states, $^1P_1$, $^3P_1$, $^3P_2$ and
$3P_0$, were also computed for $aM_0=1.0$, $2.0$ and $4.0$. The
corresponding operators used in the simulation are given in
table~\ref{ops}. As noted in~\cite{spect} we cannot resolve the spin
splittings between these states with the statistics available. Thus,
we present the results for the $^1P_1$ state and ignore the mixing
with the $^3P_1$ state.  Figure~\ref{fit1p1} shows a $n_{exp}=1$ fit
to the $C(l,2)$ correlator is stable as $t_{min}$ is varied and we
chose the fitting range $4-10$ for all $aM_0$ and $\kappa_l$. The
ground and first excited state smearing functions did provide
significantly different overlap with the lowest two states. However,
since the signal died out around $t\sim 10$ there were too many higher
states contributing to enable a $1$, $2$ or $3$ exponential fit to
$C(l,1)$ and $C(l,2)$.  Table~\ref{tabp} gives the results for
different $aM_0$ and $\kappa_l$.

\begin{figure}
\centerline{\setlength{\epsfxsize}{80mm}\epsfbox[10 60 640 570]{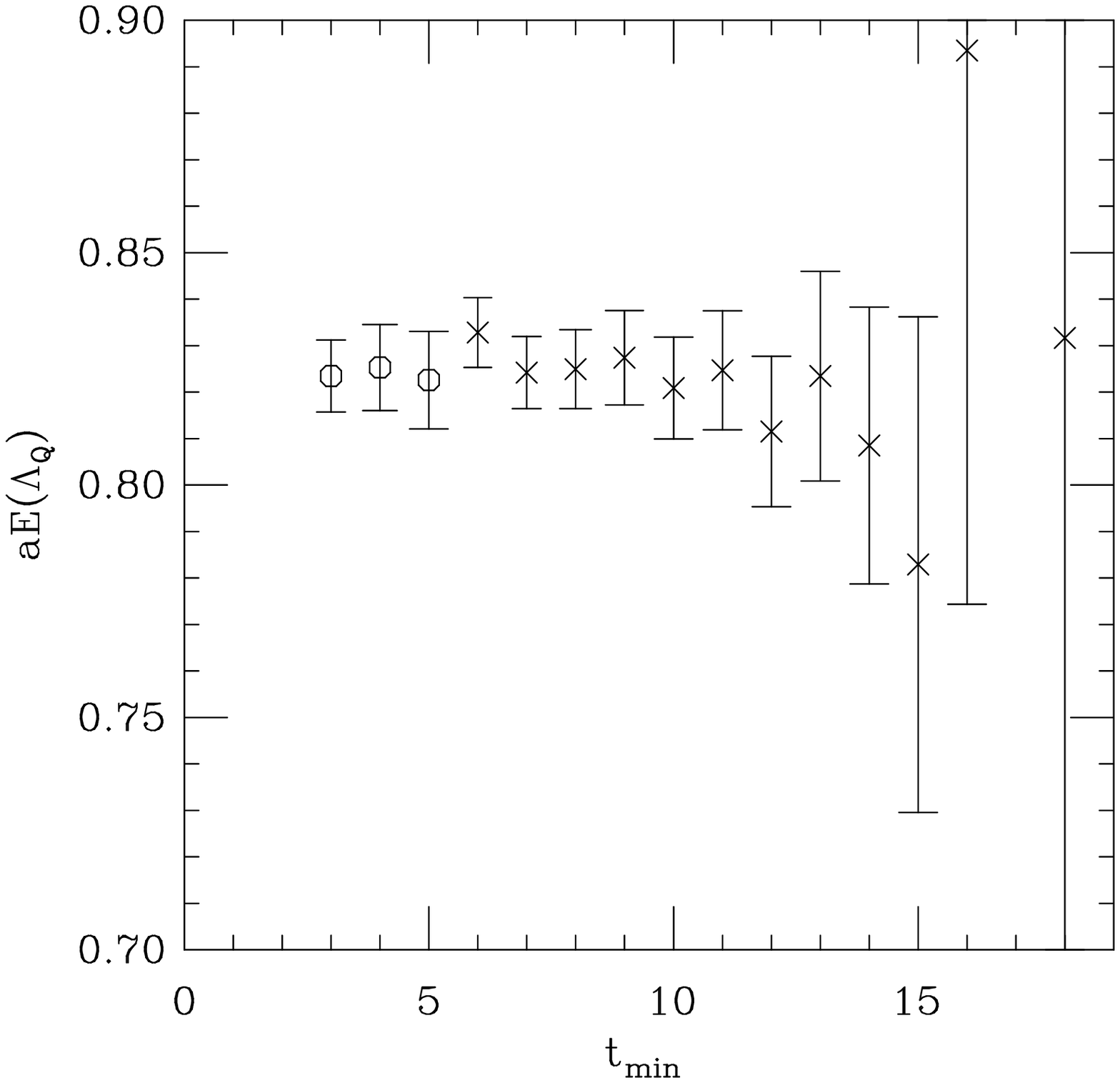}
\setlength{\epsfxsize}{80mm}\epsfbox[10 60 640 570]{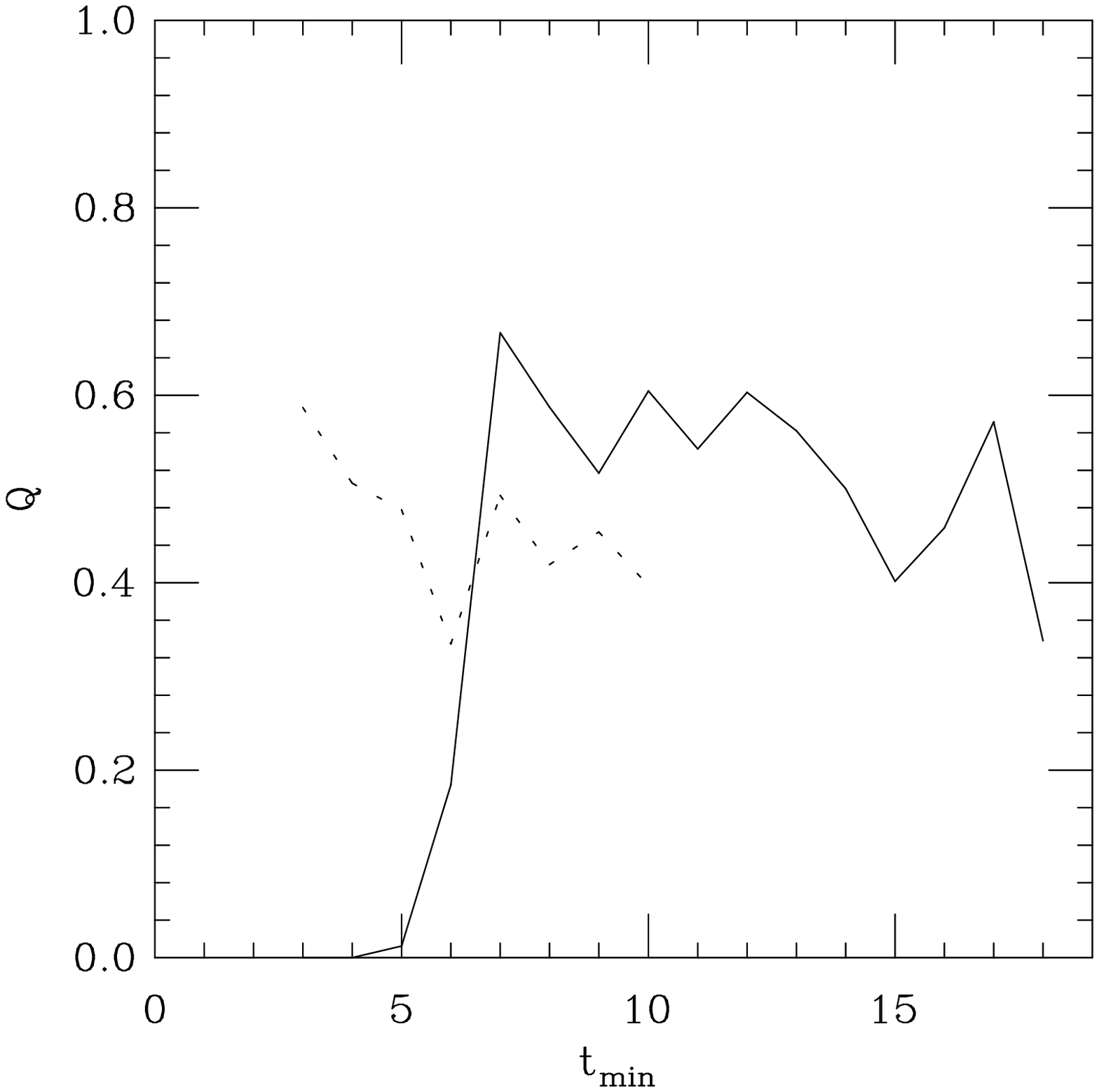}
}
\vspace{0.5cm}
\caption{The ground state energy of the $\Lambda_Q$ extracted from
a $n_{exp}=1$~(crosses) and $n_{exp}=2$~(circles) fit to the $C(l,1)$
correlator as a function of $t_{min}$ for $aM_0=1.0$ and
$\kappa_l=0.1385$; $t_{max}=20$. The corresponding values of $Q$ are
also shown as a solid~(dotted) line for a $n_{exp}=1$~(2) fit.
\label{fitbar}
}
\end{figure}

In addition to the meson spectrum the $J^P=\frac{1}{2}^+$ baryon containing
one heavy quark, $\Lambda_Q$, was computed. The lattice operator,
\begin{equation}
\epsilon_{abc}\sum_{\vec{x}_1}Q^{a}_\alpha(\vec{x}_1)\sum_{\vec{x}_2}(q^b)^T(\vec{x}_2)C\gamma_5\gamma_0q^c(\vec{x}_2)\phi(|\vec{x}_1-\vec{x}_2|)
\end{equation}
has the correct quantum numbers for this particle, where $a$, $b$ and
$c$ are the colour indices and $\alpha$ is the spinor index. $\phi$
represents the (hydrogen-like) smearing function and $C$ is the charge
conjugation matrix $\gamma_0\gamma_2$.  Figure~\ref{fitbar} presents
the energy of the ground state extracted from multi-exponential fits
to $C(l,1)$ for $aM_0$ and $\kappa_l=0.1385$. $E(\Lambda_Q)$ is stable with
$t_{min}$ for $n_{exp}=1$.  This is also the case when including
another exponential in fitting to the same correlator.  We found the
fitting range $7-20$ for the $n=1$ fit is sufficient for all $aM_0$
and $\kappa_l$, and the corresponding values of the ground state
energy are given in table~\ref{tabb}.

\subsection{Decay Constants}
\begin{figure}
\centerline{\setlength{\epsfxsize}{80mm}\epsfbox[10 60 640 570]{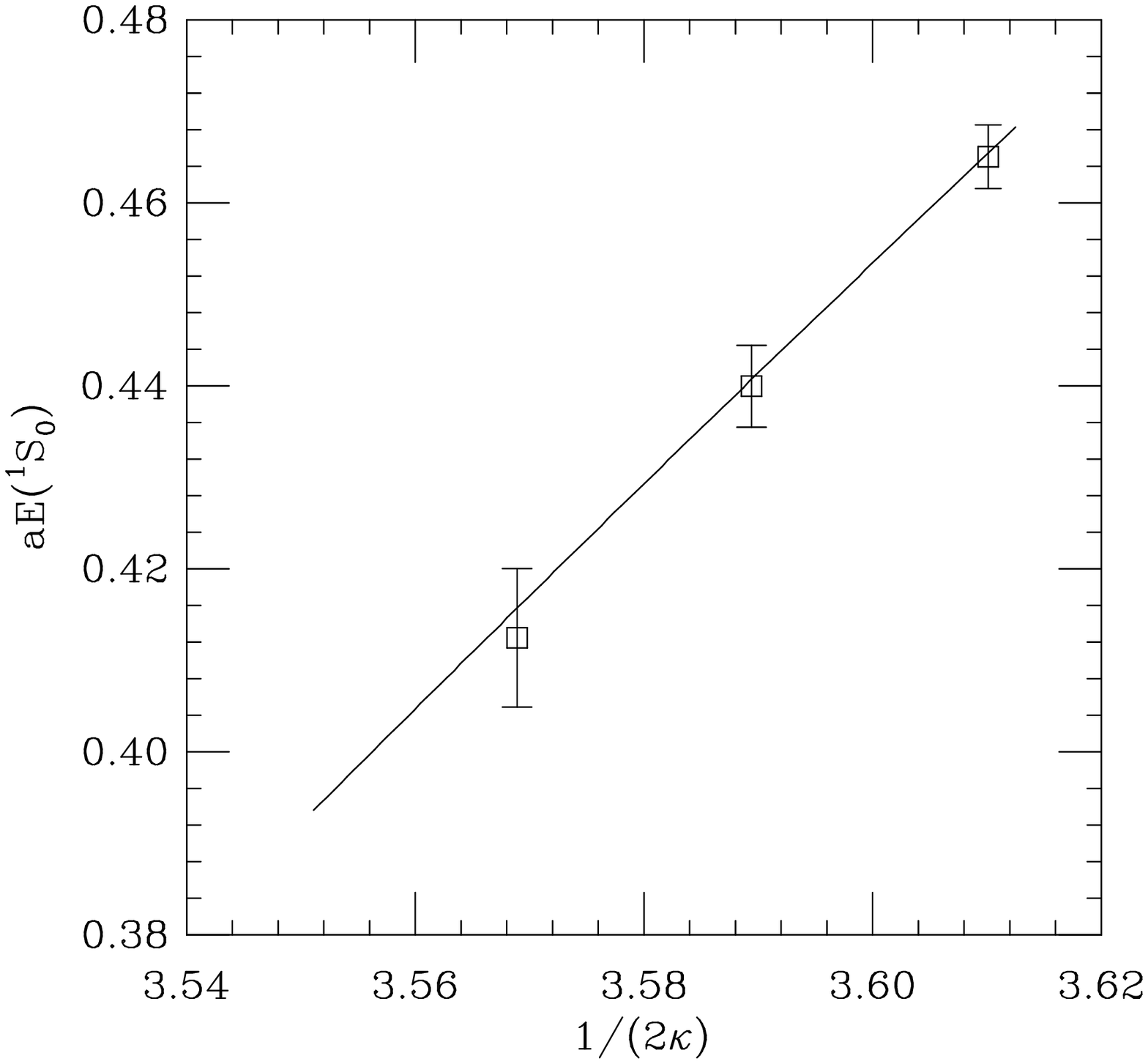}
\setlength{\epsfxsize}{80mm}\epsfbox[10 60 640 570]{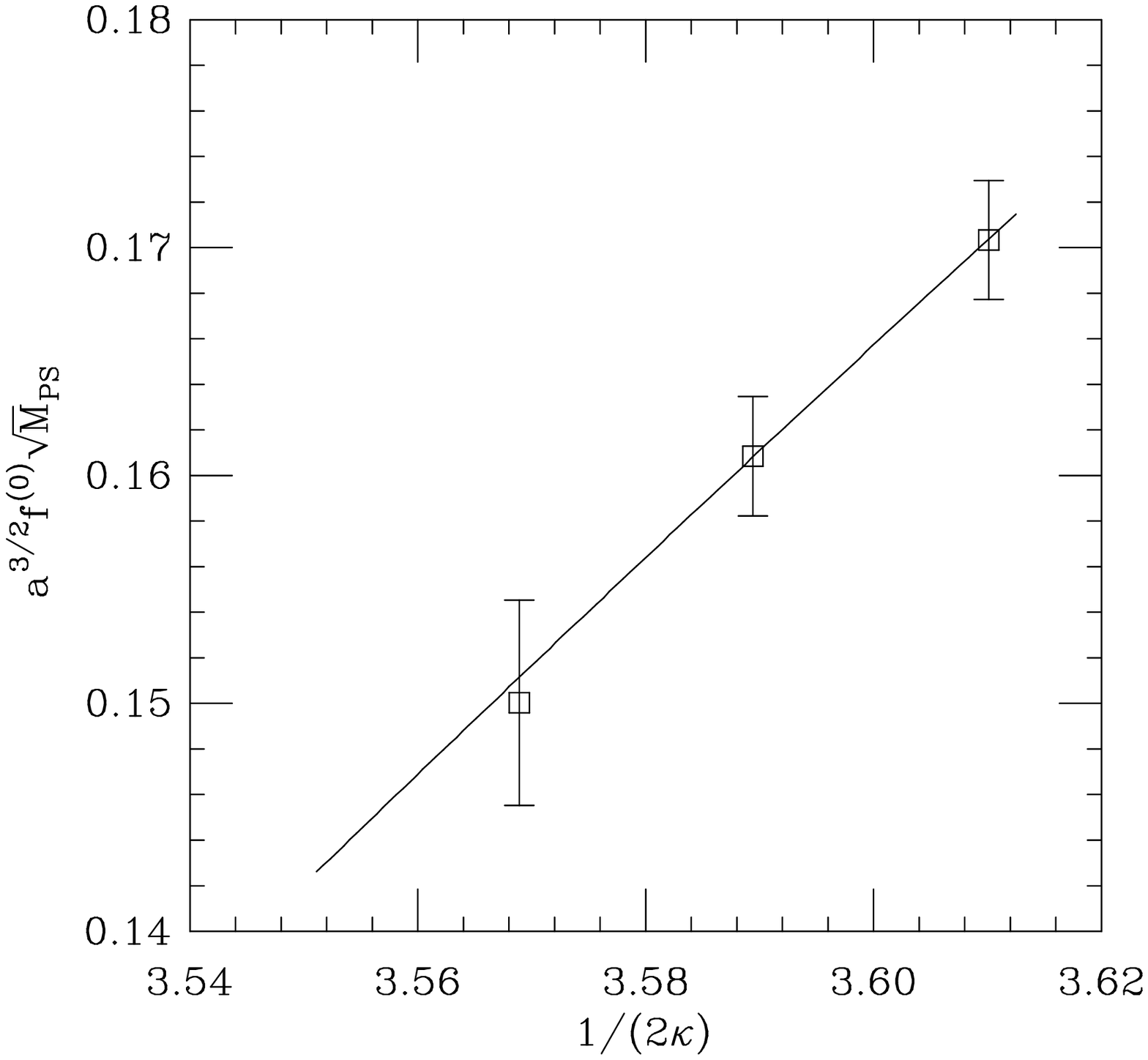}
}
\vspace{0.5cm}
\caption{The chiral extrapolation of $aE(^1S_0)$ and $a^{3/2}f^{(0)}\protect\sqrt{M_{PS}}$ in $1/\kappa$ for $aM_0=1.0$.
\label{fitchiral}
}
\end{figure}

\begin{figure}
\centerline{\setlength{\epsfxsize}{80mm}\epsfbox[10 60 640 570]{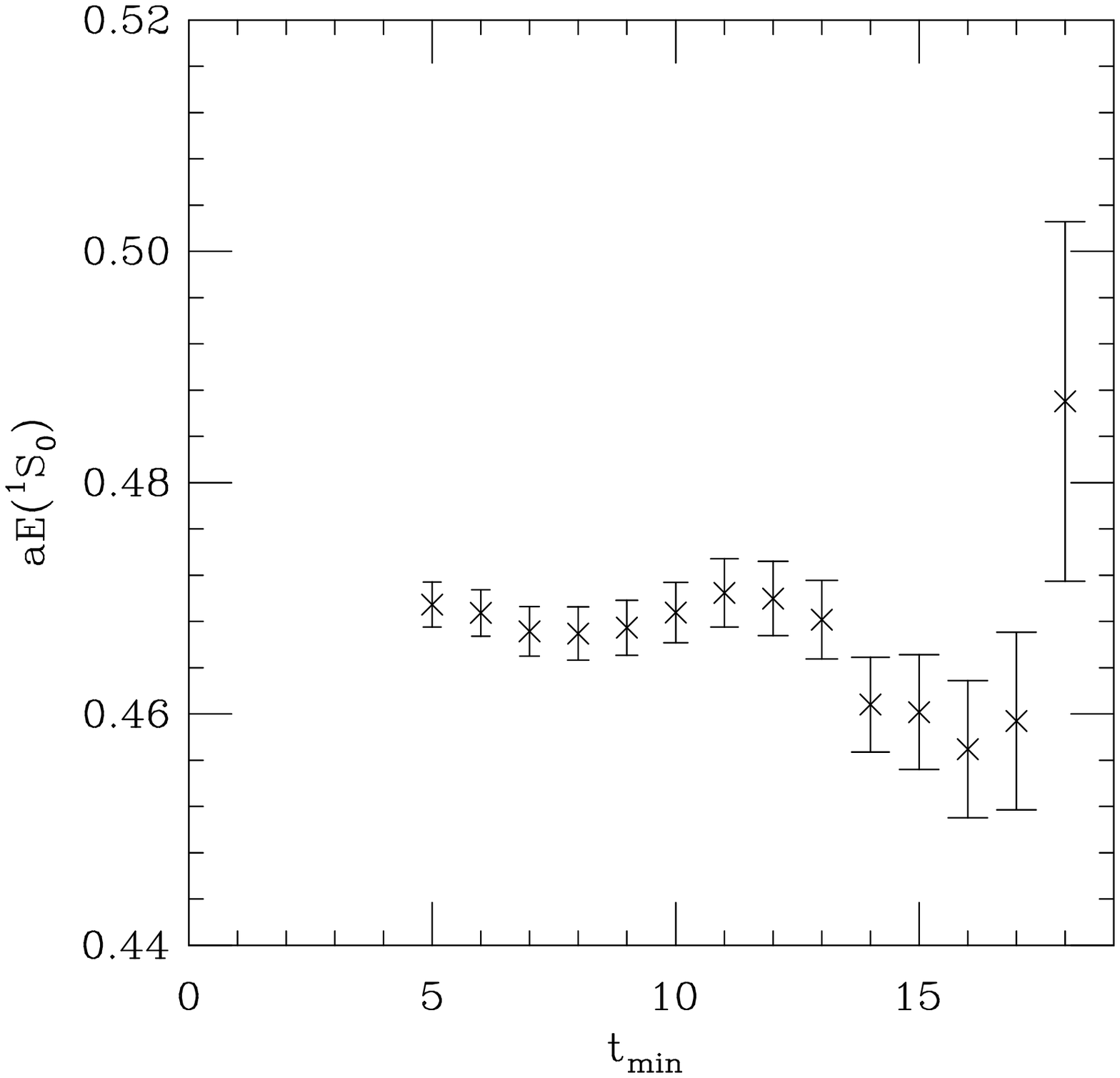}
\setlength{\epsfxsize}{80mm}\epsfbox[10 60 640 570]{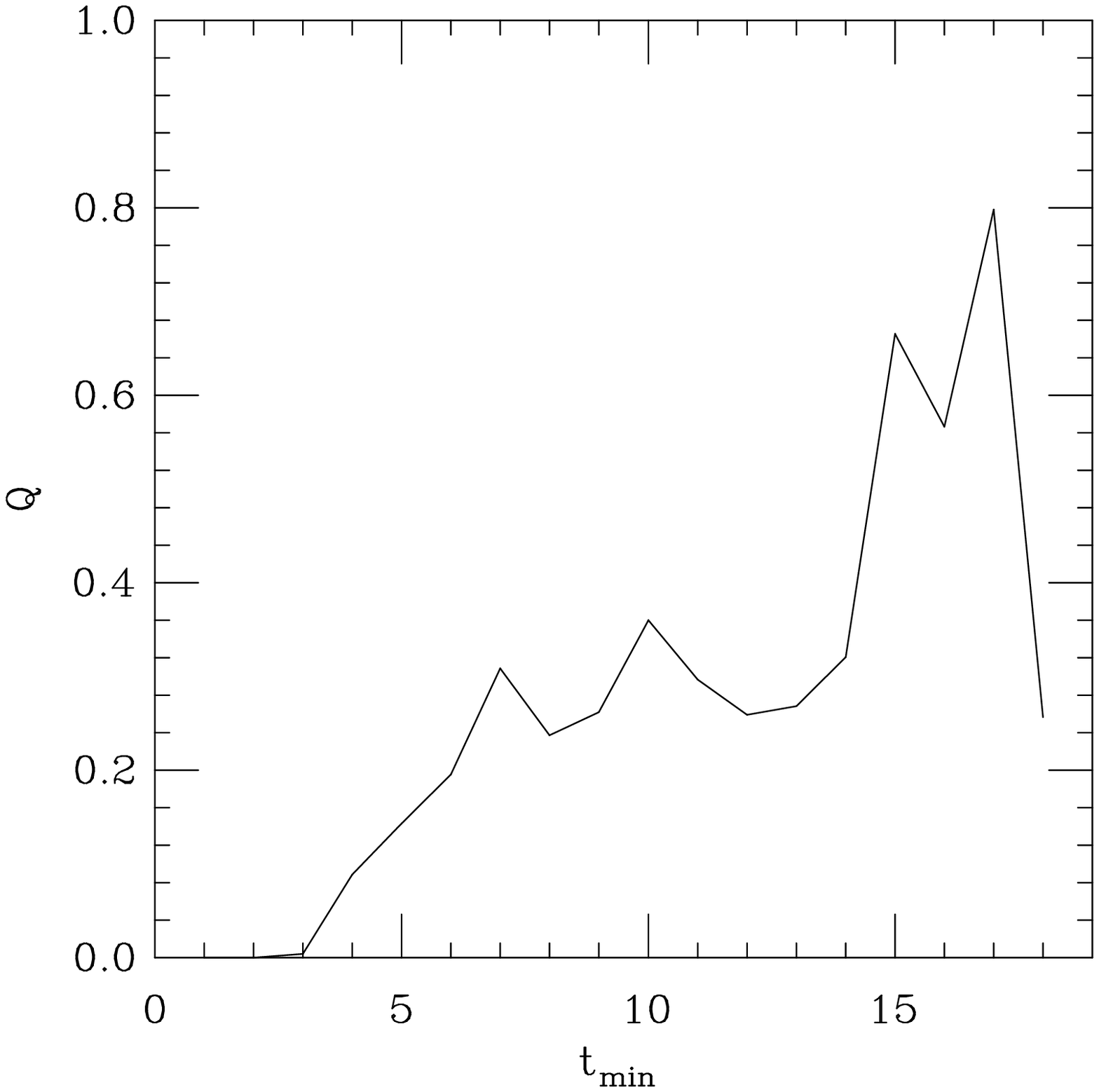}}
\vspace{0.5cm}
\caption{The ground state energy extracted from
a $n_{exp}=1$ fit to the pseudoscalar $C(l,1)$ and $C(1,1)$ meson
correlators as a function of $t_{min}$ for $aM_0=1.0$ and
$\kappa_l=0.1385$; $t_{max}=20$. The corresponding values of $Q$ are
also shown.
\label{fit1s0mat}
}
\end{figure}

The zeroth order pseudoscalar and vector decay constants,
$f^{(0)}_{PS}\sqrt{M_{PS}}$ and $f^{(0)}_V\sqrt{M_V}$ respectively, were
extracted in the standard way from simultaneously fitting to the $C(l,1)$ and
$C(1,1)$ correlators. The amplitudes of these correlators are related to the
decay constant via:
\begin{eqnarray}
C(l,1) & = & Z_{l} Z_{1} e^{-E t},\\
C(1,1) & = & Z_{1}^2 e^{-E t}.
\end{eqnarray}
where,
\begin{equation}
\sqrt{2}Z_{l} = f^{(0)}\sqrt{M}.
\end{equation}
Note that the correction to the current is not included in $C(l,1)$.  The
typical quality of the fits is illustrated in figure~\ref{fit1s0mat} for the
$PS$ meson. We found $7-20$ to be the optimal fitting range for $PS$ and $V$
mesons for all $aM_0$ and $\kappa_l$. The corresponding amplitudes obtained
from 100 bootstrap samples of the fit parameters are presented in
tables~\ref{decaytps} and~\ref{decaytv} for the $PS$ and $V$ respectively. The
results are extrapolated to zero light quark mass, an example of which is
shown in figure~\ref{fitchiral} for the $PS$ decay constant.

\begin{figure}
\centerline{\setlength{\epsfxsize}{80mm}\epsfbox[10 60 640 570]{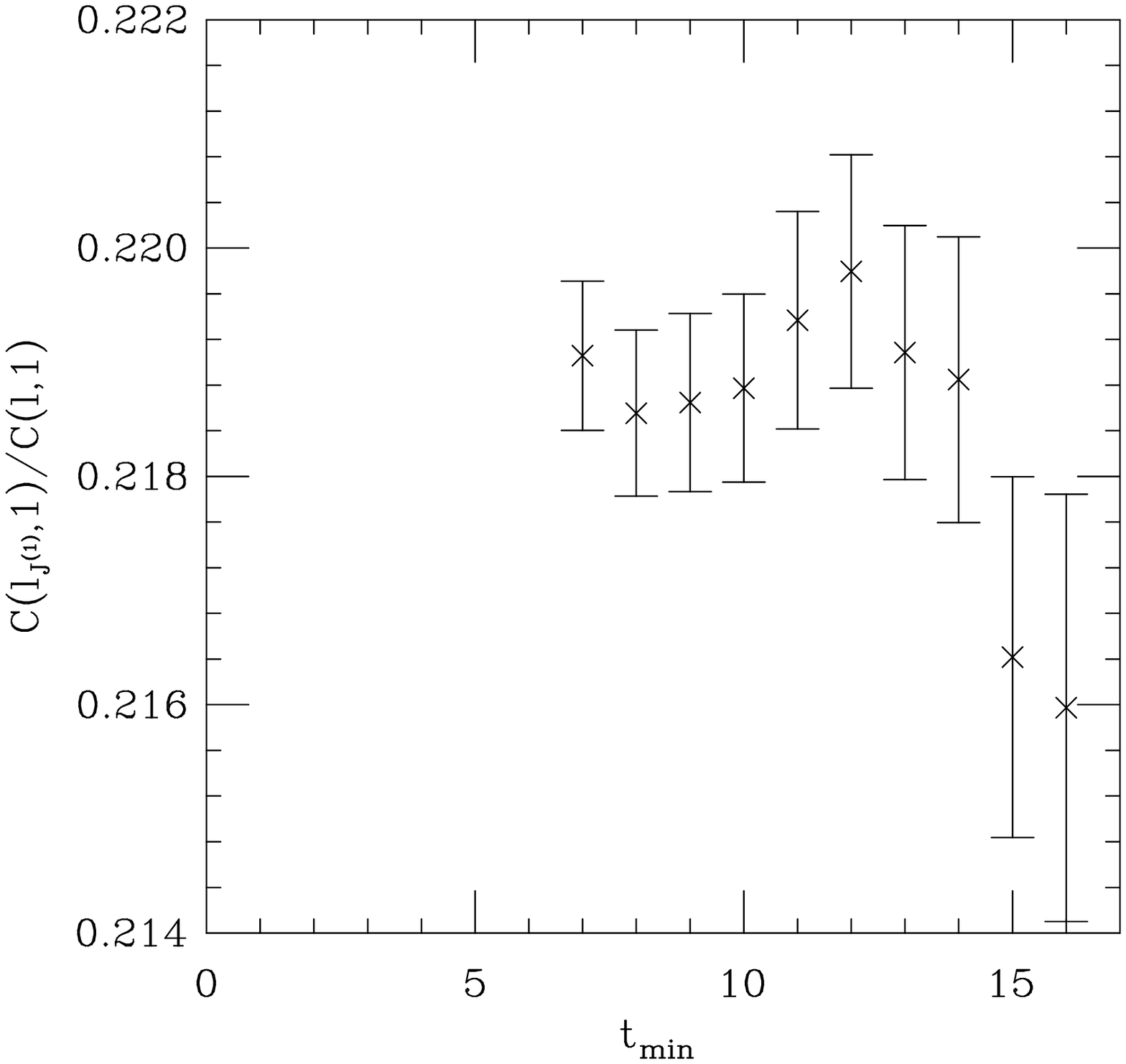}
\setlength{\epsfxsize}{80mm}\epsfbox[10 60 640 570]{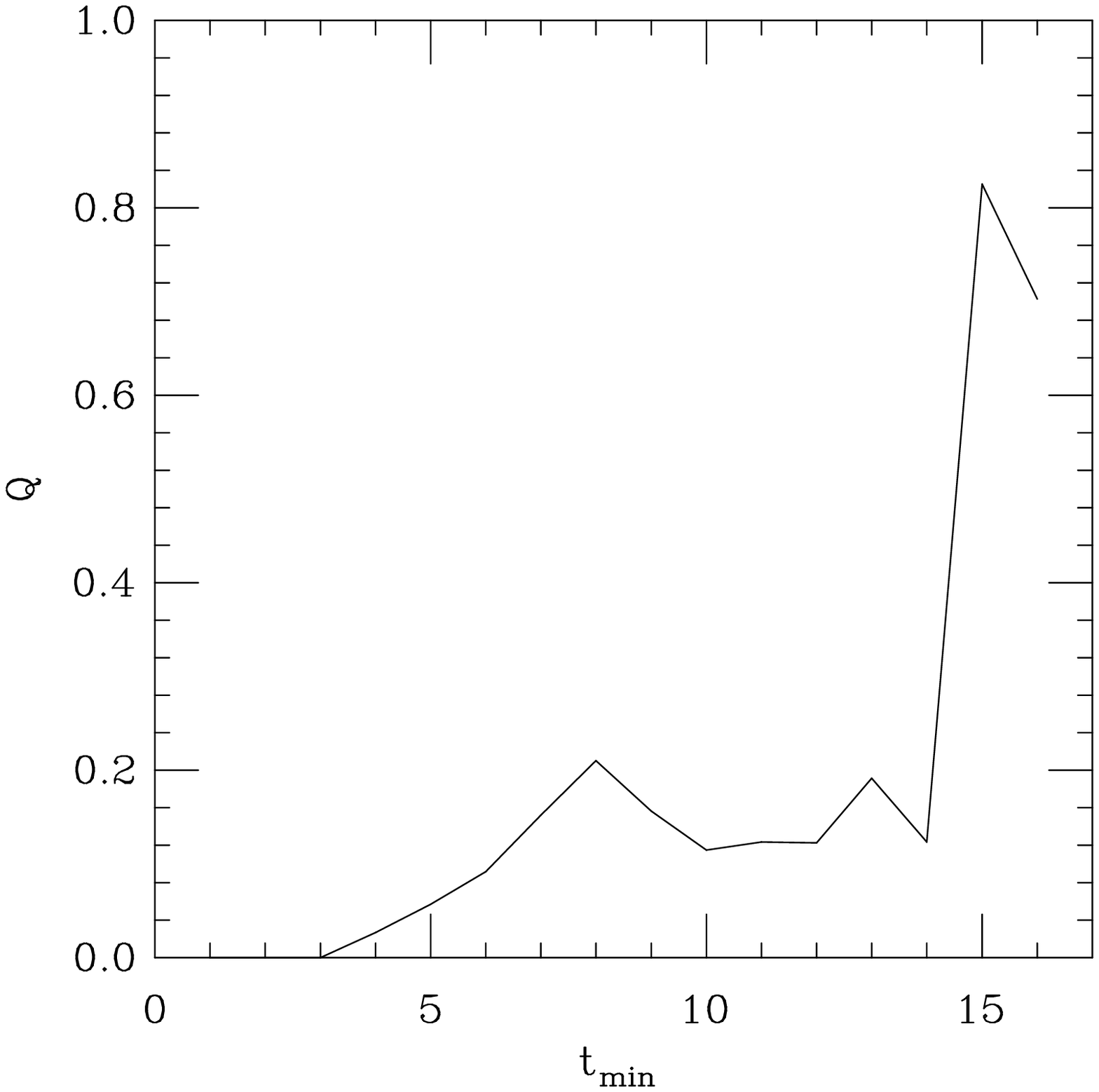}
}
\vspace{0.5cm}
\caption{The ratio $C(l_{J^{(1)}},1)/C(l,1)$ for the $PS$ meson as a function of $t_{min}$ for $aM_0=1.0$ and $\kappa_l=0.1385$; $t_{max}=20$.
\label{fitps}
}
\end{figure}

The tree-level corrections to the $PS$ and $V$ currents,
$f^{(1)}_{PS}\sqrt{M_{PS}}$ and $f^{(1)}_{V}\sqrt{M_{V}}$ respectively, are
obtained separately by computing the jackknife ratio of the correlator with
the tree-level current correction operator inserted at the
sink~($C(l_{J^{(1)}},1)$) with $C(l,1)$; $J^{(1)}$ is given
in equations~\ref{psone} and~\ref{vone} for the $PS$ and $V$ respectively. In
the limit of large times this ratio tends to
$f^{(1)}\sqrt{M}/f^{(0)}\sqrt{M}$.  From figure~\ref{fitps} we see that
$8-20$ is a reasonable fitting range, and this was found to be optimal for all
$aM_0$ and $\kappa_l$.  Tables~\ref{deltapstree} and~\ref{deltavcor} give the
resulting tree-level corrections to the decay constant for the $PS$ and $V$
mesons respectively. The only $1$-loop correction which must be calculated is 
$f^{(3)}_V\sqrt{M_V}$. This is obtained in the same way as the tree-level
correction. The results are detailed in table~\ref{deltavloop}.

%
%
\begin{table}
\caption{The systematic errors expected to dominate the predictions of mass splittings and decay constants of the $B$ meson. }
\label{syserrors}
\begin{center}
\begin{tabular}{ccc}
Quantity & Dominant Errors & Size\\\hline
$B^*{-}B$ & Pert. correction to $c_B$ & $O(\alpha_S(\Lambda_{QCD}/M))\sim 10{-}30\%$\\\hline
$B_s{-}B$ & Determination of $\kappa_s$ &  $\sim 10\%$ \\
&  Light quark discretisation & $O((a\Lambda_{QCD})^2)$,
$O(\alpha(a\Lambda_{QCD}))< 5\%$ \\
 & Finite volume & $5\%$ ? \\\hline
$B(2S){-}B(1S)$ &  Finite volume        &  $<10\%$ ?  \\
$B^{**}-B$ &   Light quark discretisation & $O((a\Lambda_{QCD})^2)$, $O(\alpha(a\Lambda_{QCD}))< 5\%$ \\
$\Lambda_B{-}B$ &     & \\\hline
$f_B$, $f_{B^{*}}$ & Light quark discretisation & $O((a\Lambda_{QCD})^2)< 5\%$ \\
 & Pert. matching factors & $O(\alpha^2)\sim 1{-}10\%$ \\
 & Truncation of NRQCD series & $O(1/M^2) \sim 6\%$ \\
 & Finite volume & $5\%$ ? \\\hline
$f_B/f_{B_s}$ & Determination of $\kappa_s$ & $\sim 2\%$ \\ 
\end{tabular}
\end{center}
\end{table}

\begin{table}
\caption{Values of the inverse lattice spacing in GeV as determined
  from light hadrons and heavy
  quarkonia~\protect\cite{nrqcdscaling} for this ensemble and a
  similar quenched ensemble at
  $\beta^{n_f=0}=6.0$~\protect\cite{nrqcdscaling,arifajunk}. 
\label{latt}
}
\begin{center}
\begin{tabular}{cccc}
 & $\rho$  & $\Upsilon(1P-1S)$ & $\Upsilon(2S-1S)$\\\hline
$\beta^{n_f=2}=5.6$ & 1.97(3)  & 2.44(7) & 2.37(10) \\
$\beta^{n_f=0}=6.0$ &  1.93(3) & 2.59(5) & 2.45(8)\\
\end{tabular}
\end{center}
\end{table}

\begin{table}
\caption{Predictions for various mass splittings 
compared to experiment. The results are converted into physical
units~(MeV) using $a^{-1}=2.0$~GeV. The first error shown is
statistical while the second is the estimate of the systematic errors
which we expect to be dominant, as detailed in
table~\protect\ref{syserrors}; the third indicates the change in the
prediction if $a^{-1}=2.4$~GeV is used and includes the change due to
the change in $M_0^b$. For $B_s{-}B$ the systematic errors are split
up between the light quark discretisation error and that due to the
uncertainty in $\kappa_s$. The systematic error for the hyperfine
splitting indicates the change in the central value if 
$c_B$ is increased by $30\%$.
\label{prediction}}
\begin{center}
\begin{tabular}{ccc}
 &   $\beta^{n_f=2}=5.6$ & expt \\\hline
$B^*-B$       &  20(12)(+6)(+6)    &   45.7(4)    \\
$B^*_s-B_s$       &  26(8)(+10)(+9)   &   47(3)    \\
$B_s-B$       &  98(8)(5)(-12)(+20)  &   90(2)    \\
$B(2S)-B(1S)$ &  540(120)(23)(+110) &   $\sim580$    \\
$B_s(2S)-B_s(1S)$ &  500(80)(20)(+100) & -  \\
$B^*(2S)-B(2S)$ &  78(32)(10)(+22) & - \\
$B^{**}-B$    & 520(40)(20)(+110)  &   $350{-}500$   \\
$\Lambda_b-B$ & 560(40)(22)(+110)  &   345(9)    \\
\end{tabular}
\end{center}
\end{table}

\begin{table}
\caption{The coefficients of various spectral quantities, as
determined from first order perturbation theory in $1/M$.
The results are in physical units, GeV and $\mbox{GeV}^2$
for $C_0$ and $C_1$ respectively, converted using
$a^{-1}=2.0$~GeV. The hyperfine term has been ignored for the $^1P_1$
state.  Unless stated otherwise, $\kappa_l=\kappa_c$.
\label{Etab}}
\begin{center}
\begin{tabular}{ccccc}
Quantity & Coefficient & Results & Expectation & Expt\\\hline
$\bar{E}(2S)-\bar{E}(1S)$ & $C_0 = \bar{\Lambda}_{2S} - \bar{\Lambda}_{1S}$ & 0.38(10) & $+\Lambda_{QCD}$  & - \\
              & $C_1 = \left<2S\right| {\cal O}_{kin} \left|2S\right>_{phys} -$ & &  & -\\
              & $\left< 1S\right| {\cal O}_{kin} \left| 1S\right>_{phys}$ & 0.60(40) & $+$ve & \\
$E(^3S_1)_{\kappa_s}- E(^1S_0)_{\kappa_s}$ & $C_0=0$ & 0.002(8)  &  0 & $\sim 0$ \\
                     & $C_1= \left< 1S\right| {\cal O}_{hyp} \left| 1S\right>_{phys}^{\kappa_s}$ & 0.08(4) & $+(\Lambda_{QCD})^2$ & $\sim 0.28$\\
$E(2^3S_1)_{\kappa_s}- E(2^1S_0)_{\kappa_s}$ & $C_0=0$ & -0.02(2) &  0 & -\\
                     & $C_1= \left<2S\right| {\cal O}_{hyp} \left|2S\right>_{phys}^{\kappa_s}$ & $0.56(28)$ & $+(\Lambda_{QCD})^2$ &- \\
$\bar{E}(^1S_0)_{\kappa_s}-\bar{E}(^1S_0)_{\kappa_c}$ & $C_0=\bar{\Lambda}_{\kappa_s}-\bar{\Lambda}_{\kappa_c}$ & 0.106(10) &  & $\sim .085$ \\
                                          & $C_1=\left< 1S\right|{\cal O}_{kin} \left| 1S\right>_{phys}^{\kappa_s}-$ & 0.020(32) & & $\sim 0.03$ \\
 & $\left< 1S\right|{\cal O}_{kin} \left| 1S\right>_{phys}^{\kappa_c}$ & & & \\
$E(\Lambda_Q) - \bar{E}(1S)$ & $C_0=\bar{\Lambda}_{1/2^+}-\bar{\Lambda}$ & 0.58(4) & $\Lambda_{QCD}$ & $\sim 0.36$ \\
                             & $C_1=\left<\frac{1}{2}^+\right|{\cal O}_{kin} \left|\frac{1}{2}^+\right>_{phys}+$  & -0.12(8) &  & $\sim -0.1$ \\
 & $\left<\frac{1}{2}^+\right|{\cal O}_{hyp} \left|\frac{1}{2}^+\right>_{phys}-$ & &  &\\
   &  $\left< 1S\right|{\cal O}_{kin} \left| 1S\right>_{phys}$ & & & \\
$E(^1P_1)-\bar{E}(1S)$ & $C_0=\bar{\Lambda}_{^1P_1}-\bar{\Lambda}$ & 0.40(2) & $\Lambda_{QCD}$  & -\\
                   & $C_1=\left<^1P_1\right|{\cal O}_{kin} \left|^1P_1\right>_{phys}-$ & 0.36(8) & & -\\
   & $\left< 1S\right|{\cal O}_{kin} \left| 1S\right>_{phys}$ &  & & \\
\end{tabular}
\end{center}
\end{table}

\begin{table}
\caption{ Predictions for various mass splittings from this work and
previous results on quenched
configurations~\protect\cite{arifaspec2}.
The experimental values are also included. The results in both cases
are converted into physical units~(MeV) using the scale from $m_\rho$,
given in table~\protect\ref{latt}. The error shown is purely
statistical, except for $B_s{-}B$, where the uncertainty in $\kappa_s\
$ is shown. The central value for this splitting is set using
$\kappa_s\ $ from the $K$ meson.
\label{pred_pev}}
\begin{center}
\begin{tabular}{cccc}
 &  $\beta^{n_f=0}=6.0$ & $\beta^{n_f=2}=5.6$ & expt \\\hline
$B^*-B$       &  24(5) & 20(12)    &   45.7(4)    \\
$B^*_s-B_s$       &  27(3) & 26(8)    &   47(3)    \\
$B_s-B$       &  87(9)(+20)& 98(8)(-12)  &   90(2)    \\
$B(2S)-B(1S)$ &  602(86)     & 540(120) &   $\sim580$    \\
$B_s(2S)-B_s(1S)$ & 559(55)   & 500(80) &  -  \\
$B^{**}-B$    & 474(32)& 520(40)  &   350{-}500    \\
$\Lambda_b-B$ & 388(68) & 560(40)  &   345(9)    \\
\end{tabular}
\end{center}
\end{table}

\begin{table}
\caption{The zeroth order decay amplitude of the $PS$ 
meson, $a^{3/2}f^{(0)}_{PS}\protect\sqrt{M_{PS}}$, for all $aM_0$ and $\kappa_l$.
\label{decaytps}
}
\begin{center}
\begin{tabular}{ccccc}
$aM_0$ & 0.1385 & 0.1393 & 0.1401 & $\kappa_c$\\\hline
0.8  &  0.203(3)  &  0.190(3)  &  0.176(6)  &  0.167(4)\\
1.0  &  0.214(3)  &  0.200(3)  &  0.185(6)  &  0.175(5)\\
1.2  &  0.221(4)  &  0.207(4)  &  0.191(6)  &  0.181(6)\\
1.7  &  0.236(4)  &  0.220(4)  &  0.203(6)  &  0.192(5)\\
2.0  &  0.243(4)  &  0.227(4)  &  0.210(7)  &  0.197(6)\\
3.0  &  0.262(4)  &  0.244(5)  &  0.227(7)  &  0.212(7)\\
3.5  &  0.268(4)  &  0.251(5)  &  0.234(5)  &  0.219(6)\\
4.0  &  0.274(4)  &  0.256(6)  &  0.243(8)  &  0.224(8)\\
7.0  &  0.295(5)  &  0.275(6)  &  0.270(19)  &  0.239(8)\\
10.0  &  0.309(6)  &  0.288(7)  &  0.278(26)  &  0.249(8)\\
\end{tabular}
\end{center}
\end{table}

\begin{table}
\caption{The tree-level correction to the $PS$ current, $a^{3/2}f^{(1)}_{PS}\protect\sqrt{M_{PS}}=a^{3/2}f^{(2)}_{PS}\protect\sqrt{M_{PS}}$, for all $aM_0$
and $\kappa_l$.
\label{deltapstree}
}
\begin{center}
\begin{tabular}{ccccc}
$aM_0$ & 0.1385 & 0.1393 & 0.1401 & $\kappa_c$\\\hline
0.8  & -0.0520(10) & -0.0494(9) & -0.0462(17) & -0.0446(13)\\
1.0  & -0.0467(9) &  -0.0443(8) & -0.0414(15) & -0.0398(12)\\
1.2  & -0.0422(9) &  -0.0399(8) & -0.0372(13) & -0.0358(12)\\
1.7  & -0.0340(7) &  -0.0321(7) & -0.0299(10) & -0.0286(10)\\
2.0  & -0.0308(5) &  -0.0291(6) & -0.0271(10) & -0.0258(10)\\
3.0  & -0.0234(4) &  -0.0220(5) & -0.0207(7) & -0.0196(8)\\
3.5  & -0.0209(4) &  -0.0197(5) & -0.0185(4) & -0.0175(5)\\
4.0  & -0.0189(4) &  -0.0178(5) & -0.0170(6) & -0.0159(6)\\
7.0  & -0.0120(2) &  -0.0113(3) & -0.0112(9) & -0.0100(4)\\
10.0  &-0.0089(2) & -0.0084(3) & -0.0082(8) & -0.0074(3)\\
\end{tabular}
\end{center}
\end{table}

\begin{table}
\caption{The zeroth order decay amplitude of the $V$ 
meson, $a^{3/2}f^{(0)}_{V}\protect\sqrt{M_{V}}$, for all $aM_0$ and $\kappa_l$.
\label{decaytv}
}
\begin{center}
\begin{tabular}{ccccc}
$aM_0$ & 0.1385 & 0.1393 & 0.1401 & $\kappa_c$\\\hline
0.8  &  0.163(3)  &  0.152(3)  &  0.140(5)  &  0.131(4)\\
1.0  &  0.176(3)  &  0.164(3)  &  0.151(5)  &  0.142(4)\\
1.2  &  0.186(3)  &  0.174(3)  &  0.159(6)  &  0.150(4)\\
1.7  &  0.206(3)  &  0.191(4)  &  0.177(4)  &  0.165(5)\\
2.0  &  0.215(3)  &  0.201(4)  &  0.186(4)  &  0.173(5)\\
3.0  &  0.239(4)  &  0.223(5)  &  0.208(8)  &  0.194(6)\\
3.5  &  0.248(4)  &  0.232(5)  &  0.216(6)  &  0.202(7)\\
4.0  &  0.255(5)  &  0.239(6)  &  0.224(8)  &  0.209(7)\\
7.0  &  0.283(6)  &  0.264(6)  &  0.254(27)  &  0.230(10)\\
10.0  &  0.300(6)  &  0.279(7)  &  0.272(23)  &  0.241(9)\\
\end{tabular}
\end{center}
\end{table}

\begin{table}
\caption{The tree-level correction to the $V$ current,
$a^{3/2}f^{(1)}_{V}\protect\sqrt{M_{V}}$, for all
$aM_0$ and $\kappa_l$.
\label{deltavcor}
}
\begin{center}
\begin{tabular}{ccccc}
$aM_0$ & 0.1385 & 0.1393 & 0.1401 & $\kappa_c$\\\hline
0.8  & 0.0136(3)  &  0.0128(3)  &  0.0119(5)  &  0.0114(5)\\
1.0  & 0.0126(3)  &  0.0119(3)  &  0.0111(5)  &  0.0106(4)\\
1.2  & 0.0117(3)  &  0.0110(3)  &  0.0102(4)  &  0.0097(3)\\
1.7  & 0.0098(2)  &  0.0092(2)  &  0.0086(3)  &  0.0082(3)\\
2.0  & 0.0090(1)  &  0.0085(2)  &  0.0079(3)  &  0.0075(3)\\
3.0  & 0.0071(1)  &  0.0067(2)  &  0.0063(3)  &  0.0060(2)\\
3.5  & 0.0064(1)  &  0.0061(2)  &  0.0057(2)  &  0.0054(2)\\
4.0  & 0.0059(1)  &  0.0055(2)  &  0.0052(2)  &  0.0050(2)\\
7.0  & 0.0038(1)  &  0.0036(1)  &  0.0035(4)  &  0.0032(1)\\
10.0  & 0.0029(1)  &  0.0027(1)  &  0.0027(2)  &  0.0024(1)\\
\end{tabular}
\end{center}
\end{table}

\begin{table}
\caption{The one-loop correction to the $V$ current,
$a^{3/2}f^{(3)}_{V}\protect\sqrt{M_{V}}=a^{3/2}f^{(4)}_V\protect\sqrt{M_V}$, for all $aM_0$ and $\kappa_l$.
\label{deltavloop}
}
\begin{center}
\begin{tabular}{ccccc}
$aM_0$ & 0.1385 & 0.1393 & 0.1401 & $\kappa_c$\\\hline
0.8  & -0.0164(3)  &  -0.0154(3)  &  -0.0144(5)  &  -0.0136(6)\\
1.0  & -0.0144(3)  &  -0.0136(3)  &  -0.0126(5)  &  -0.0119(4)\\
1.2  & -0.0129(3)  &  -0.0121(3)  &  -0.0112(4)  &  -0.0107(4)\\
1.7  & -0.0103(2)  &  -0.0097(2)  &  -0.0091(3)  &  -0.0086(3)\\
2.0  & -0.0094(1)  &  -0.0088(2)  & -0.0082(3)  &  -0.0078(3)\\
3.0  & -0.0072(1)  &  -0.0068(2)  & -0.0064(3)  &  -0.0061(2)\\
3.5  & -0.0065(1)  &  -0.0061(2)  & -0.0058(2)  &  -0.0055(2)\\
4.0  & -0.0059(1)  &  -0.0056(2)  & -0.0053(2)  &  -0.0050(2)\\
7.0  & -0.0038(1)  &  -0.0036(1)  & -0.0035(4)  &  -0.0032(1)\\
10.0  &- 0.0029(1)  &  -0.0027(1)  & -0.0027(2)  &  -0.0024(1)\\
\end{tabular}
\end{center}
\end{table}

\begin{table}
\caption{The tree-level and renormalised $PS$ decay constants in
  lattice units, $a^{3/2}f_{PS}\protect\sqrt{M_{PS}}$, for all $aM_0$ and
  $\kappa_l=\kappa_c$.  $aM_0$ has been used for the argument of the
  logarithms appearing in the matching coefficients.
\label{deltaps}
}
\begin{center}
\begin{tabular}{cccc}
$aM_0$ & tree-level & $q^*=\pi$ & $q^*=1.0$  \\\hline
0.8  & 0.122(3)   &  0.121(3)&  0.120(3)  \\
1.0  & 0.135(4)   &  0.130(4)&  0.126(3)  \\
1.2  & 0.145(5)   &  0.137(4)&  0.132(4)  \\
1.7  & 0.163(5)   &  0.150(4)&  0.141(3)   \\
2.0  & 0.172(5)   &  0.155(4)&  0.145(4)  \\
3.0  & 0.192(5)   &  0.169(5)&  0.154(5)  \\
3.5  & 0.202(5)   &  0.175(5)&  0.158(4)   \\
4.0  & 0.208(7)   &  0.179(6)&  0.161(5)  \\
7.0  & 0.229(8)   &  0.195(6)&  0.174(5)  \\
10.0  & 0.242(9)    &  0.206(7)&  0.184(6) \\
\end{tabular}
\end{center}
\end{table}

\begin{table}
\caption{The tree-level and renormalised $V$ decay constants in
lattice units, $a^{3/2}f_{V}\protect\sqrt{M_{V}}$, for all $aM_0$ and $\kappa_l=\kappa_c$.
$aM_0$ has been used for the argument of the logarithms
appearing in the matching coefficients.
\label{deltav}
}
\begin{center}
\begin{tabular}{cccc}
$aM_0$ & tree-level & $q^*=\pi$ & $q^*=1.0$  \\\hline
0.8  & 0.143(4)  &  0.134(4)&  0.128(4)   \\
1.0  & 0.152(4)    &  0.139(4)&  0.131(4) \\
1.2  & 0.159(5)  &  0.143(4)&  0.134(4)   \\
1.7  & 0.173(5)    &  0.152(5)&  0.139(4)  \\
2.0  & 0.181(6)  &  0.156(5)&  0.141(4)   \\
3.0  & 0.200(6)    &  0.168(5)&  0.148(5) \\
3.5  & 0.207(7)  &  0.173(6)&  0.151(6)   \\
4.0  & 0.214(8)    &  0.177(6)&  0.154(6) \\
7.0  & 0.232(9)   &  0.190(8)&  0.164(7)  \\
10.0  & 0.244(9)    &  0.199(7)&  0.170(6) \\
\end{tabular}
\end{center}
\end{table}

\begin{table}
\caption{The corrections to the zeroth order pseudoscalar and vector
  matrix elements expressed as a ratio to $\left<J^{(0)}\right>$.  The
  order at which each correction contributes is indicated.  $aq^*=1.0$
  and $aM_0=2.0$, which is close to the bare $b$ quark mass for
  $a^{-1}=2.0$~GeV. Note that $\zeta_A$ and $\zeta_V$ are set to zero.
  The statistical errors are less than $1\%$.
\label{contribs}}
\begin{center}
\begin{tabular}{ccccccc}
 & $\frac{\left<J^{(1)}\right>}{\left<J^{(0)}\right>}$ &
 $\alpha\rho_0$ &
 $\frac{\alpha\rho_1\left<J^{(1)}\right>}{\left<J^{(0)}\right>}$ &
 $\frac{\alpha\rho_2\left<J^{(2)}\right>}{\left<J^{(0)}\right>}+\frac{\alpha\left<J^{disc}\right>}{\left<J^{(0)}\right>}$ & $\frac{C_3\left<J^{(3)}\right>}{\left<J^{(0)}\right>}$ & $\frac{C_4\left<J^{(4)}\right>}{\left<J^{(0)}\right>}$ \\
& $O(1/M)$ & $O(\alpha)$ & $O(\alpha/M)$ & $O(\alpha/M)+O(a\alpha)$ & $O(\alpha/M)$ & $O(\alpha/M)$ \\\hline
$f_{PS}\sqrt{M_{PS}}$& -13 & -9 & +1 & -7 & - & - \\
$f_{V}\sqrt{M_{V}}$ &+4  & -14 & +1 & -6 & -1 & -1 \\
\end{tabular}
\end{center}
\end{table}

\begin{table}
\caption{The pseudoscalar and vector decay constants in MeV,
calculated with and without renormalisation and converted to physical
units using $a^{-1}=2.0$ and $2.4$~GeV. Note that $aM_0^b=2.0$ for
$a^{-1}=2.0$~GeV, while $aM_0^b=1.8$ for $a^{-1}=2.4$~GeV.
Statistical errors, only, are shown except for those involving
$f_{B_s}$ and $f_{B_s^*}$, where the error due to the uncertainty in
$\kappa_s$ is also given. }
\label{work}
\begin{center}
\begin{tabular}{ccccc}
 & $a^{-1}$~GeV  & tree-level & $q^*=\pi$ & $q^*=1.0$\\\hline
$f_{B}$ & $2.0$ & 217(6)     &   192(5)        &   180(5)       \\
        & $2.4$  &  269(8)         &   244(8)        &   231(8)       \\
$f_{B_s}$ & 2.0 &  247(5)(-5)  &   222(2)(-5)   & 208(2)(-5)   \\
          & 2.4 &  310(6)(-6)  &   284(4)(-6)   & 266(4)(-6)   \\
$f_{B^*}$ & 2.0 &   223(6)     &  189(6)         & 173(5)         \\
          & 2.4 &   280(8)    &  248(8)         &  224(6)        \\
$f_{B^*_s}$ & 2.0 &  256(4)(-3) & 224(4)(-4) & 202(4)(-4) \\
          & 2.4 &  323(6)(-6) &  285(5)(-7)  & 261(5)(-5)  \\
$f_{B_s}/f_{B}$ & - &   1.14(2)(-2)        &   1.14(2)(-2)                & 1.14(2)(-2)           \\
$f_{B_s^*}/f_{B^*}$ & - & 1.14(2)(-2) & 1.14(2)(-2) & 1.14(2)(-2)\\
\end{tabular}
\end{center}
\end{table}

\begin{table}
\caption{The dependence of the $O(1/M)$ slope of
$a^{3/2}f\protect\sqrt{M_{PS}}$ on the light quark mass in lattice units.
\label{fcoef}}
\begin{center}
\begin{tabular}{cc}
$\kappa_l$ & $aC_1/C_0$\\\hline 
$0.1385$ & 1.7(5) \\
$0.1393$ & 1.0(4)\\
$0.1401$ & 1.1(3)\\
$\kappa_c$ & 1.1(3)\\
\end{tabular}
\end{center}
\end{table}

\begin{table}
\caption{The slope of $a^{3/2}f\protect\sqrt{M}_{PS}$ in lattice units at
tree-level and when renormalisation is included. }
\label{sloperenorm}
\begin{center}
\begin{tabular}{cccc}
 & $a^{3/2}C_0$ & $a^{5/2}C_1$ & $aC_1/C_0$\\\hline
tree-level & 0.27(1) & -0.3(1) & -1.1(3)\\
$q^*=\pi$ & 0.192(8) & -0.09(2) & -0.5(1) \\
$q^*=1.0$ & 0.150(6) & -0.00(2) & -0.0(1) \\
\end{tabular}
\end{center}
\end{table}

\begin{table}
\caption{The $O(1/M)$ slope found for various combinations of the
tree-level $PS$ and $V$ decay constants in lattice units.
$\kappa_l=\kappa_c$.
\label{ftab}}
\begin{center}
\begin{tabular}{ccc}
Quantity & $aC_1/C_0$ & Results\\\hline 
$f_{PS}\sqrt{M_{PS}}$ &
$G_{kin}+6G_{hyp}+G_{corr}/2$ & $-1.1$(3) \\
$\overline{f\sqrt{M}}$ & $G_{kin}$ & $-0.9$(3) \\
$\frac{f_{PS}^{(0)}\sqrt{M_{PS}}}{f_V^{(0)}\sqrt{M_V}}$ & $8G_{hyp}$ & 0.36(3)
\\ 
$\frac{f_{PS}\sqrt{M_{PS}}}{f_V\sqrt{M_V}}$ &
$8G_{hyp}+2G_{corr}/3$ & $-0.14$(4) \\ \hline
Quantity & $aC_0$ &
Results\\ \hline
$\frac{2M_0f^{(1)}_{PS}\sqrt{M_{PS}}}{f^{(0)}_{PS}\sqrt{M_{PS}}}$
& $-G_{corr}$ & $-0.62$(1) \\
\end{tabular}
\end{center}
\end{table}

\begin{table}
\caption{A comparison of the decay constants of the $B$ and $B_s$
  mesons from this work and those on quenched configurations at
  $\beta^{n_f=0}=6.0$~\protect\cite{arifajunk}.  The results have been
  converted into physical units~(MeV) using $a^{-1}$ from $m_\rho$.
  The errors are purely statistical, with the exception of
  $f_{B_s}/f_B$ where the second error comes from the uncertainty in
  $\kappa_s$. Note that the $O(1/M^2)$ current corrections included in
  reference~\protect\cite{arifajunk} have been omitted from the
  quenched results in order to better compare with the $n_f=2$ results.
\label{comp_pev}}
\begin{center}
\begin{tabular}{cccc}
 & & $\beta^{n_f=0}=6.0$ & $\beta^{n_f=2}=5.6$ \\\hline
$f_B$ & $q^*=1.0$ &  145(10) &  180(5)    \\
    & $q^*=\pi$ &   154(10) &  192(5)    \\
    & tree-level &  176(11)      &  217(6)  \\
$f_{B_s}$ & $q^*=1.0$ &175(6)(+7)  & 208(2)(-4)   \\
    & $q^*=\pi$ & 186(6)(+7)   &   222(2)(-4)  \\
    & tree-level &  205(6)(+7)      &  247(2)(-4)  \\
$f_{B_s}/f_{B}$ &   &  1.20(4)(+4) &    1.14(2)(-2)  \\
\end{tabular}
\end{center}
\end{table}

\begin{table}
\caption{The operators and corresponding quantum numbers 
used for the computation of the spectrum of the heavy-light mesons.
\label{ops}
}
\begin{center}
\begin{tabular}{cc}
$^{2S+1}L_J$ ($J^{PC}$)& $q^\dagger \Gamma Q$ \\\hline
 ${^1S}_0\;(0^{-+})$ & $q^\dagger\hat{I}Q$  \\
 ${^3S}_1\;(1^{--})$ & $q^\dagger\sigma_i Q $   \\
 ${^1P}_1\;(1^{+-})$ & $q^\dagger\Delta_i Q $  \\
 ${^3P}_0\;(0^{++})$ & $q^\dagger\sum_j  \Delta_j  \sigma_j Q$   \\
 ${^3P}_1\;(1^{++})$ & $ q^\dagger(\Delta_i \sigma_j - \Delta_j \sigma_i )Q$  \\
 ${^3P}_2\;(2^{++})$ & $ q^\dagger(\Delta_i \sigma_i - \Delta_j \sigma_j )Q$  \\
                     & $ q^\dagger(\Delta_i \sigma_j + \Delta_j \sigma_i )Q$  $\qquad$ ($i \neq j$)  \\
\end{tabular}
\end{center}
\end{table}

\begin{table}
\caption{The ground state pseudoscalar meson energies, $aE(^1S_0)$, extracted from
  $n_{exp}=2$ fits to $C(l,1)$ and $C(l,2)$ correlators. The mass shift needed
  to convert the simulation energies to the meson mass is also given.
\label{energies}
}
\begin{center}
\begin{tabular}{cccccc}
$aM_0$ & $\Delta$      &  $0.1385$ & $0.1393$ & $0.1401$ & $\kappa_c$\\\hline
0.8  &0.89(2)&  0.420(4)  &  0.395(4)  & 0.367(7)  & 0.347(7)  \\
1.0  &1.09(3)&  0.465(4)  &  0.440(5)  & 0.412(8)  & 0.394(8) \\
1.2  &1.27(2)&  0.492(4)  &  0.467(5)  & 0.439(8)  & 0.421(9) \\
1.7  &1.76(2)&  0.524(5)  &  0.500(5)  & 0.473(8)  & 0.453(9) \\
2.0  &2.07(2)&  0.534(5)  &  0.509(5)  & 0.483(8)  & 0.463(10)\\
3.0  &2.58(2)&  0.548(5)  &  0.524(6)  & 0.499(9)  & 0.478(7) \\
3.5  &3.10(3)&  0.551(5)  &  0.528(6)  & 0.503(9)  & 0.483(9) \\
4.0  &3.64(3)&  0.553(5)  &  0.530(7)  & 0.507(9)  & 0.486(9) \\
7.0  &6.61(12)&  0.558(7)  &  0.536(7)  & 0.515(10)  & 0.492(10)\\
10.0 &9.43(22)&  0.560(8)  &  0.536(8)  & 0.515(11)  & 0.491(12)\\
\end{tabular}
\end{center}
\end{table}

\begin{table}
\caption{The first excited state pseudoscalar meson energies, $aE(2^1S_0)$, extracted from $n_{exp}=2$ fits to $C(l,1)$ and $C(l,2)$ correlators.
\label{energies2}
}
\begin{center}
\begin{tabular}{ccccc}
$aM_0$       & $0.1385$ & $0.1393$ & $0.1401$ & $\kappa_c$\\\hline
0.8  &   0.70(5)  & 0.68(5)  & 0.68(6)  & 0.68(8) \\
1.0  &   0.73(4)  & 0.72(5)  & 0.71(5)  & 0.70(7)\\
1.2  &   0.75(3)  & 0.74(4)  & 0.72(5)  & 0.71(6)\\
1.7  &   0.77(3)  & 0.76(4)  & 0.74(5)  & 0.73(6)\\
2.0  &   0.78(3)  & 0.77(4)  & 0.74(5)  & 0.73(5)\\
3.0  &   0.78(3)  & 0.77(3)  & 0.74(4)  & 0.73(5)\\
3.5  &   0.78(3)  & 0.76(3)  & 0.74(5)  & 0.73(5)\\
4.0  &   0.77(3)  & 0.76(3)  & 0.74(4)  & 0.73(5)\\
7.0  &   0.76(3)  & 0.75(3)  & 0.75(4)  & 0.73(5)\\
10.0 &   0.74(3)  & 0.73(3)  & 0.73(5)  & 0.72(5)\\
\end{tabular}
\end{center}
\end{table}

\begin{table}
\caption{The ground state vector meson energies, $aE(^3S_1)$, extracted from $n_{exp}=2$ fits to $C(l,1)$ and $C(l,2)$ correlators.
\label{energiesV}
}
\begin{center}
\begin{tabular}{ccccc}
$aM_0$       &  $0.1385$ & $0.1393$ & $0.1401$ & $\kappa_c$\\\hline
0.8  &  0.448(4) &  0.422(5) &  0.391(8) &  0.372(9) \\
1.0  &  0.490(4) &  0.463(5) &  0.432(9) &  0.415(9) \\
1.2  &  0.513(4) &  0.487(6) &  0.456(9) &  0.440(9) \\
1.7  &  0.541(4) &  0.515(6) &  0.485(9) &  0.470(9) \\
2.0  &  0.548(4) &  0.523(5) &  0.493(9) &  0.477(8) \\
3.0  &  0.558(4) &  0.534(5) &  0.507(9) &  0.488(8) \\
3.5  &  0.560(5) &  0.536(5) &  0.510(8) &  0.490(7) \\
4.0  &  0.561(5) &  0.537(5) &  0.512(8) &  0.492(8) \\
7.0  &  0.564(6) &  0.540(7) &  0.518(11) &  0.494(9)\\
10.0 &  0.565(10) &  0.541(9) &  0.519(12) &  0.495(13) \\
\end{tabular}
\end{center}
\end{table}

\begin{table}
\caption{The first excited state vector meson energies, $aE(2^3S_1)$, extracted from $n_{exp}=2$ fits to $C(l,1)$ and $C(l,2)$ correlators.
\label{energiesV2}
}
\begin{center}
\begin{tabular}{ccccc}
$aM_0$  & $0.1385$ & $0.1393$ & $0.1401$ & $\kappa_c$\\\hline
0.8  &  0.75(5) &  0.74(6) &  0.71(6) &  0.69(7)\\
1.0  &  0.78(5) &  0.77(6) &  0.74(6) &  0.72(7)\\
1.2  &  0.80(4) &  0.79(5) &  0.75(5) &  0.72(6)\\
1.7  &  0.81(4) &  0.81(4) &  0.76(5) &  0.75(6)\\
2.0  &  0.82(3) &  0.81(4) &  0.76(5) &  0.73(6)\\
3.0  &  0.81(3) &  0.80(4) &  0.76(5) &  0.74(6)\\
3.5  &  0.80(3) &  0.79(4) &  0.76(4) &  0.74(5)\\
4.0  &  0.80(3) &  0.79(4) &  0.76(4) &  0.75(5)\\
7.0  &  0.77(3) &  0.76(3) &  0.75(4) &  0.75(5)\\
10.0 &  0.74(3) &  0.74(4) &  0.74(4) &  0.73(5)\\
\end{tabular}
\end{center}
\end{table}

\begin{table}
\caption{The ground state energy of the $^1P_1$ state, $aE(^1P_1)$, for various $aM_0$
and all $\kappa_l$.
\label{tabp}
}
\begin{center}
\begin{tabular}{ccccc}
$aM_0$ & 0.1385 & 0.1393 & 0.1401 & $\kappa_c$\\\hline
1.0 & 0.79(1) & 0.77(1) & 0.75(1) & 0.74(1)\\
2.0 & 0.78(1) & 0.76(1) & 0.75(1) & 0.74(1)\\
4.0 & 0.76(1) & 0.75(1) & 0.74(1) & 0.71(1)\\
\end{tabular}
\end{center}
\end{table}

\begin{table}
\caption{The ground state energy of the $\Lambda_Q$, $aE(\Lambda_Q)$, extracted from
$n_{exp}=1$ fits to $C(l,1)$ for all $aM_0$ and $\kappa_l$.
\label{tabb}
}
\begin{center}
\begin{tabular}{ccccc}
$aM_0$ & 0.1385 & 0.1393 & 0.1401 & $\kappa_c$\\\hline
0.8 &  0.783(9) &  0.727(13) &  0.669(17) &  0.621(20)\\
1.0 &  0.826(10) &  0.773(13) &  0.716(17) &  0.671(18)\\
1.2 &  0.851(10) &  0.799(12) &  0.744(17) &  0.701(17)\\
1.7 &  0.880(9) &  0.832(11) &  0.779(17) &  0.739(18)\\
2.0 &  0.889(9) &  0.841(10) &  0.789(19) &  0.751(18)\\
3.0 &  0.902(9) &  0.858(10) &  0.804(17) &  0.772(18)\\
3.5 &  0.905(9) &  0.862(10) &  0.808(18) &  0.778(18)\\
4.0 &  0.907(9) &  0.865(12) &  0.811(19) &  0.782(17)\\
7.0 &  0.893(17) &  0.873(13) &  0.821(18) &  0.801(28)\\
10.0 &  0.886(23) &  0.878(12) &  0.828(24) &  0.819(26)\\
\end{tabular}
\end{center}
\end{table}

\vspace{5mm}

\newpage

\newpage

\newpage

\newpage

\newpage
\newpage

\newpage

\newpage

\newpage

\end{document}